\documentclass[a4paper,11pt]{article}
\pdfoutput=1 % if your are submitting a pdflatex (i.e. if you have
\usepackage{jheppub} % for details on the use of the package, please
\usepackage[T1]{fontenc} % if needed

\usepackage{ulem}  % \sout{ text }
\preprint{
\begin{minipage}{5cm}
KYUSHU-HET-150\\
KYUSHU-RCAPP-2015-03\\
EPHOU-15-012\\
UME-PP-001
\end{minipage}
}

\title{\boldmath Phenomenology of NMSSM in TeV scale mirage mediation}

\author[a]{Kei~Hagimoto,}
\author[b]{Tatsuo~Kobayashi,}
\author[a]{Hiroki~Makino,}
\author[a]{Ken-ichi~Okumura,}
\author[c]{and Takashi~Shimomura}

\affiliation[a]{Department of Physics, Kyushu University,\\ Fukuoka 812-8581,
Japan}
\affiliation[b]{Department of Physics, Hokkaido University,\\ Sapporo 060-0810, Japan}
\affiliation[c]{Faculty of Education and Culture, University of Miyazaki,\\
Miyazaki, 889-2192, Japan}

\abstract{
We study the next-to-minimal supersymmetric standard model (NMSSM) 
with the TeV scale mirage mediation, which is known 
as a solution for the little hierarchy problem in supersymmetry.
Our previous study showed that 125 GeV Higgs boson is realized
 with $\it O$(10)\% fine-tuning for 1.5 TeV gluino (1 TeV stop) mass. 
The $\mu$ term could be as large as 500 GeV without sacrificing the fine-tuning thanks to a cancellation mechanism.
The singlet-doublet mixing is suppressed by $\tan\beta$.
In this paper, we further extend this analysis. 
We argue that approximate scale symmetries play a role behind the suppression of
 the singlet-doublet mixing.
They reduce the mixing matrix to a simple form that is
 useful to understand the results of the numerical analysis.
We perform a comprehensive analysis of the fine-tuning including the singlet sector by introducing a simple formula for the fine-tuning measure. 
This shows that the singlet mass of the least fine-tuning
 is favored by the LEP anomaly for moderate $\tan\beta$.
We also discuss prospects for
 the precision measurements of the Higgs couplings at LHC and ILC
 and direct/indirect dark matter searches in the model.
}

\begin{document} 
\maketitle
\flushbottom

\section{Introduction}
\label{sec:intro}

The first run of the LHC has finished successfully.
The Higgs boson, the final missing piece of the 
standard model (SM), was discovered by ATLAS \cite{Aad:2012tfa} and CMS \cite{Chatrchyan:2012ufa} and its mass was measured as 
$m_h \approx 125$ GeV \cite{Aad:2015zhl}.
Subsequently, its spin, parity \cite{Chatrchyan:2012jja,Aad:2013xqa,Khachatryan:2014kca,Aad:2015rwa} and couplings to the SM bosons \cite{Moriond2015-Higgs-coupling-ATLAS,Khachatryan:2014jba} were measured 
 and appeared to have the property predicted by the SM
within the current experimental accuracy.
Now any new physics model must predict this almost SM-like Higgs boson.
On the other hand, no sign of the new particles beyond the SM was discovered.
This casts a serious doubt on the naturalness of
 the electroweak (EW) symmetry breaking and new physics models based on it.

In particular, supersymmetry (SUSY) has been one of the most 
popular physics beyond the SM to solve the hierarchy problem.
In the context of SUSY, the LHC Run I did not find expected superpartners, 
instead put the lower bound of superpartner masses as 
about 1.4 TeV for gluino, 1 TeV for the first/second generation squarks
 \cite{gluino_ATLAS, gluino_CMS, gluino3_ATLAS, gluino3_CMS} and 700 GeV for the third generation squarks \cite{stop_ATLAS, stop_CMS, sbottom_ATLAS, sbottom_CMS} in simplified models.
Thus, the little hierarchy between 
the Higgs mass and superpartner masses now becomes manifest.
Within the framework of the 
minimal supersymmetric standard model (MSSM), 
the $Z$-boson mass, $m_Z$, is obtained as 
\begin{equation}
m^2_Z  \simeq - 2 m_{H_u}^2 + \frac{2}{\tan^2 \beta} m_{H_d}^2
  - 2\mu^2 \   ,
\label{eq:mZ}
\end{equation}
where $m^2_{H_u}$ and $m^2_{H_d}$  are the soft SUSY breaking 
scalar mass squared of 
the up-sector and down-sector Higgs fields and 
$\mu$ is the supersymmetric mass.
For example, in the constrained MSSM the radiative 
corrections on $m^2_{H_u}$ are dominated by the gluino mass $M_3$, 
and obtained as $m^2_{H_u} \sim - M_3^2$.
Thus, if the gluino and stop masses are of ${\cal O}(1)$ TeV, 
we need fine-tuning among $m^2_{H_u}$, $m^2_{H_d}$ and $\mu^2$ 
to realize the experimental value of $m_Z$.
Furthermore, the stop mass is required to be of ${\cal O}(1)$ TeV or 
larger to lead to the Higgs mass $m_h \approx 125$ GeV.
If we further rely on the naturalness as a guiding principle,
 we need a mechanism of SUSY breaking which reconciles
 this little hierarchy with the notion of naturalness.  

The mirage mediation is one of the interesting mediation mechanisms 
of SUSY breaking \cite{Choi:2004sx,Choi:2005uz,Endo:2005uy} (also see \cite{Falkowski:2005ck}).
It is a mixture of the modulus mediation \cite{Kaplunovsky:1993rd} 
and the anomaly mediation \cite{Randall:1998uk}
with a certain ratio.
In the mirage mediation, the radiative corrections and 
the anomaly mediation contributions cancel each other at a 
certain energy scale, where SUSY spectrum appears as that of the pure 
modulus mediation.
Such an energy scale is called the mirage scale.
TeV scale mirage mediation sets this scale at TeV scale
and it was pointed out that the TeV scale mirage mediation 
can significantly ameliorate the above fine-tuning problem of 
the MSSM \cite{Choi:2005hd,Kitano:2005wc,Choi:2006xb}.\footnote{
Certain relations among the soft SUSY breaking parameters at the unification scale may be useful to ameliorate fine-tuning \cite{Chan:1997bi, Kane:2002ap, Abe:2007kf, Horton:2009ed}.
%Non-universal gaugino masses with a certain ratio may be useful to 
%ameliorate fine-tuning \cite{Abe:2007kf}.
}

Although the TeV scale mirage mediation is an attractive scenario
 to solve the little hierarchy problem, there are
 two unsatisfactory features in the MSSM.
First, a typical mass scale of the $B$-term is 
the gravitino mass ($10-100$ TeV)  
and a delicate cancellation between 
 different contributions is required to realize
 the correct EW symmetry breaking \cite{Choi:2005uz, Choi:2005hd, Nakamura:2008ey}.
Second, $125$ GeV Higgs mass is still difficult
 to achieve by ${\it O}$(10) \% tuning of parameters
 because the $A$-term for the top Yukawa coupling
 is fixed by the model, which prevents the stop mixing from taking the optimal value known as the maximal mixing.  
A simple solution for these problems is an extension to
 the next-to-minimal supersymmetric standard model (NMSSM) \cite{Choi:2005uz}.

The NMSSM is the minimal extension of the MSSM by adding a singlet 
superfield $S$ \cite{Fayet:1974pd} 
(see for review e.g. \cite{Ellwanger:2009dp, Maniatis:2009re}).
Here we impose the $Z_3$ symmetry, which does not allow 
the $\mu$-term, $\mu H_u H_d$, in the superpotential, 
where $H_u$ and $H_d$ denote the up and down-sector Higgs 
superfields.
Instead, the term such as $\lambda S H_u H_d$ is allowed.
After the scalar component of $S$ develops its vacuum expectation 
value (VEV), the effective $\mu$-term is generated as 
$\mu = \lambda \langle S \rangle $.
In the $Z_3$ symmetric NMSSM, all of the dimensionful parameters 
are originated from SUSY breaking.
Hence, the value of $\mu$ is also obtained by SUSY breaking effects.
That gives us a solution for the so-called $\mu$-problem  \cite{Kim:1983dt}.
The $B$-term in the MSSM is replaced with the $A$-term, $A_\lambda$ which 
has a scale of gaugino mass in the mirage mediation
 and solves the first problem in the MSSM.
The NMSSM has an additional Higgs quartic coupling, $\lambda^2$, 
in the scalar potential, 
which is helpful to increase the tree-level Higgs mass 
compared with one in the MSSM. This ameliorates the second problem
 if $\tan\beta$ is small.
Also the mixing with the singlet and doublet helps
 to raise the Higgs mass if the singlet is light,
 although it tightens the constraint on the singlet
 from the LEP Higgs search \cite{Schael:2006cr}
(For recent studies on the mixing effect in NMSSM,
 see \cite{Kang:2012sy, Cao:2012fz, Cao:2012yn, Kobayashi:2012ee, Jeong:2012ma, Jeong:2014xaa,Agashe:2012zq, Choi:2012he, Kowalska:2012gs, Gherghetta:2012gb,Gherghetta:2014xea,Barbieri:2013hxa,Badziak:2013bda,Farina:2013fsa}). 

The TeV scale mirage mediation was applied to 
the NMSSM in Ref. \cite{Kobayashi:2012ee}.
It was found that the fine-tuning is improved 
in the parameter region to realize $m_h \approx 125$ GeV.
The effective $\mu$ can be considerably larger than the EW scale
 without sacrificing the fine-tuning due to a cancellation mechanism, 
 although it is assumed to be around the EW scale
 in conventional natural SUSY models \cite{natural-SUSY}
\footnote{As a different approach in this direction, see \cite{Cohen:2015ala}.}.
The mixing between the singlet and doublet Higgs bosons is suppressed.
In addition, the Higgs sector as well as the neutralino sector 
has a rich structure compared with one in the MSSM by 
adding the singlet.
That leads to interesting aspects as  shown 
in Ref. \cite{Kobayashi:2012ee}.\footnote{See also Ref.~\cite{Asano:2012sv}.}
In this paper, we further extend the analysis in \cite{Kobayashi:2012ee}
including new phenomenological studies.
We show that approximate scale symmetries play an important roll to suppress
 the singlet-doublet mixing in the NMSSM. 
We discuss the suppression holds as far as $\kappa\approx 0$ and
$m_{S,H_u}^2<< m_{H_d}^2$.
% $\mu \approx A_\lambda \sin 2\beta/2$ is satisfied even
% if the symmetry is badly broken by the singlet soft mass as in our model. 
We introduce a simple formula for the fine-tuning measures and study the fine-tuning of the EW symmetry breaking in detail.
We show that the singlet mass of the least fine-tuning
 is favored by the LEP anomaly \cite{Schael:2006cr,Belanger:2012tt,Kobayashi:2012ee,Drees:2005jg,Dermisek:2005gg} when $\tan\beta$ is moderately large.
We also study other phenomenological aspects such as the precision measurement of the Higgs couplings and the dark matter search.

This paper is organized as follows.
In section 2, we give a brief review on 
the mirage mediation, and the TeV scale mirage mediation.
We introduce the $Z_3$ symmetric NMSSM in section 3 and
 apply the TeV scale mirage mediation in section 4.
In section 5, we study phenomenological aspects 
such as the mass spectrum, fine-tuning
 in the EW symmetry breaking, Higgs couplings and dark matter.
Section 6 is devoted to conclusion and discussion.
In Appendix A, we show explicitly 
initial conditions of soft parameters, 
which are induced through the mirage mediation  
in the NMSSM.

\section{TeV-scale mirage mediation}

Here we give a brief review on the mirage mediation \cite{Choi:2004sx}.
The mirage mediation is the mixture of the modulus mediation 
and the anomaly mediation with a certain ratio, 
which would be determined by the modulus stabilization mechanism 
and SUSY breaking mechanism.
In the mirage mediation, the gaugino masses are written by 
\begin{eqnarray}\label{eq:gaugino-mass}
M_a(M_{GUT}) = M_0 +\frac{m_{3/2}}{8 \pi^2}b_a g_a^2,
\end{eqnarray}
where $g_a$ and $b_a$ are the gauge couplings 
and their $\beta$ function coefficients, and $m_{3/2}$ denotes 
the gravitino mass.
We assume that the initial conditions of our 
SUSY breaking parameters are input at the GUT scale, 
$M_{GUT} =2\times 10^{16}$ GeV.
The first term, $M_0$, in the right hand side denotes 
the gaugino mass due to the pure modulus mediation, 
while the second term corresponds to the anomaly mediation 
contribution.
In addition, we can write the soft scalar mass $m_i$ of 
matter fields $\phi^i$  and 
the so-called $A$-terms of $\phi^i \phi^j \phi^k$ corresponding to the
Yukawa couplings $y_{ijk}$ as 
%\begin{eqnarray}
\begin{subequations}
\label{eq:A-m}
\begin{align}
A_{ijk}(M_{GUT}) & = a_{ijk}M_0 - (\gamma_i + \gamma_j + \gamma_k)\frac{m_{3/2}}{8\pi^2}, %\nonumber 
\\
m_i^2(M_{GUT}) & = c_i M_0^2 - \dot{\gamma_i}(\frac{m_{3/2}}{8\pi^2})^2
				- \frac{m_{3/2}}{8\pi^2} M_0 \theta_i,
\end{align}
\end{subequations}
%\end{eqnarray}
where
%\begin{eqnarray}
\begin{subequations}
\begin{align}
\gamma_i & = 2\sum_a g_a^2 C_2^a(\phi^i) - \frac{1}{2} \sum_{jk} |y_{ijk}|^2, %\nonumber 
\\
\theta_i & = 4\sum_a g_a^2 C_2^a(\phi^i) - \sum_{jk} a_{ijk} |y_{ijk}|^2, %\nonumber 
\\
\dot{\gamma_i} & = 8\pi^2 \frac{d\gamma_i}{d \ln \mu_R}.
\end{align}
\end{subequations}
%\end{eqnarray}
Here, $\gamma_i$ denotes the anomalous dimensions of $\phi^i$ and 
$C_2^a(\phi^i)$ denotes the quadratic Casimir 
corresponding to the representation of the matter field $\phi^i$.
In the right hand side, $a_{ijk}M_0$  and $c_i M_0^2$ 
denote the A-term and soft scalar masses squared due to 
the pure modulus mediation.
These coefficients, $a_{ijk}$ and $c_i$, are determined by 
modulus-dependence of the K\"ahler metric of $\phi^i$, 
 $\phi^j$ and $\phi^k$ as well as Yukawa couplings.
Indeed, by using the tree-level K\"ahler metric, 
the coefficient $c_i$ is explicitly calculated as a fractional number 
such as $0,1,1/2,1/3$, etc. \cite{Kaplunovsky:1993rd,Abe:2005rx,Choi:2006xb}.
We would have ${\cal O}(1/8\pi^2)$ of corrections on 
$c_i$ due to the one-loop corrections on the K\"ahler metric \cite{Choi:2008hn}.
Such a correction would be important when $c_i =0 $, 
but that is model-dependent.
Here, we consider the case with 
\begin{equation}
a_{ijk} = c_i + c_j +c_k.
\end{equation}
Such a relation is often satisfied for ${\cal O}(1)$ of Yukawa
couplings in explicit string-derived supergravity models 
\cite{Kaplunovsky:1993rd}.\footnote{See also \cite{Kawamura:1997cw}.}

It is convenient to use the following parameter \cite{Choi:2005uz},
\begin{equation}
\alpha \equiv \frac{m_{3/2}}{M_0 \ln(M_{pl}/m_{3/2})},
\end{equation}
to represent the ratio of the anomaly mediation 
to the modulus mediation. Here $M_{pl}$ is the reduced Planck scale.

The mirage mediation has a very important energy scale, that is, 
the mirage scale defined by, 
\begin{equation}
M_{\rm mir} = \frac{M_{GUT}}{(M_{pl}/m_{3/2})^{\alpha/2}} .
\end{equation}
The above spectrum of the gaugino masses at $M_{GUT}$ leads 
to  \cite{Choi:2005uz},
\begin{equation}\label{eq:gaugino-mir}
M_a(M_{\rm mir}) = M_0,
\end{equation}
at the mirage scale.
That is, the anomaly mediation contributions 
and the radiative corrections cancel each other, 
and the pure modulus mediation appears at the mirage scale.
Furthermore, the $A$-terms and the scalar mass squared
also satisfy 
\begin{equation}\label{eq:A-m-mir}
A_{ijk}(M_{\rm mir}) = (c_i+c_j+c_k)M_0, \qquad m^2_i(M_{\rm mir}) =c_iM_0^2,
\end{equation}
if the corresponding Yukawa couplings are small enough or 
if the following conditions are satisfied, 
\begin{equation}\label{eq:mir-condition}
a_{ijk}=c_i+c_j+c_k=1,
\end{equation}
for non-vanishing Yukawa couplings, $y_{ijk}$ \cite{Choi:2005uz}.

When $\alpha =2$, the mirage scale $M_{\rm mir}$ is 
around $1$ TeV.
Then, the above spectrum \eqref{eq:gaugino-mir} and 
\eqref{eq:A-m-mir} is obtained around the TeV scale.
That is the TeV scale mirage mediation scenario.
In particular, there would appear a large gap 
between $M_0$ and the scalar mass $m_i$ with $c_i \approx 0$.
We will apply the TeV scale mirage scenario to the NMSSM in section 4.

In the TeV scale mirage scenario, the stop mass squared 
becomes negative at high energy \cite{Lebedev:2005ge}, 
while it is positive 
at low energy below $10^{6}$ GeV.
Thus, the vacuum which breaks the EW symmetry
at the EW scale might be a local minimum, but instead  
there would be a color and/or charge breaking vacuum 
with  field values larger than $10^{6}$ GeV.
Here, we assume the thermal history of the Universe 
such that field values remain around the origin 
until the EW scale temperature.
In addition, we need to confirm that the tunneling rate is 
small enough, i.e. less than the Hubble expansion rate.
In Refs.~\cite{Riotto:1995am}, it has been shown that 
such a rate is small enough, as long as 
the squark/slepton masses squared are vanishing or positive 
around $10^4$ GeV.
This condition is satisfied in our TeV scale mirage mediation
scenario.

\section{NMSSM}

Here, we briefly review on the NMSSM, in particular its Higgs sector.
We extend the MSSM by adding a singlet chiral multiplet $S$ 
and imposing a $Z_3$ symmetry.
Then, the superpotential of the Higgs sector is written as 
\begin{equation}
W_{\rm Higgs} = - \lambda S H_u H_d + \frac{\kappa}{3}S^3.
\end{equation}
Here and hereafter, for $S$, $H_u$ and $H_d$ we use the convention that 
the superfield and its lowest component are 
denoted by the same letter. 
The full superpotential also includes the Yukawa coupling terms 
between the matter fields and the Higgs fields, which are 
the same as those in the MSSM.

The following soft SUSY breaking terms in the Higgs sector are 
induced, 
\begin{equation}
V_{\rm soft}= m^2_{H_u}|H_u|^2 + m^2_{H_d}|H_d|^2 
+m^2_S |S|^2-\lambda A_\lambda SH_uH_d + \frac{\kappa}{3}A_\kappa S^3+ h.c.
\end{equation} 
Then, the scalar potential of the neutral Higgs fields is 
given as 
\begin{eqnarray}
V &=& \lambda^2|S|^2(|H^0_d|^2+|H^0_u|^2)+|\kappa S^2- \lambda
H^0_dH^0_u|^2 +V_D \nonumber \\
& & +m^2_{H_u}|H_u|^2 + m^2_{H_d}|H_d|^2 
+m^2_S |S|^2-\lambda A_\lambda SH_uH_d + \frac{\kappa}{3}A_\kappa S^3+ h.c.,
\label{eq:higgs-potential}
\end{eqnarray}
with
\begin{equation}
V_D = \frac18 (g^2_1 + g^2_2)(|H^0_d|^2 - |H^0_u|^2)^2,
\end{equation}
where $g_1$ and $g_2$ denote the gauge couplings of U(1)$_{\rm Y}$  
and SU(2).
Similarly, there appear the soft SUSY breaking terms 
including squarks and sleptons as well as gaugino masses.
These are the same as those in the MSSM.

The minimum of the Higgs potential is obtained by analyzing 
the stationary conditions of the Higgs potential,
\begin{subequations}
\begin{align}
\frac{\partial V}{\partial H^0_d} &= \lambda^2 v \cos\beta ( s^2 + v^2 \sin^2 \beta )
  - \lambda \kappa v s^2 \sin\beta + \frac{1}{4} g^2 v^3 \cos\beta \cos2 \beta \nonumber \\
 &\quad + m_{H_d}^2 v \cos\beta - \lambda A_\lambda v s \sin\beta = 0, \label{eq:1sub1} \\
\frac{\partial V}{\partial H^0_u} &= \lambda^2 v \sin\beta ( s^2 + v^2 \cos^2 \beta )
  - \lambda \kappa v s^2 \cos\beta - \frac{1}{4} g^2 v^3 \sin\beta \cos2 \beta \nonumber \\
 &\quad + m_{H_u}^2 v \sin\beta - \lambda A_\lambda v s \cos\beta = 0, \label{eq:1sub2} \\
\frac{\partial V}{\partial S} &= \lambda^2 s v^2 + 2 \kappa^2 s^3 - \lambda \kappa v^2 s \sin 2\beta
 + m_S^2 s - \frac{1}{2} \lambda A_\lambda v^2 \sin 2\beta + \kappa A_\kappa s^2 = 0, \label{eq:1sub3}
\end{align}
\label{eq:1}
\end{subequations}
where $g^2=g^2_1+g^2_2$.
Here, we denote VEVs as 
\begin{equation}
v^2 = | \langle H^0_d \rangle |^2 + | \langle H^0_u \rangle |^2, \qquad \tan \beta = \frac{\langle H^0_u \rangle }{\langle H^0_d \rangle}, 
\qquad s = \langle S \rangle.
\end{equation}
Using the above stationary conditions, we obtain the $Z$ boson mass $m_Z^2=\frac12 g^2v^2$ as
\begin{equation}
m_Z^2 = \frac{1 - \cos 2\beta}{\cos 2\beta} m_{H_u}^2 - \frac{1 + \cos 2\beta}{\cos 2\beta} m_{H_d}^2
  - 2\mu^2, 
\end{equation}
where $\mu = \lambda s$.
For $\tan\beta \gg 1$, this equation becomes eq.~\eqref{eq:mZ}.
That is, 
this relation is the same as the one in the MSSM.
Thus, the natural values of $|m_{H_u}|$ and $|\mu|$ would be 
of ${\cal O}(100)$ GeV.
Furthermore, the natural value of $|m_{H_d}|/\tan \beta$ would be 
of  ${\cal O}(100)$ GeV or smaller.
Alternatively, $|\mu|$ and $|m_{H_d}|/\tan \beta$ could be larger 
than ${\cal O}(100)$ GeV when 
$\mu^2$ and $m_{H_d}^2/\tan^2 \beta$ are canceled each other 
in the above relation at a certain level.
Even in such a case, $|m_{H_u}|$ would be naturally of ${\cal O}(100)$ GeV.
On the other hand, other sfermion masses as well as gaugino 
masses must be heavy as the recent LHC results suggested.
To realize such a spectrum, we apply the 
TeV scale mirage mediation in the next section, 
where we take $c_{H_u}=0$ to realize a suppressed value of 
$|m_{H_u}|$ compared with $M_0$.

\section{TeV scale mirage mediation in NMSSM}

Here, we apply the TeV scale mirage mediation scenario 
to the NMSSM and 
study its phenomenological aspects.

\subsection{Model}

Soft SUSY breaking terms are obtained 
through the generic formulas \eqref{eq:gaugino-mass} 
and \eqref{eq:A-m} with taking $\alpha =2$.
For concreteness, we give explicit results of 
all the soft SUSY breaking terms  for 
the NMSSM in Appendix~\ref{app:soft}.
We concentrate on the Higgs sector as well as gauginos and stops.

A concrete model in the mirage mediation is fixed by choosing $c_i$.
We consider the following values of $c_i$ \cite{Kobayashi:2012ee},
\begin{equation}\label{eq:ci}
c_{H_d}=1, \qquad  c_{H_u}=0, \qquad c_S = 0, \qquad c_{t_L}=c_{t_R}=\frac12, 
\end{equation}
up to one-loop corrections 
for $H_d$, $H_u$, $S$, and left and right-handed (s)top fields, respectively.
This is the same assignment as the pattern II in Ref. \cite{Choi:2006xb} 
for the MSSM except for $c_S$.
Then, the soft parameters due to only modulus mediation 
contribution are given by 
\begin{eqnarray}
& & (A_t)_{\rm modulus}=(A_\lambda )_{\rm modulus}=M_0, \qquad (A_\kappa )_{\rm modulus} =0, \nonumber \\
& & (m^2_{H_d})_{\rm modulus}=M^2_0, \qquad (m^2_{\tilde t_L})_{\rm modulus} =
(m^2_{\tilde t_R})_{\rm modulus} =\frac12 M^2_0,  \\ 
& &  (m^2_{H_u})_{\rm modulus }=(m^2_S)_{\rm modulus}=0,   \nonumber 
\end{eqnarray}
up to one-loop corrections.
The above assignment of $c_i$ \eqref{eq:ci}
satisfies the condition, \eqref{eq:mir-condition} for 
the top Yukawa coupling and the coupling $\lambda$, but 
not for the coupling $\kappa$.
However, we do not consider a large value of $\kappa$ 
to avoid the blow-up of $\kappa$ and $\lambda$ as will be shown later.
Thus, we obtain the following values,
\begin{eqnarray}\label{eq:soft-at-mirage-1}
& & A_t \approx A_\lambda \approx M_0, \\ \nonumber 
&  &  m^2_{H_d} \approx M^2_0, \qquad m^2_{\tilde t_L} \approx m^2_{\tilde t_R} \approx \frac12 M^2_0, 
\end{eqnarray}
up to ${\cal O}(\kappa^2/8\pi^2)$ at the TeV scale.

Similarly, at the TeV scale we can obtain 
\begin{equation}\label{eq:soft-at-mirage-2}
m_{H_u}^2 \approx 0, \qquad 
m_{S}^2 \approx 0,
\end{equation}
up to ${\cal O}(M_0^2/8\pi^2)$, and 
\begin{equation}
A_\kappa \approx 0, 
\end{equation}
up to ${\cal {O}}(M_0/8\pi^2)$.
That is, values of $|A_\kappa|^2$, $m_{H_u}^2$ and 
$m_{S}^2$  are suppressed compared with 
$M_0^2$, and their explicit values depend on the one-loop corrections 
on the K\"ahler metric.
Thus, we use $A_\kappa$ 
as a free parameter, which must be small 
compared with $M_0$.
In addition, we determine the values of 
$m_{H_u}^2$, $m_S^2$ and $\mu~(=\lambda s)$ at the EW scale 
from the stationary conditions,  
\eqref{eq:1}, where we use the experimental value 
$m_Z = \frac{1}{\sqrt{2}} g v = 91.19$ GeV and $\tan \beta$ as a free parameter.

Through the above procedure, the parameters, $m^2_{H_u}$, $m^2_S$ and $\mu$, 
at the EW scale are expressed by $\tan\beta$, $m_{H_d}^2$, $A_\lambda$ as follows
\footnote{We are only interested in the natural spectrum. Thus we discard another solution of the quadratic equation, $\mu \approx A_\lambda \tan\beta/ 2$.}, 
%\begin{eqnarray}
\begin{subequations}
\label{eq:minimum}
\begin{align}
\mu & = \lambda \langle S \rangle = \frac{A_\lambda \tan\beta
}{2\left(1-\frac{\kappa}{\lambda}\tan\beta\right)}   \left\{1
-\sqrt{1-4 X}\right\} ,
%\nonumber
\\
m^2_S & = -2\left(\frac{\kappa}{\lambda}\right)^2\mu^2
-\left(\frac{\kappa}{\lambda}\right) A_\kappa \mu
+\frac{\lambda^2}{g^2}m_Z^2
\left\{\left(\frac{A_\lambda}{\mu}+2\frac{\kappa}{\lambda}\right)\sin 2\beta -2\right\},
\\
m_{H_u}^2 & =
\frac{\tan^2\beta-1}{\tan^2\beta}\left(\frac{m_{H_d}^2}{\tan^2\beta-1}-\mu^2-\frac{m_Z^2}{2}\right), %\nonumber
\end{align}
\end{subequations}
%\end{eqnarray}
where,   
\begin{equation}
X = \frac{m_{H_d}^2\left(1-\frac{\kappa}{\lambda}\tan\beta\right)}{ A_\lambda^2 \tan^2\beta}\left\{1 +\frac{\tan^2\beta}{\tan^2\beta+1}
\left(\frac{2\lambda^2}{g^2}-\frac{\tan^2\beta-1}{2\tan^2\beta}\right)\frac{m_Z^2}{m_{H_d}^2}\right\}. 
\end{equation}
For $\tan\beta \gg {\rm max}( 1, \kappa/\lambda)$, these parameters are approximated as 
\begin{subequations}
\begin{align}
& \mu = \lambda \langle S \rangle \sim   \frac{m^2_{H_d}}{A_\lambda \tan \beta},  \label{eq:app-1}\\  
& m^2_S \sim -2 \left(\frac{\kappa}{\lambda}\right)^2\left(\frac{m_{H_d}^2}{A_\lambda\tan\beta}\right)^2-\left(\frac{\kappa}{\lambda}\right)A_\kappa \left(\frac{m_{H_d}^2}{A_\lambda\tan\beta}\right)+2\frac{\lambda^2}{g^2}\frac{A_\lambda^2}{m_{H_d}^2} m_Z^2, \label{eq:app-2}\\
& m_{H_u}^2 \sim  \frac{m^2_{H_d}}{\tan^2 \beta} -
\frac{m^4_{H_d}}{A^2_\lambda \tan^2 \beta}- \frac{m^2_Z}{2} .
\label{eq:app-3}
\end{align}
\label{eq:app}
\end{subequations}
These formulas tell that the leading terms in the expansion by $\tan\beta$
 are ${\cal O}(m_Z^2)$ for $m_S^2$ and $m_{H_u}^2$ and the next-to-leading terms are ${\cal O}(M_0^2/\tan^2\beta)$ because $m_{H_d} \simeq A_\lambda \simeq M_0$.
If $\tan \beta ={\cal O}(10)$, the values of $\mu$ and $|m_{H_u}|$ could be of ${\cal O}(100)$ GeV 
while the other masses of the superpartners are of ${\cal O}(M_0)={\cal O}(1)$ TeV.
Thus, the fine-tuning problem can be ameliorated.
Actually, the first and the second terms in the last equation  
cancel each other for our choice of $c_i$. 
The next leading contributions are of ${\cal O}(m^2_{H_d}/\tan^4\beta)$ or
${\cal O}(m^2_{H_d}\mu/\tan^2\beta A_\lambda)$. 
Then, $m^2_Z$ is almost determined by $m^2_{H_u}$ alone and
insensitive to $\mu\approx M_0/\tan\beta$. 
This means that $\tan\beta \approx 3$ is enough to obtain
the fine-tuning of $|\partial \ln m_Z^2/\partial \ln m_{H_u}^2|^{-1}=m_Z^2/2m_{H_u}^2={\cal O}(100)$\% for $M_0\approx 1$ TeV.
Therefore $\mu$ can be as heavy as ${\cal O}(400)$ GeV without sacrificing the fine-tuning.
This cancellation originates in the structure of the doublet mass matrix,
\begin{equation}
{\cal L}_M = -\left( H_d, H_u^\ast\right)
{\cal M}^2_H
\left(
\begin{array}{c}
H_d^\ast \\
H_u
\end{array}
\right),
\end{equation}
where 
\begin{equation}
{\cal M}^2_H =
\left( 
\begin{array}{cc}
m_{H_d}^{2} +\mu^2 & -A_\lambda \mu \\
-A_\lambda \mu & m_{H_u}^2+\mu^2
\end{array}
\right)
\approx
\left( 
\begin{array}{cc}
M_0^2 +\mu^2 & -M_0 \mu \\
-M_0 \mu & \mu^2
\end{array}
\right).
\label{eq:doublet-mass-matrix}
\end{equation}
Calculating the determinant of the mass matrix, it can be easily checked that  
the modulus mediated contribution $M_0$ cancels,  
${\rm det}({\cal M}^2_H) \approx \mu^4$ 
while the trace of the mass matrix is $M_0^2+2\mu^2$. 
This means that the heavy mode has mass of
 ${\cal O}(M_0)$ and the mass of 
the light mode is suppressed as $\mu^2/M_0\approx \mu/\tan\beta$.
Therefore a flat direction appears along $H_u/H_d \approx M_0/\mu \approx \tan\beta$.
This mechanism was previously observed in \cite{Choi:2006xb}
in the context of the MSSM where $A_\lambda$ is replaced with the $B$-term.
In the mirage mediation, $B$-term is a remnant of
the fine-tuned cancellation between the terms of ${\cal O}(m_{3/2})$.
Thus the relation $m_{H_d} \approx B$ is subject to uncontrolled corrections.
However, in the NMSSM, the relation is well controlled up to the leading contribution of the modulus mediation. 

This cancellation is also important for raising the SM-like Higgs boson mass radiatively.
In the NMSSM, singlino-higgsino loop with $\lambda={\it O}(1)$ could contribute to the radiative correction of the Higgs boson mass in addition to the top quark loop \cite{Nakayama:2011iv, Jeong:2012ma}. 
If the singlet scalar is as light as the singlino or higgsino, this correction is canceled by the scalar loops because of SUSY. However, the structure of the mass matrix \eqref{eq:doublet-mass-matrix} tells the coupling $|S|^2 h^2$ $(\sim \mu^2 h^2)$ vanishes due to the SUSY breaking effect. This means the correction could be sizable even if the singlet scalar is light as in our case.
The renormalization group equation in the effective theory roughly gives 
\begin{eqnarray}
\Delta m_h^2 &\simeq& 
\frac{v^2}{16\pi^2}\left[
-4\lambda^2 \frac{m_h^2}{v^2} +8\lambda^4
-4\lambda^4\left(1-\frac{A_\lambda^2}{m_H^2} \right)\right.
\nonumber\\
&&
\left.
\phantom{\frac{v^2}{16\pi^2}}
-16\lambda^4\frac{A_\lambda^2}{m_H^2}
\left(1-\frac{A_\lambda^2}{m_H^2}\right)\frac{1}{\tan^2\beta}
\right]\ln\left(\frac{m_H}{m_h}\right)\nonumber\\
&\simeq&\frac{v^2}{16\pi^2}\left[
-4\lambda^2 \frac{m_h^2}{v^2} +8\lambda^4\right]\ln\left(\frac{m_H}{m_h}\right),
\nonumber\\
&\simeq&\left(56 \,{\rm GeV}\right)^2 \lambda^2\left(\lambda^2-0.26\right)\left(1+1.1 \log_{10}\left[\frac{m_H}{1 \,{\rm TeV}}\right]\right), 
\label{eq:nmssm-higgs-radiative-correction}
\end{eqnarray}
where $m_h$ ($m_H$) is the mass of the light (heavy) doublet. In our case, $m_H\approx A_\lambda\approx M_0$.
This correction amounts $\sim 5$ GeV for $m_h = m_Z$ and $\lambda=0.7$.

%*****
Potentially, the mixing between the light doublet and the singlet may invalidate the above discussion on the hierarchy of the Higgs boson masses. In models with $m_S^2 \approx 0$ and $\kappa << 1$, this mixing is automatically suppressed by $\lambda v/A_\lambda /\tan\beta$ if we neglect the gauge interaction (This suppression is noticed in \cite{Kang:2012sy} without a reason.).  
In this limit, the Higgs action has the following approximate scaling symmetry:
%\begin{eqnarray}
\begin{subequations}
\label{eq:scale}
\begin{align}
H_u(x) & = e^{2\phi} H_u^\prime (e^\phi x) %\nonumber
\\
H_d(x) & = e^{2\phi} H_d^\prime (e^\phi x) %\nonumber
\\
S(x) & = S^\prime (e^\phi x),
\end{align}
\end{subequations}
%\end{eqnarray}
where $\phi$ is a free scaling parameter. This symmetry is explicitly broken by the K\"ahler potential,
\begin{equation}
{\cal K}_S = S^\dag S, 
\end{equation}
the D-term potential and all the kinetic terms.
After the Higgs bosons develop the vacuum expectation values, the symmetry is spontaneously broken and the Nambu-Goldstone boson appears, which corresponds to the light doublet\footnote{Similar symmetry can be defined in MSSM, which forbids all the quartic potential. In such a case, the non-trivial VEV is not possible unless the mass parameters are tuned so that a massless mode appears in that direction.}. 
Thus it is inherently mass eigenstate without the singlet component.
The symmetry breaking K\"ahler potential generates the F-term potential: 
\begin{equation}
V_F^S = \lambda^2 | H_u |^2 | H_d |^2, 
\end{equation}
which does not affect the singlet sector at all.
This potential gives additional terms in the doublet mass matrix which 
we neglected before:
\begin{equation}
\frac{\partial^2 V_F^{S}}{\partial H_u^\ast \partial H_u} = \lambda^2 |H_u|^2,\qquad
\frac{\partial^2 V_F^{S}}{\partial H_d^\ast \partial H_d} = \lambda^2 |H_d|^2,\qquad
\frac{\partial^2 V_F^{S}}{\partial H_d^\ast \partial H_u} = \lambda^2 H_u^\ast H_d.
\label{eq:symmetry-breaking}
\end{equation}
The contributions to the diagonal elements are eliminated by the minimization conditions \eqref{eq:1sub1}, \eqref{eq:1sub2}.
The CP even part of the mass matrix is given by 

\begin{eqnarray}
&&{\cal M}^2_S \equiv\frac{\partial^2 V}{\partial \phi_i \partial \phi_j}  \nonumber\\
&&=
\left(
\begin{array}{ccc}
A_\lambda^2\sin^2\beta & (-A_\lambda^2 +  2\lambda^2 v^2) \sin\beta\cos\beta & \lambda A_\lambda v \cos 2\beta \sin\beta \\
(-A_\lambda^2 +  2 \lambda^2 v^2) \sin \beta\cos\beta & A_\lambda^2\cos^2\beta & -\lambda A_\lambda v \cos 2\beta \cos\beta \\
\lambda A_\lambda v \cos 2\beta \sin\beta &-\lambda A_\lambda v \cos 2\beta \cos\beta & \lambda^2v^2  
\end{array}
\right) \nonumber\\
&&=
\left(
\begin{array}{ccc}
\cos\beta & -\sin\beta & 0 \\
\sin\beta & \cos\beta & 0 \\
0 & 0 & 1
\end{array}
\right)
\left(
\begin{array}{ccc}
0 & 0 & 0 \\
0 & A_\lambda^2 & -\lambda A_\lambda v \cos 2\beta \\
0 & -\lambda A_\lambda v \cos 2\beta & \lambda^2 v^2
\end{array}
\right)
\left(
\begin{array}{ccc}
 \cos\beta & \sin\beta & 0 \\
 -\sin\beta & \cos\beta & 0 \\
0 & 0 & 1
\end{array}
\right)\nonumber\\
&&\phantom{=}+
\left(
\begin{array}{ccc}
0 & \lambda^2 v^2 \sin 2\beta & 0 \\
\lambda^2 v^2 \sin 2\beta & 0 & 0 \\
0 & 0  & 0 
\end{array}
\right), 
\label{eq:CP-even-mass-matrix}
\end{eqnarray}
where $\phi_i = \sqrt{2}(Re H_d -\langle H_d\rangle,\, Re H_u -\langle H_u \rangle,\, Re S -\langle S \rangle )$ and
we used the minimization condition \eqref{eq:1sub3}, e.g. $\mu = \frac{1}{2}A_\lambda \sin 2 \beta$ which is not affected by the gauge coupling.
The last term which corresponds to the off-diagonal contribution in \eqref{eq:symmetry-breaking} mixes the light and heavy doublets and slightly changes the diagonalization angle of the CP even mass matrix, $\beta^\prime\equiv \pi/2-\alpha$ from the direction of the Nambu-Goldstone boson, $\beta$ as
\begin{equation}
\frac{1}{2} \tan 2 \beta^\prime = \frac{1}{2} \tan 2\beta \left(1-2\frac{\lambda^2 v^2 }{A_\lambda^2}\right),\qquad
\frac{2 \delta \beta}{\sin 2\beta \cos 2\beta} \approx -2 \frac{\lambda^2 v^2}{A_\lambda^2}, 
\end{equation}
where $\delta \beta \equiv \beta^\prime -\beta$.
This small rotation generates a non-zero entry in the singlet-light doublet mixing terms (1-3, 3-1 elements) in the CP even Higgs mass matrix.   
The 2-3 (3-2) element is $-\lambda A_\lambda v \cos 2\beta$. Thus the rotation yields $\approx 2 \lambda^3 v^3 /A_\lambda \sin 2\beta \cos^2 2\beta$ in 1-3 (3-1) element. The mass of the light doublet and that of the singlet are order of $\lambda v$. Then the mixing angle between the light doublet and the singlet is order of $\lambda v/A_\lambda/\tan\beta$.  
Similar discussion applies for the D-term potential.
The diagonalization matrix ${\cal S}$ of ${\cal M}^2_S$ is given by 
\begin{eqnarray}
{\cal S} {\cal M}^2_S {\cal S}^\dag&=& 
 \left(
\begin{array}{ccc}
m_Z^2\cos^2 2\beta + \lambda^2 v^2 \sin^2 2\beta & 0 & 0 \\ 
0 & A_\lambda^2 +(m_Z^2-\lambda^2 v^2)\sin^22\beta & 0 \\ 
0 & 0 & \lambda^2 v^2 \sin^2 2\beta 
\end{array}
\right)\nonumber\\
&&
+{\cal O} \left(\frac{v}{A_\lambda}\right)^4, \\
{\cal S}^\dag &=& 
\left(
\begin{array}{ccc}
\displaystyle
\cos\beta & -\sin\beta 
& \lambda \frac{v}{A_\lambda} \sin\beta
\\
\sin\beta & \cos\beta &
\lambda \frac{v}{A_\lambda} \cos\beta\\
-\lambda \frac{v}{A_\lambda} \sin 2\beta
 & -\lambda \frac{v}{A_\lambda} \cos 2\beta & 
1
\end{array}
\right)
+{\cal O} \left(\frac{v}{A_\lambda}\right)^2
,
\end{eqnarray}
at tree level.
Note that the mass of the light doublet originates in the symmetry breaking terms.
Non-vanishing $\kappa$ and $m_S^2$ are expected to add corrections of order $\kappa$ and $m_S^2/\lambda^2 v^2$ to ${\cal S}$.
Radiative corrections to the Higgs potential are another source of the singlet-doublet mixing because the kinetic terms and the quark-lepton Yukawa couplings also break the scaling symmetry.
 Actually, this is not the end of story. 
The Higgs action has another approximate scale symmetry if $\kappa=m_{H_u}^2=0$, in which the transformations of $H_u$ and $S$ are interchanged in eq.\eqref{eq:scale}. The VEVs of $S$ and $H_d$ spontaneously break this symmetry and the Nambu-Goldstone boson appears, which consists of a combination of $S$ and $H_d$ corresponding to the singlet-like Higgs boson. 
In the $\kappa=m^2_{S,H_u}=\langle H_d\rangle=0$ limit, the real parts of $S$ and $H_u$ are the Nambu-Goldstone bosons (mass eigenstates). After we turn on the symmetry breaking terms, the 1-3 and 3-2 elements of the mixing matrix ${\cal S}$ must break the both symmetries because $S$ or $H_u$ is still the Nambu-Goldstone boson if one of the symmetries is preserved. 
In addition, the mixing between $S$ and $H_u$ must vanish in the Lagrangian, eq.\eqref{eq:higgs-potential} if $H_d$ decouples \cite{Cao:2012fz}. Therefore the leading contributions to $({\cal S}^\dag)_{31}$, $({\cal S}^\dag)_{23}$ are given by $\kappa\langle S \rangle/M_0$, $\langle H_d \rangle/M_0$, $\langle H_u \rangle\langle S \rangle/M_0^2$, $\langle H_u\rangle m_{H_u}^2/M_0^3$, $\langle S\rangle m_S^2/M_0^3$ and $m_{H_u}^2 m_S^2/M_0^4$ at tree level. 
Our argument on the suppression of the singlet-doublet mixing generally holds for models with small $\kappa$ and $m_{S,\,H_u}^2<< m_{H_d}^2$. 
%In our model the first condition is actually badly broken due to the ${\it O}(M_0^2/8\pi^2)$ corrections. However, the requirement for the mixing pattern eq.(\ref{eq:CP-even-mass-matrix}) is $\mu = \frac{1}{2} A_\lambda \sin2\beta$ \cite{Kang:2012sy,Kobayashi:2012ee,Agashe:2012zq} and not directly related to $m_S^2$.  In our model, the condition $m_{H_d}^2=A_\lambda^2$ ensures $\mu \approx A_\lambda/\tan\beta$. Thus the singlet-doublet mixing is suppressed even for $m_S^2 \sim m_Z^2$ unless $\tan\beta \simeq 1$.

The CP-odd part of the mass matrix is diagonalized as 
\begin{eqnarray}
{\cal M}_P^2 &=& \frac{\partial^2 V}{\partial \eta_i \partial \eta_j} =
{\cal P}^\dag
\left(
\begin{array}{ccc}
0 & 0 & 0 \\
0 & A_\lambda^2+\lambda^2 v^2 & 0 \\
0 & 0 & 0 \\
\end{array}
\right)
{\cal P}, \\
{\cal P}^\dag &=& 
\left(
\begin{array}{ccc}
\cos\beta & \sin\beta \cos\gamma & \sin\beta \sin\gamma \\
-\sin\beta & \cos\beta \cos\gamma & \cos\beta \sin\gamma \\
0 & -\sin\gamma & \cos\gamma
\end{array}
\right),
\end{eqnarray}
in the $\kappa=0$ limit. Here $\eta_i = \sqrt{2} ( Im(H_d), Im(H_u), Im(S))$ 
 and $\tan\gamma = \lambda v/A_\lambda$. Two Nambu-Goldstone bosons appear for the broken electroweak and Pecci-Quinn symmetries. The former is absorbed by $Z$ boson and the latter acquires mass with non-vanishing $\kappa$.
In the following section, we show numerically the spectrum 
of our model.

\section{Spectrum and phenomenological aspects }

Here, we study numerically the spectrum and phenomenological aspects 
of our model. 
Our numerical calculation is performed with \texttt{NMSSMTools} package \cite{Ellwanger:2005dv}.
We choose input parameters as
 the squark and slepton masses, their tri-linear couplings, and parameters in the Higgs sector ($\lambda$, $\kappa$, $A_\lambda$, $A_\kappa$, $m_{H_d}$ and $\tan\beta$).
We first minimize the one-loop effective potential of our model 
at the SUSY scale and calculate the effective $\mu$-term.  
Then use it as an input of the \texttt{NMSSMTools} instead of $m_{H_d}$.
\texttt{NMSSMTools} allows various levels of sophistication in the calculation of the Higgs boson mass. To keep the consistency with the input, we choose the default option (\texttt{BLOCK MODSEL 8 0}) which corresponds to the one-loop calculation at SUSY scale. 
The small parameters $m^2_S$, $m^2_{H_u}$ are calculated so that we can obtain
 the correct EW symmetry breaking.
In the following, we only adopt the parameters which predict $|m^2_S|,\,|m^2_{H_u}|\lesssim M_0^2/8\pi^2$ within the range of the model ambiguity.
We choose the small parameter $A_\kappa$ as an input because the number of minimization conditions is not enough to predict it. 

\subsection{Higgs Mass Spectrum}

We first discuss the mass spectrum of the Higgs sector.
As described in the previous section, our model is in the decoupling region where the heavy doublet $\cos\beta H_u-\sin\beta H_d^\ast$ orthogonal to the flat direction has a mass much larger than the EW scale. 
Then this heavy doublet approximately consists of the heaviest CP-even, CP-odd and Charged Higgs bosons and they have degenerate masses $\approx M_0$ up to the effect of the EW symmetry breaking. 
Remaining light degrees of freedom are the first and second lightest CP-even Higgs bosons and the lightest CP-odd Higgs boson.

\begin{figure}[tbp]
%\begin{center}
\centering
\begin{tabular}{l @{\hspace{5mm}} r}
\includegraphics[height=48mm]{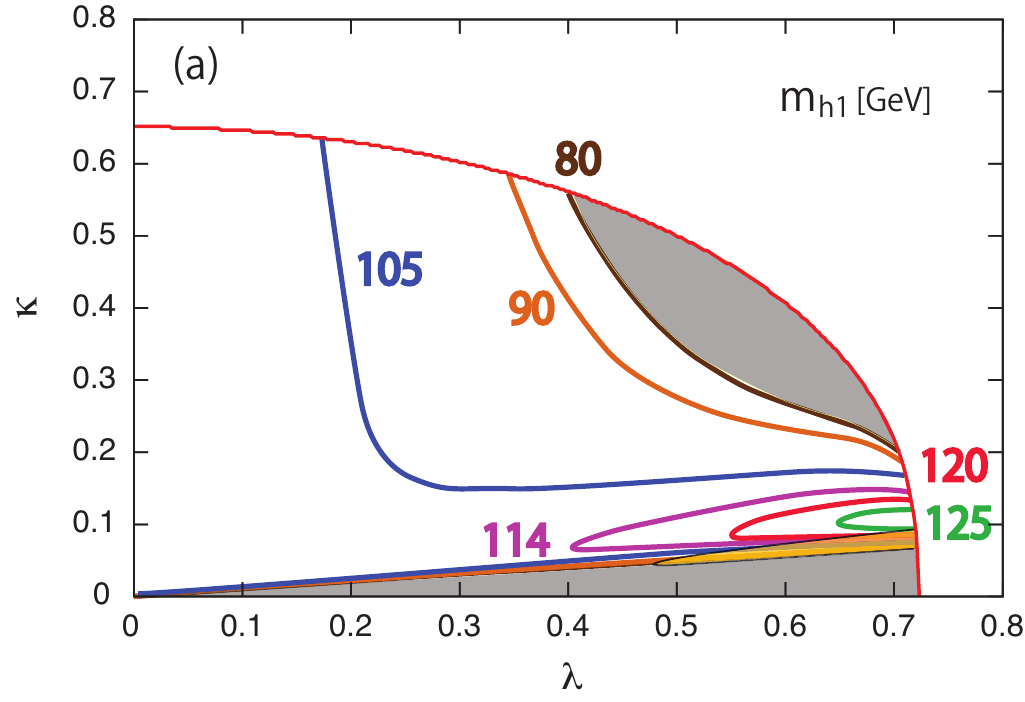} &
\includegraphics[height=48mm]{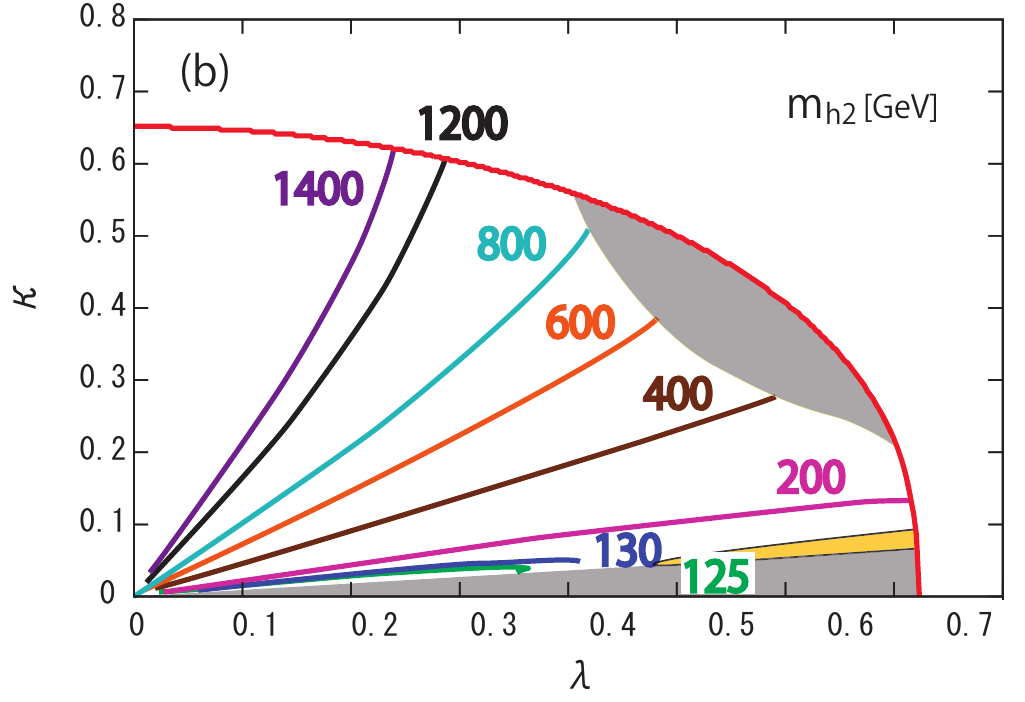} \\
\rule{0cm}{10mm} & \\
\includegraphics[height=48mm]{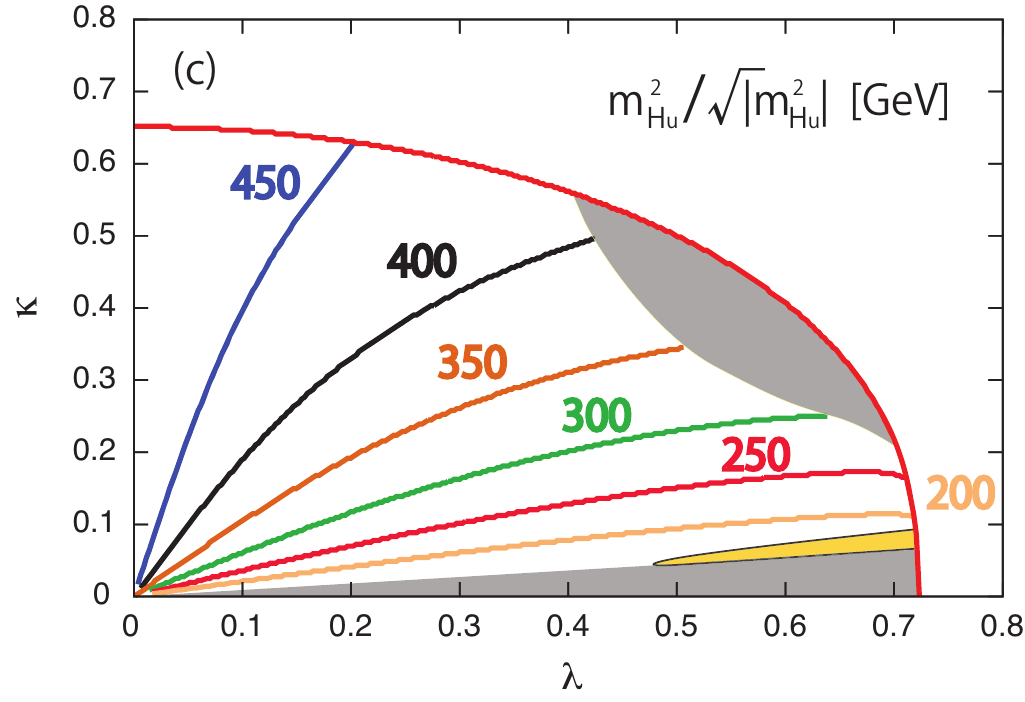} & 
\includegraphics[height=48mm]{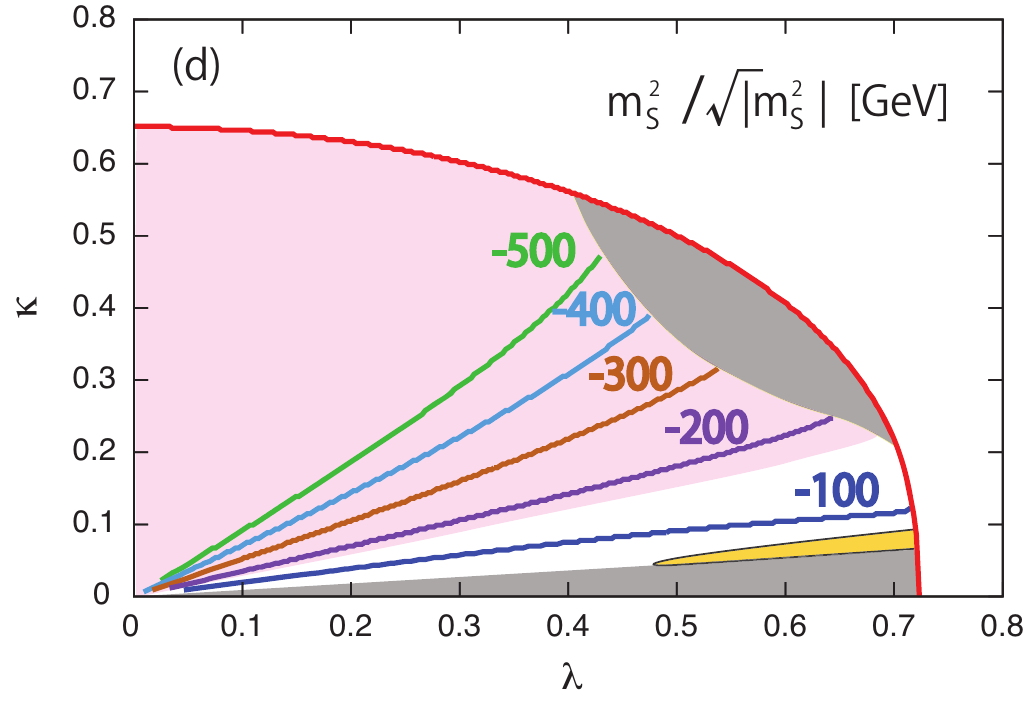} \\
\rule{0cm}{10mm} & \\
\includegraphics[height=48mm]{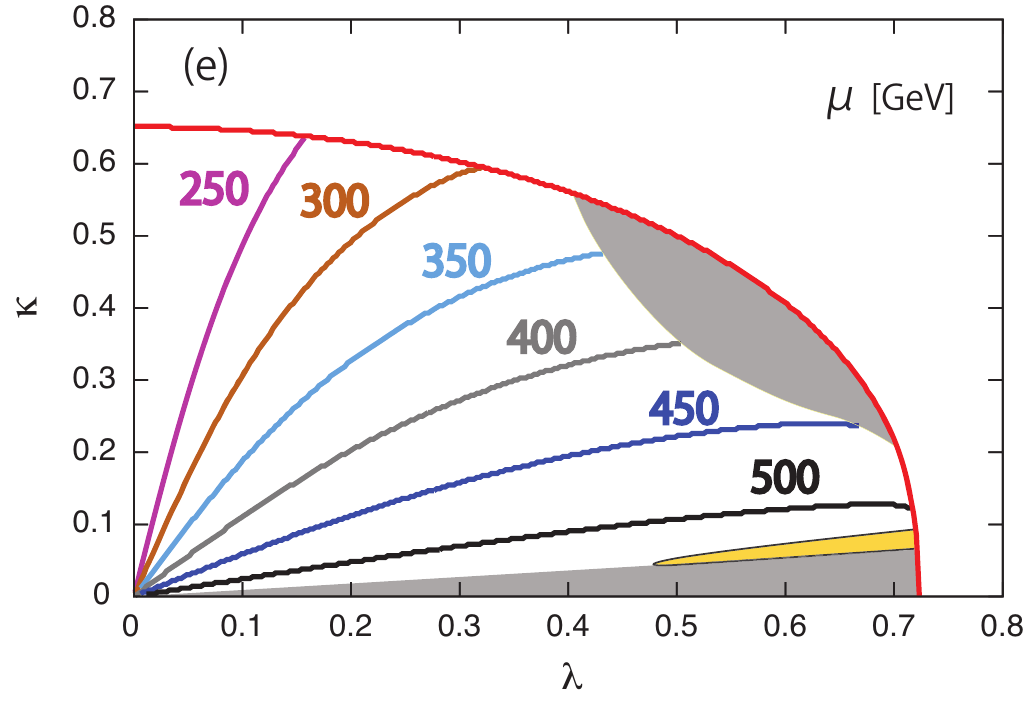} &
\includegraphics[height=48mm]{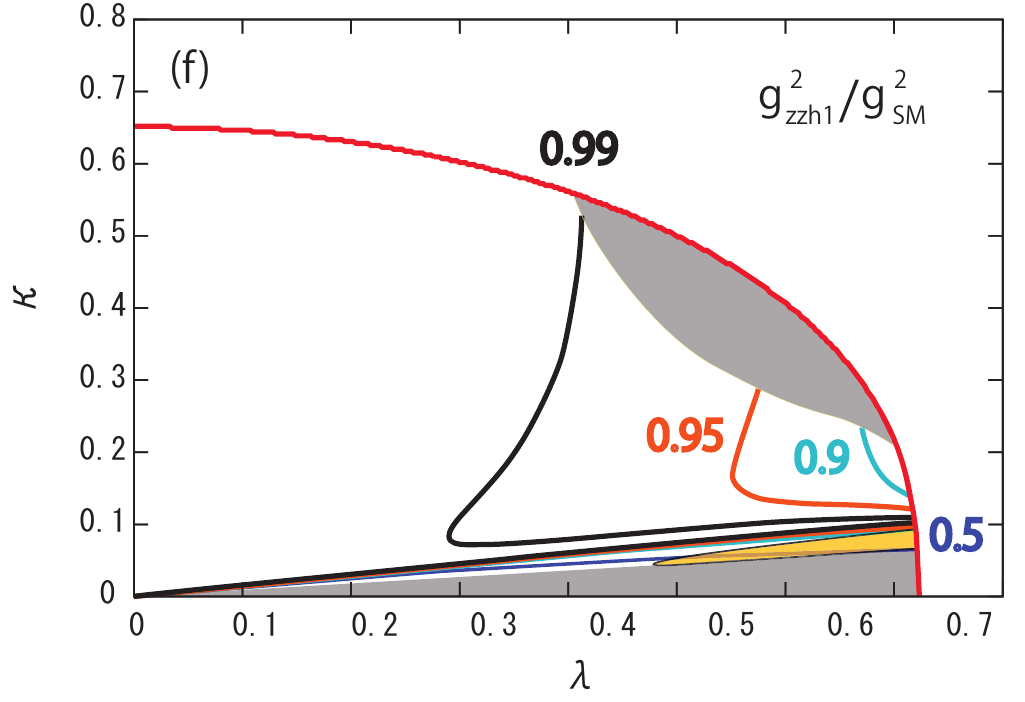} 
\end{tabular}
%\end{center}
\caption{The CP-even Higgs masses and the Higgs mass parameters for $\tan\beta=3$, $M_0=1500$ GeV, $A_\kappa=-100$ GeV. \label{fig:tanb3}}
\end{figure}

In figure~\ref{fig:tanb3}, we show the mass spectrum of the light CP-even Higgs bosons as a function of $\lambda$ and $\kappa$ for $\tan\beta=3$ and $M_0 = 1500$ GeV. 
In addition we present $m^2_{H_u}$, $m^2_S$ and $\mu$ to check whether the parameters fit in the range favored by the mirage mediation or not.
We also plot the coupling squared of the lightest CP-even Higgs boson to the gauge boson normalized with its SM value. This measures the amount of the doublet component in the lightest CP-even Higgs boson.
The red curve in the figure indicates the boundary where $\lambda$ and $\kappa$ blow up at the Planck scale. Inside this curve the model remains perturbative until the Planck scale. 
The gray shaded area around the red curve shows the region where our iterative method fails to calculate the minimum of the effective potential. This is due to the tachyonic tree-level Higgs mass which we use as the initial condition of the iteration. 
The quantum correction may lift it, however, we do not pursue this possibility because the region is already excluded by the LEP Higgs mass limit. 
The gray shaded area in the small $\kappa$ shows the region where the lightest CP-even Higgs becomes tachyonic. 
The yellow region is disfavored due to the false vacuum developed far beyond the EW scale as studied in \cite{Kanehata:2011ei}.

As we see in the upper left panel, the lightest CP-even Higgs boson mass can reach $125$ GeV for large value of $\lambda$ ($\approx 0.7$) and small value of $\kappa$ ($\approx 0.1$). This region is favored by the mirage mediation as shown in the middle left and middle right panels where the corresponding small mass parameters satisfy $m_s^2, \, m_{H_u}^2\lesssim M_0^2/8\pi^2$ for the observed Higgs boson mass. The lower right panel also indicates $h_1$ is almost doublet in this region. This is because the mixing between the doublet and the singlet in the mass matrix reduces the smaller eigenvalue while raises the larger eigenvalue \cite{Kang:2012sy, Cao:2012fz, Cao:2012yn, Kobayashi:2012ee, Jeong:2012ma, Jeong:2014xaa, Agashe:2012zq, Choi:2012he, Kowalska:2012gs, Gherghetta:2012gb,Gherghetta:2014xea,Barbieri:2013hxa,Badziak:2013bda,Farina:2013fsa}. Then the mass of the lightest Higgs boson is maximized when the mixing is minimized. 
The second lightest CP-even Higgs is around $200$ GeV in the upper right panel and dominantly consists of the singlet. The higgsino mass parameter $\mu$ is around $500$ GeV in the lower left panel which is much heavier than the EW scale, however, the cancellation mechanism described in the previous section protects the model from the fine-tuning which we will see numerically later.
It is noted that the region is naturally realized in the model where $\lambda$ has a strong dynamics origin at the Planck scale and approximate Pecci-Quinn (PQ) symmetry moderately suppresses $\kappa$.

\begin{figure}[tbp]
\centering
%\begin{center}
\begin{tabular}{l @{\hspace{5mm}} r}
\includegraphics[height=48mm]{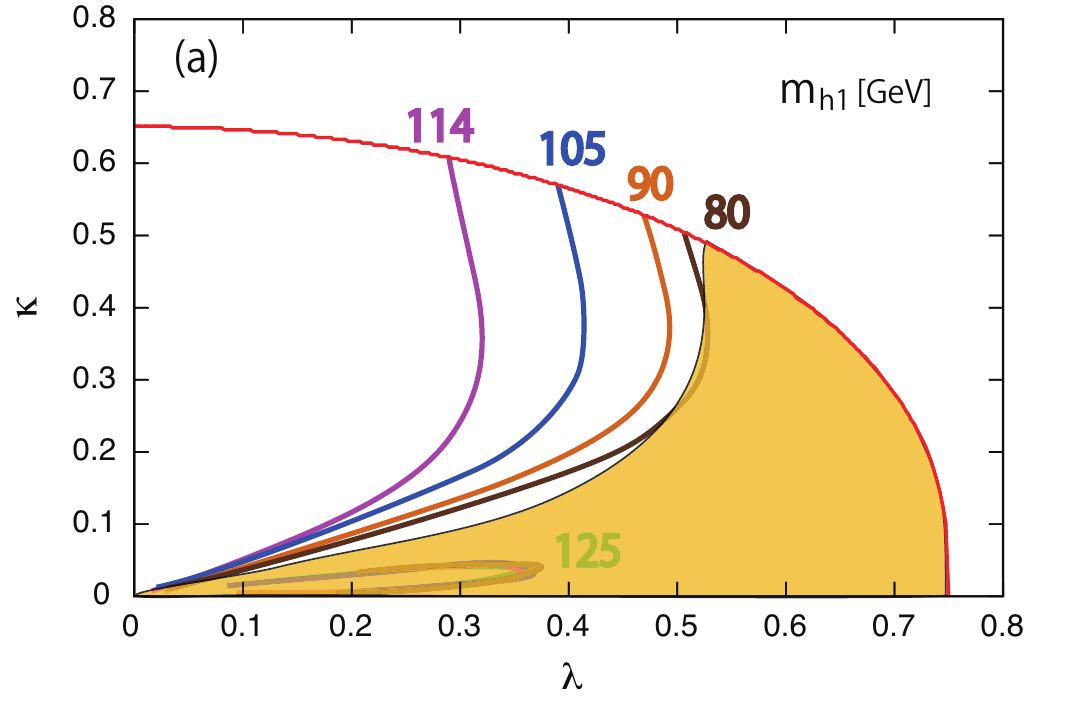} &
\includegraphics[height=48mm]{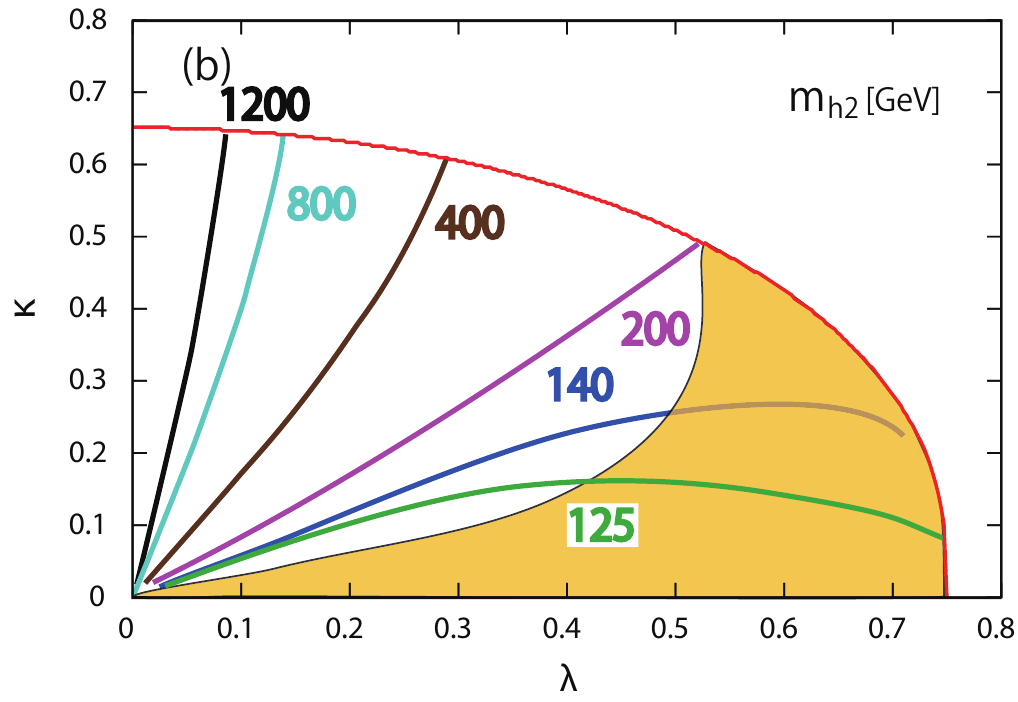} \\
\rule{0cm}{10mm} & \\
\includegraphics[height=48mm]{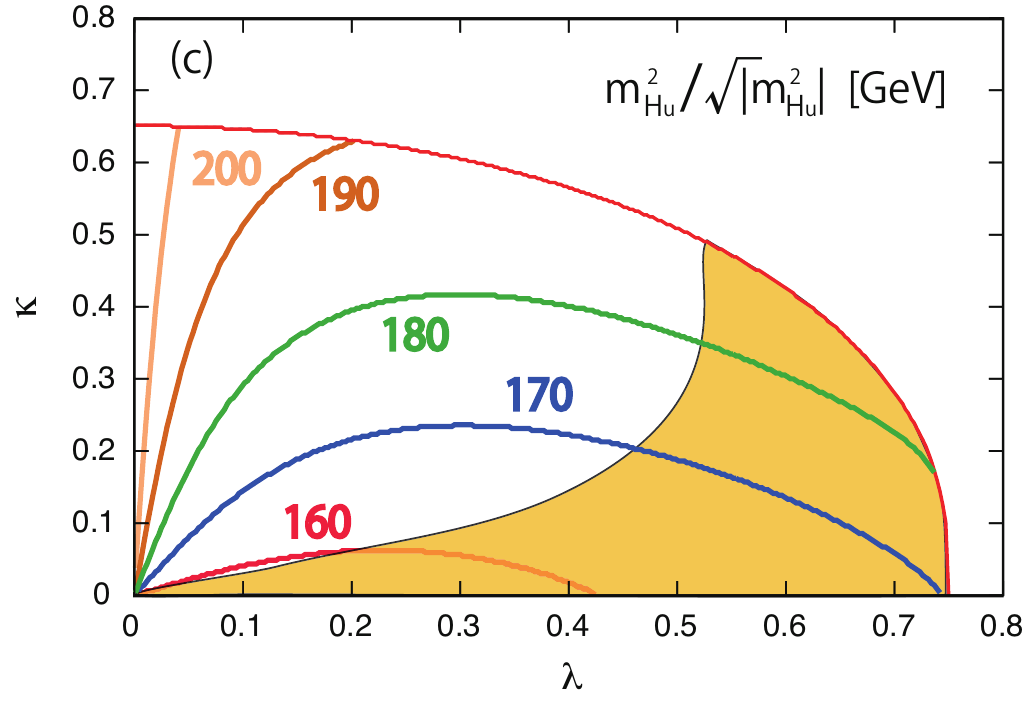} &
\includegraphics[height=48mm]{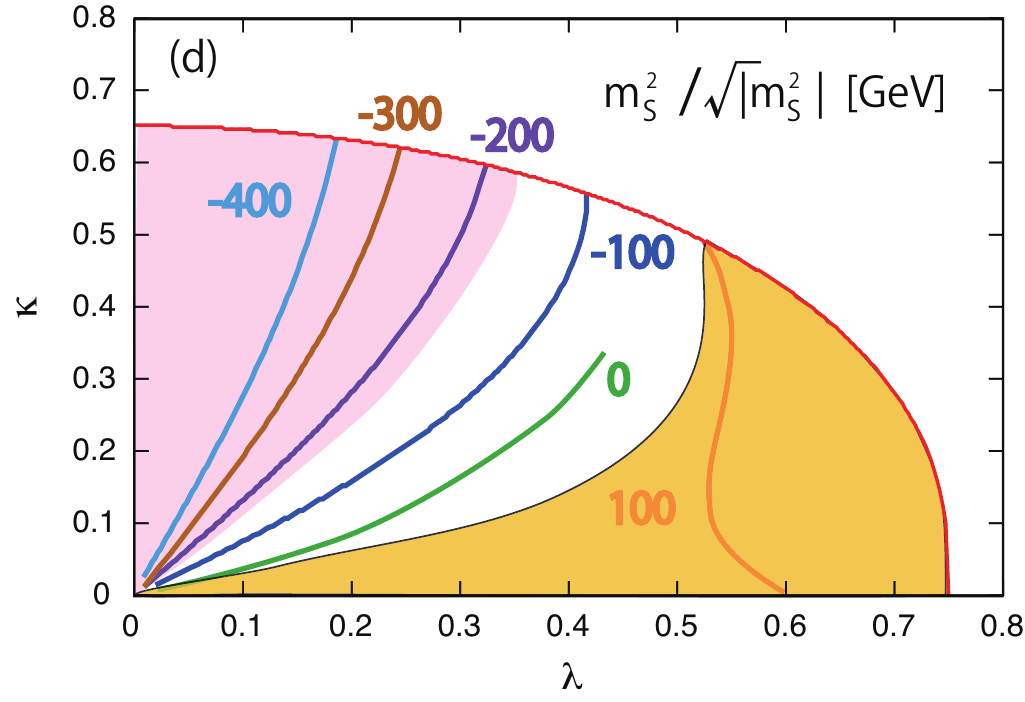} \\ 
\rule{0cm}{10mm} & \\
\includegraphics[height=48mm]{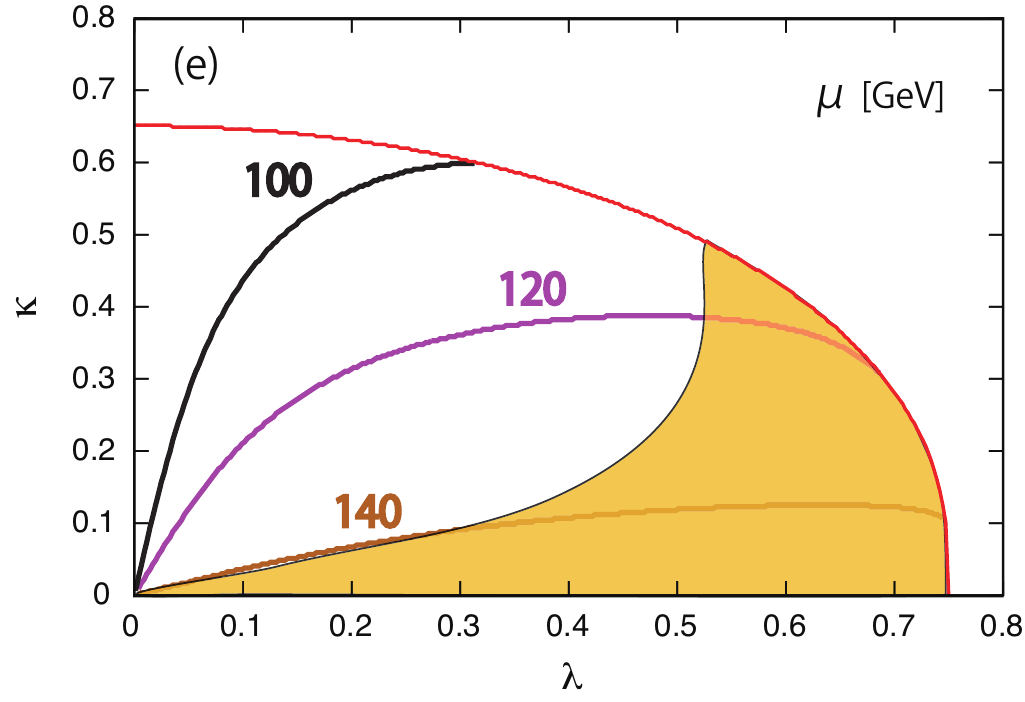} &  
\includegraphics[height=48mm]{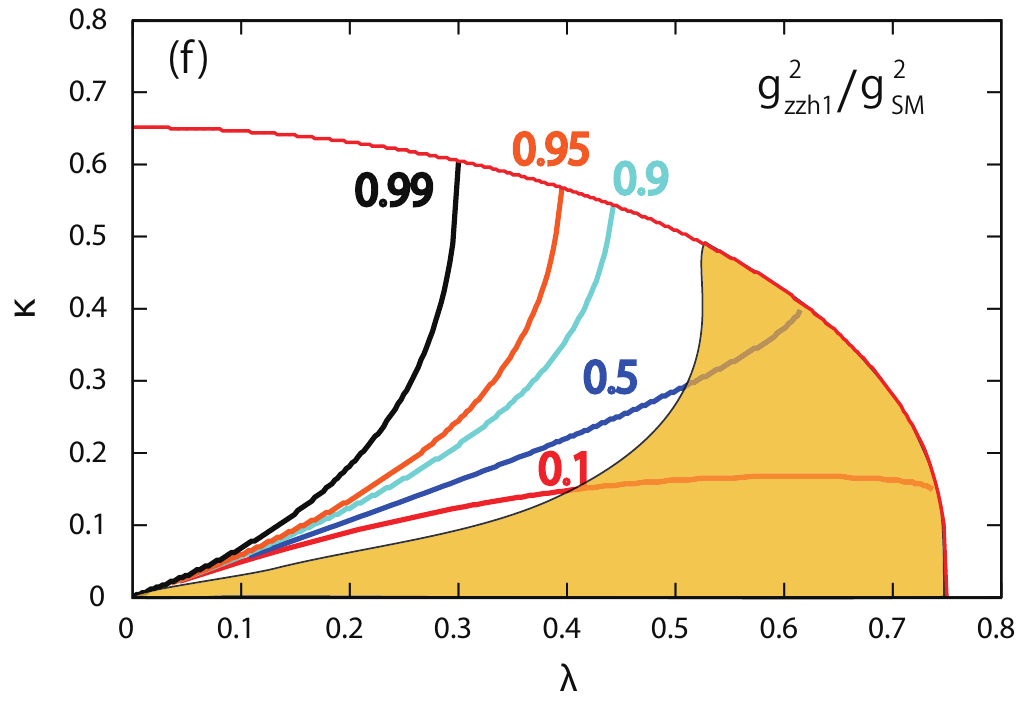}   
\end{tabular}
%\end{center}
\caption{The CP-even Higgs mass and the Higgs mass parameters for $\tan\beta=10$, $M_0=1500$ GeV, $A_\kappa=-100$ GeV. \label{fig:tanb10}}
\end{figure}

In figure~\ref{fig:tanb10}, we plot the same observables for $\tan\beta=10$. The other input parameters are fixed as in figure~\ref{fig:tanb3}. Now the lightest CP-even Higgs boson mass cannot reach $125$ GeV. Instead, the second lightest CP-even Higgs boson can be as light as $125$ GeV. This region is favored by the mirage mediation according to the contours of the small parameters, $m_{H_u}$ and $m_S$.
In this region, the coupling squared to the gauge boson indicates that $h_1$ almost consists of the singlet and hence $h_2$ is almost doublet. 
Now the mixing between the doublet and the singlet in the mass matrix
 is non-negligible because the radiative correction is not enough to push the doublet to $125$ GeV and the mixing is required to raise the larger eigenvalue.
The F-term contribution to the doublet mass from $\lambda$ is suppressed due to the large $\tan\beta$.
This mixing imposes a rather strong constraint on the model from the LEP Higgs boson search via $e^+e^- \to Z^\ast \to Z h$.
The region around $m_{h_1} \gtrsim 100$ GeV still survives and is an interesting possibility in light of the observed excess around 98 GeV \cite{Schael:2006cr,Belanger:2012tt,Kobayashi:2012ee,Drees:2005jg}.
If we raise the SUSY scale, the radiative correction takes the place of the mixing in realizing the 125 GeV Higgs boson mass and the constraint is relaxed.
The higgsino mass is around $M_0/\tan\beta \approx 150$ GeV as expected.
The LEP chargino mass bound implies $\tan\beta$ cannot exceed $\approx M_0/(100 {\rm GeV})$.

\begin{figure}[tbp]
\centering
%\begin{center}
\begin{tabular}{l @{\hspace{5mm}} r}
\includegraphics[height=48mm]{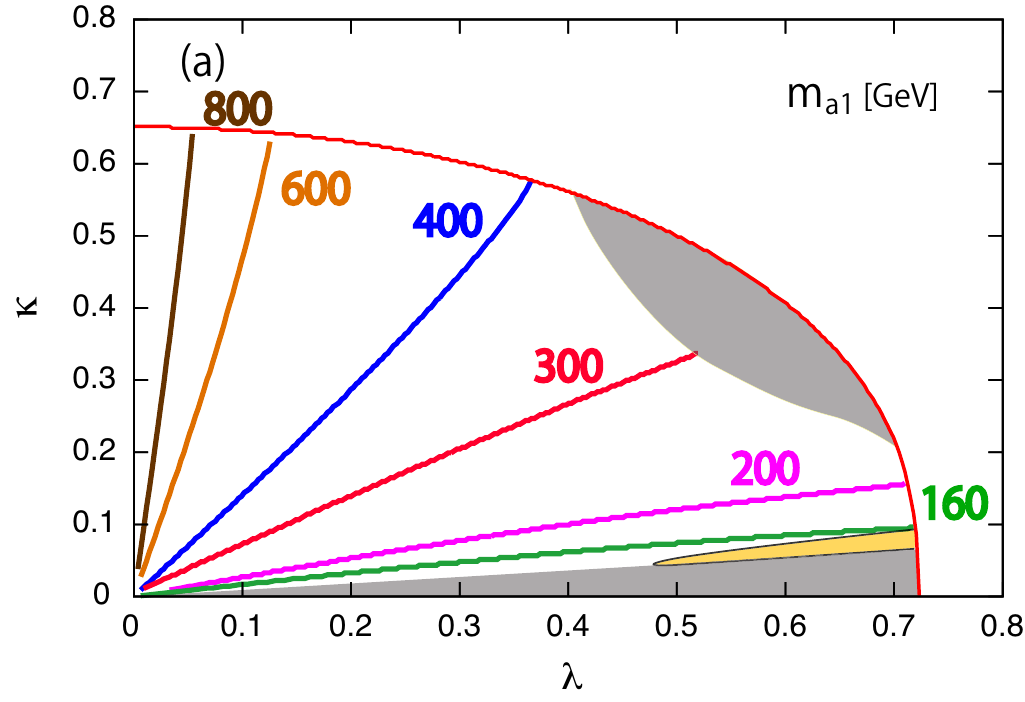} &
\includegraphics[height=48mm]{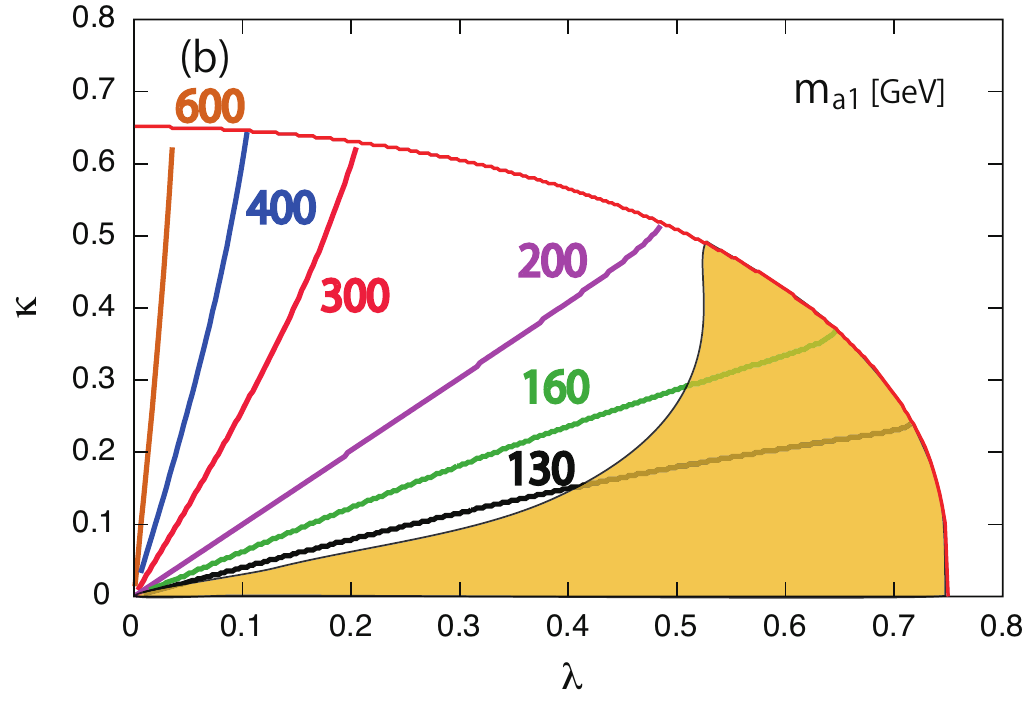} 
\end{tabular}
%\end{center}
\caption{The lightest CP-odd Higgs mass for $\tan\beta=3$ (left) and $\tan\beta=10$ (right). The other input parameters are same as figure~\ref{fig:tanb3}.\label{fig:A1}}
\end{figure}

For completeness, we present the contours for the lightest CP-odd Higgs boson mass in figure~\ref{fig:A1}. The mass decreases as $\kappa$ is reduced as expected from the approximated PQ symmetry. The contours are almost similar to those of the second lightest CP-even Higgs boson.  $125$ GeV CP-even Higgs boson corresponds to $m_{a_1}\simeq 200$ GeV for $\tan\beta=3$ and $m_{a_1} \simeq 150$ GeV for $\tan\beta=10$.

\subsection{Fine-tuning in the electroweak symmetry breaking}
Here we numerically estimate the degree of fine-tuning of the EW symmetry breaking in our model. Following the standard lore \cite{FT,Dimopoulos:1995mi,FTLEP,Casas:2003jx,Cassel:2009ps} we define the fine-tuning measure of a observable $y$ against an input parameter $x$ as 
\begin{equation}
\Delta_x^y = \frac{\partial \ln(y)}{\partial \ln(x)}.
\end{equation}
To evaluate the EW symmetry breaking, the simplest choice of $y$'s are the three Higgs VEVs $v_i=(\langle H_u\rangle, \langle H_d \rangle, \langle S \rangle)$. Instead of them, we choose $m_Z^2$, $\tan\beta$ and $\mu$, extending the standard choice, $y=m_Z^2$ in the literature \cite{King:1995vk,Dermisek:2005ar,Dermisek:2005gg}.
Here we emphasize that in principle all the VEVs must be examined to estimate the degree of fine-tuning once we extend the EW sector of the SM model, including usually disregarded $\tan\beta$. 
The maximum of these measures, $\Delta = {\rm max}|\Delta_x^y|$ could be regarded as the measure of the EW fine-tuning in the model. 
As $x$, we take $\lambda$, $\kappa$ and the small parameters, $m_{H_u}$, $m_S$ and $A_\kappa$ at the SUSY scale. Note that the large parameters such as $m_{H_d}$, $A_\lambda$ are not free parameters in our model and fixed by the ultraviolet physics. 

These measures are easily calculated by the potential minimization conditions eq.\eqref{eq:1} \cite{Casas:2003jx}.
If we vary the input parameters $x_a$ with $\Delta x_a$, the Higgs VEVs $v_i$
shift with $\Delta v_i$ following the potential minimum and they are related as 
\begin{equation}
\sum_k \frac{\partial^2 V}{\partial \phi_i \partial \phi_k} (\sqrt{2} \Delta v_k) + \sum_a \frac{\partial^2 V}{\partial \phi_i \partial x_a}\Delta x_a = 0.
\end{equation}
Noticing that the coefficient of the first term is given by the CP even Higgs mass matrix defined in eq.\eqref{eq:CP-even-mass-matrix}, this leads to 
\begin{equation}
\frac{\partial\, v_i }{\partial \ln x_a}=
-\sum_{k}\frac{1}{\sqrt{2}}\left({\cal M}_S^2\right)_{ik}^{-1} \frac{\partial^2 V}{\partial \phi_k \partial \ln x_a}.
\end{equation}
The fine-tuning measures can be constructed from these derivatives as 
%\begin{eqnarray}
\begin{subequations}
\begin{align}
\Delta^{m_Z^2}_x & = \frac{2}{v}\left(\cos\beta \frac{\partial\, v_d}{\partial \ln x} +\sin\beta \frac{\partial\, v_u}{\partial \ln x}\right), %\nonumber
\\
\Delta^{\tan\beta}_x & = \frac{1}{v_u}\frac{\partial\, v_u}{\partial \ln x} - \frac{1}{v_d}\frac{\partial\, v_d}{\partial \ln x}, %\nonumber
\\
\Delta^\mu_x & = \frac{\lambda}{\mu}\frac{\partial\, v_s}{\partial \ln x}+\delta_{x\lambda}.
\end{align}
\end{subequations}
%\end{eqnarray}
%
This expression has clear physical meaning \footnote{Essentially the same formula for $y=m_Z^2$ but different presentation is given in the appendix C of \cite{Gherghetta:2014xea}.}.
It is obvious that the measures increase if a certain Higgs mass is much lighter than the mass parameters in the potential. 
In \texttt{NMSSMTools}, the Higgs masses are calculated including the quantum corrections,
 however, the fine-tuning measures are calculated using the tree-level formula.
This overestimates the fine-tuning if any of the tree-level masses vanish although the quantum corrections lift them.
We calculate the measure using the above formula with the tree-level potential $V$ and the quantum corrected mass matrix ${\cal M}_S^2$.

\begin{figure}[tbp]%[p]
\centering
%\begin{center}
\begin{tabular}{l @{\hspace{10mm}} r}
\includegraphics[height=60mm]{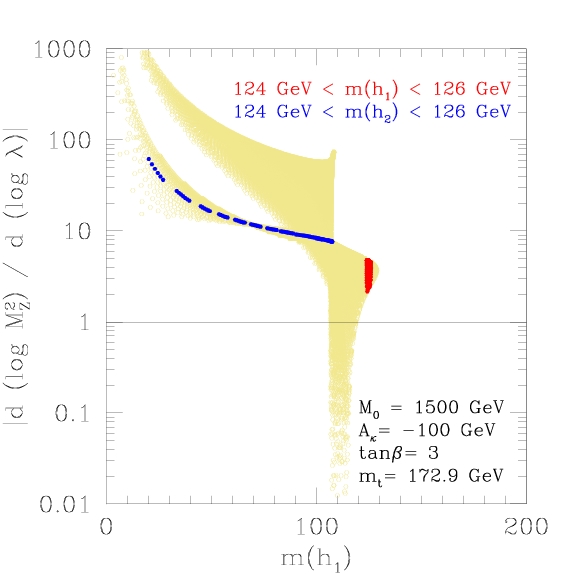} &
\includegraphics[height=60mm]{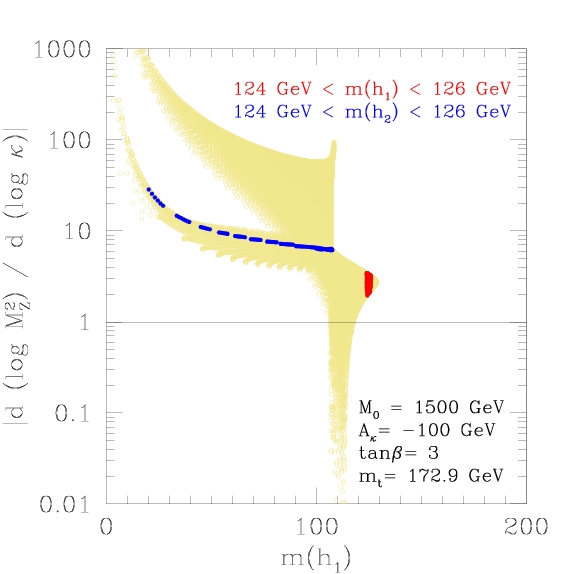} \\
\rule{0cm}{5mm} & \\
\includegraphics[height=60mm]{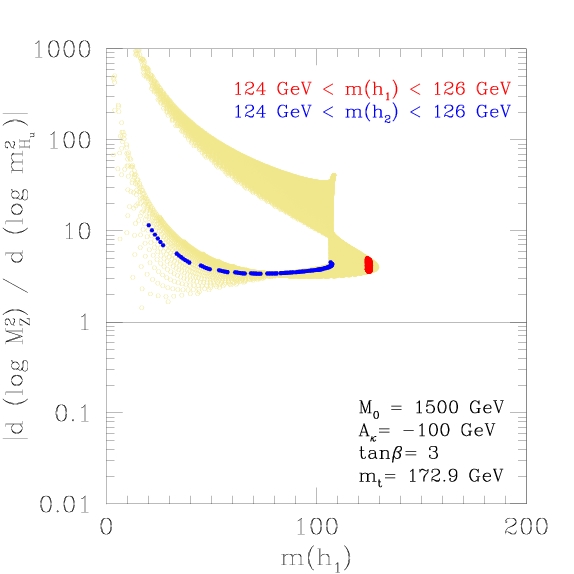} &
\includegraphics[height=60mm]{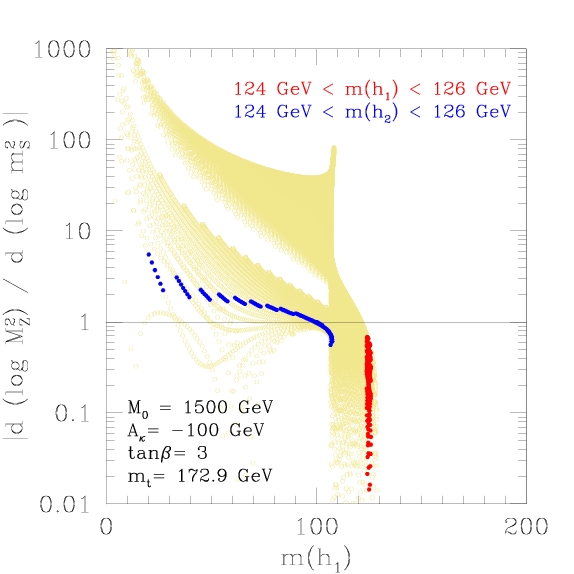} \\ 
\rule{0cm}{5mm} & \\
\includegraphics[height=60mm]{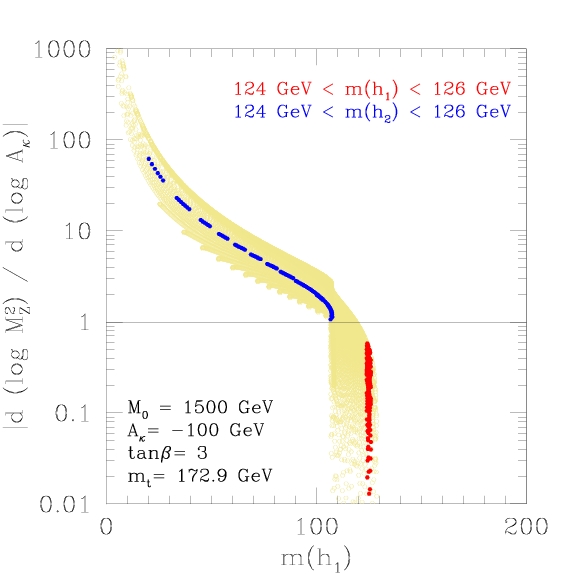} &  
\end{tabular}
%\end{center}
\caption{The fine-tuning measures of $m_Z^2$ for $\tan\beta=3$, $M_0=1500$ GeV, $A_\kappa=-100$ GeV. $\lambda$ and $\kappa$ are scanned over the white and yellow regions in figure~\ref{fig:tanb3}. \label{fig:FT-tanb3}}
\end{figure}

\begin{figure}[tbp]%[p]
\centering
%\begin{center}
\begin{tabular}{l @{\hspace{10mm}} r}
\includegraphics[height=60mm]{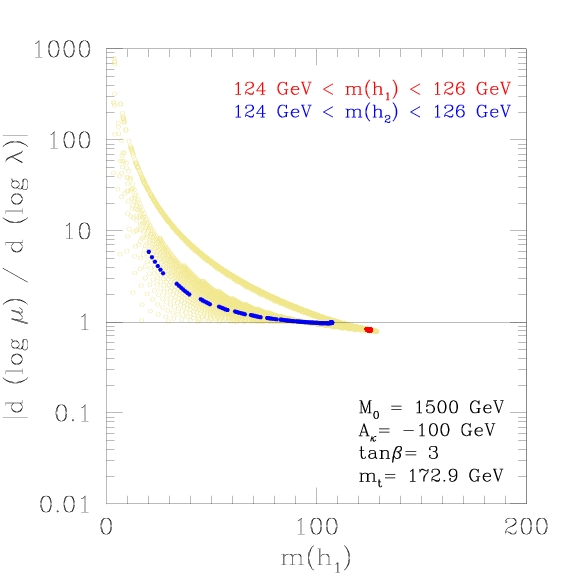} &
\includegraphics[height=60mm]{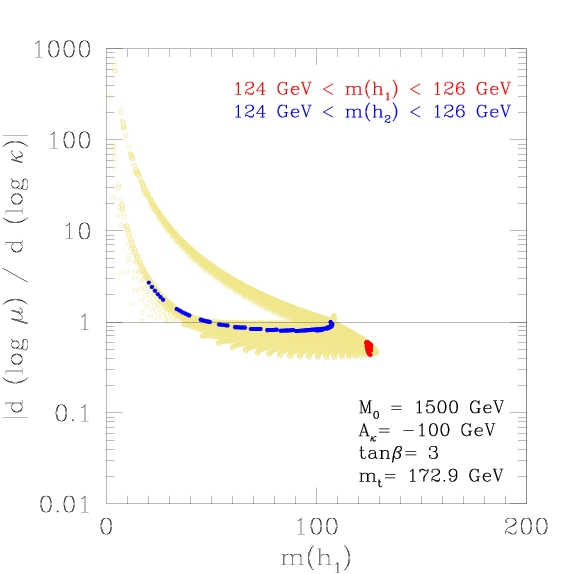} \\
\rule{0cm}{5mm} & \\
\includegraphics[height=60mm]{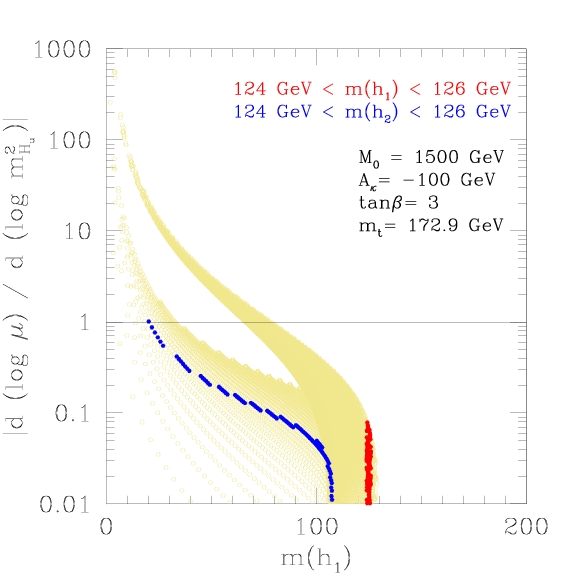} &
\includegraphics[height=60mm]{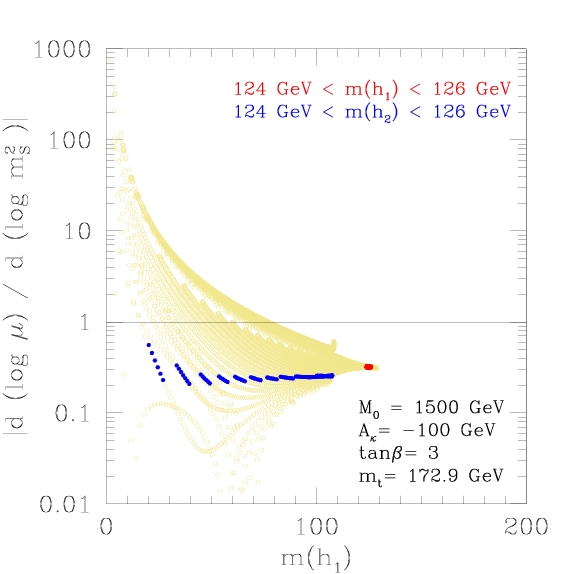} \\ 
\rule{0cm}{5mm} & \\
\includegraphics[height=60mm]{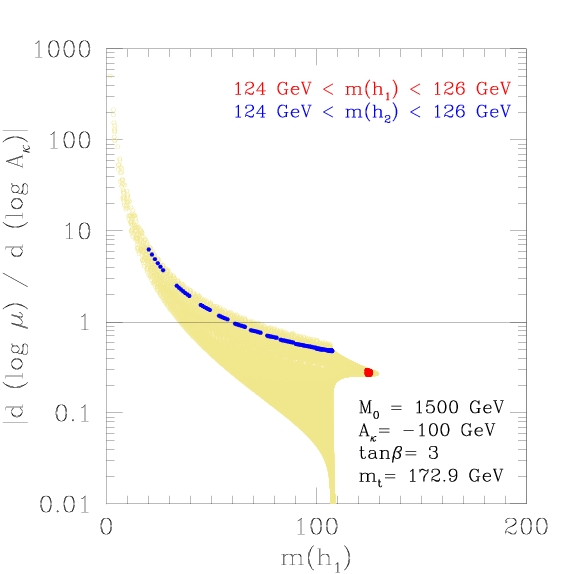} &  
\end{tabular}
%\end{center}
\caption{The fine-tuning measures of $\mu$ for $\tan\beta=3$, $M_0=1500$ GeV, $A_\kappa=-100$ GeV. \label{fig:FT-tanb3-2}}
\end{figure}

In figure~\ref{fig:FT-tanb3}, we show the fine-tuning measures, $\Delta_x^{m_Z^2}$ for $\tan\beta=3$, $M_0=1500$ GeV, $A_\kappa=-100$ GeV. The red dots satisfy the Higgs mass condition, $124\, {\rm GeV} < m_{h_1} < 126\, {\rm GeV}$.
We also plot blue dots representing the region where $124\, {\rm GeV} < m_{h_2} < 126\, {\rm GeV}$.  
Both of them are favored by the mirage mediation within the 1-loop uncertainty.
Here we choose the width, $125\pm 1$ GeV as a guide to the eye.  Note that currently the theoretical error of the SM-like Higgs mass in the NMSSM could reach $5$ GeV \cite{Staub:2015aea} and actual bound is somewhat weaker than the bands shown in the figure \footnote{We thank F.~Staub for pointing out this issue.}.  
The upper left and upper right panels show the case for $\lambda$ and $\kappa$, respectively. 
Because the effective $\mu$-term is roughly proportional to $\kappa/\lambda$ as seen in the figure~\ref{fig:tanb10}, severe fine-tuning is expected for these parameters. 
However, $\Delta^{m_Z^2}_{\lambda} \lesssim 5$ and $\Delta^{m_Z^2}_{\kappa} \lesssim 4$  are realized for $m_{h_1}\approx 125$ GeV in our model. 
Also $\Delta^{m_Z^2}_{\lambda, \kappa} \lesssim 10$ is satisfied for $m_{h_2}\approx 125$ GeV case unless $m_{h_1}$ is not too light.
This is not trivial because the khaki circles which do not necessarily satisfy the conditions required by the mirage mediation reach $\Delta^{m_Z^2}_{\lambda, \kappa}\sim 100$ even with a moderate value of $m_{h_1}$.
We confirmed that the fine-tuning measures for these dimensionless 
parameters do not deteriorate even if we increase $M_0$ to $5$ TeV.
The middle left and middle right panels present the results for $m_{H_u}^2$ and $m_S^2$. It is found that $\Delta_{m_{H_u}^2}^{m_Z^2}\simeq 5$ and $\Delta_{m_S^2}^{m_Z^2} \lesssim 1$ for
 the observed Higgs mass and a moderate value of $m_{h_1}$. Again, this is not trivial because the khaki circles spread over $\Delta_{m_{H_u}^2, m_S^2}^{m_Z^2} \sim 100$.
These measures for the dimensionful parameters are sensitive to the SUSY scale, $M_0$.
The lower left panel shows the case for $A_\kappa$ and we find $\Delta^{m_Z^2}_{A_\kappa}\lesssim 0.4$ for $m_{h_1}\approx 125$ GeV and $\Delta^{m_Z^2}_{A_\kappa}\lesssim 10$ for $m_{h_2}\approx 125$ GeV with a moderate value of $m_{h_1}$. 
This measure appears to be not sensitive to the SUSY scale, $M_0$.

In figure~\ref{fig:FT-tanb3-2}, we show the fine-tuning measures for $\mu$ with the same input parameters. 
All of them are smaller than ${\it O}(1)$ for the observed Higgs mass.  On the other hand, $\Delta_x^{\tan\beta}\approx -\Delta_x^{\mu}$ holds well in our model because of the relation eq.\eqref{eq:app-1}. Then $\Delta_x^{\tan\beta}$ do not work as independent measures.
Thus the worst measure is $\Delta^{m_Z}_\lambda$, which is below $10$ unless $h_1$ is extremely light (such a region is actually excluded by phenomenological constraints). Therefore the model requires at most $10 \%$ tuning ($\Delta^{-1}=0.1$) for $\tan\beta=3$ and $M_0=1500$ GeV. 
%Therefore all the tuning measures are below $10$ and the model requires at most $10$ \% tuning for $M_0 = 1500$ GeV unless $h_1$ is extreemly light. 
In our model the stop mass is $\sqrt{c_{t_{L,R}}}\, M_0 \approx 1$ TeV. 
Thus we numerically confirmed the model exhibits surprisingly low level of fine-tuning compared to the conventional SUSY breaking models.

\begin{figure}[tbp]%[t]
\centering
%\begin{center}
\begin{tabular}{l @{\hspace{10mm}} r}
\includegraphics[height=65mm]{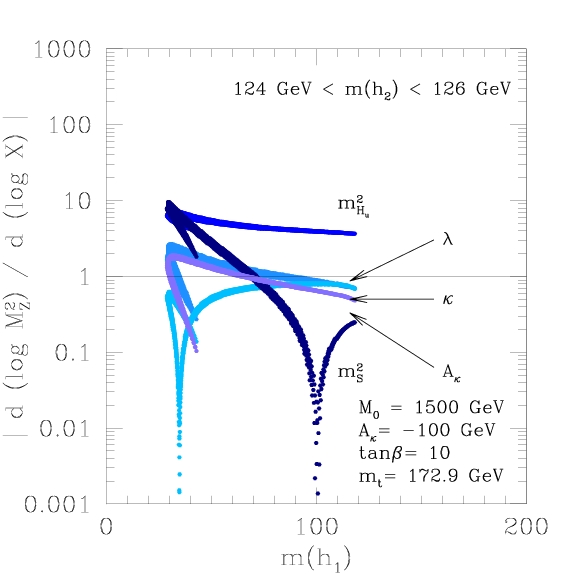} &
\includegraphics[height=65mm]{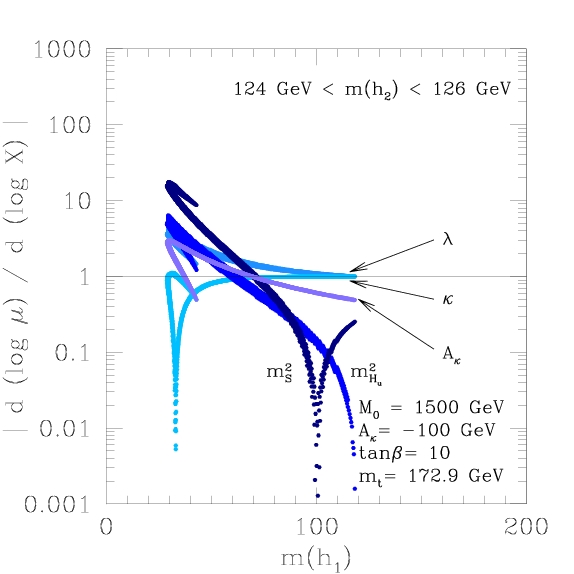} 
\end{tabular}
%\end{center}
\caption{The fine-tuning measures of the EW symmetry breaking for $\tan\beta=10$, $M_0=1500$ GeV, $A_\kappa=-100$ GeV. $\lambda$ and $\kappa$ are scanned over the white and yellow regions in figure~\ref{fig:tanb10}. Only the points satisfying $124 {\rm GeV} < m_{h_2} < 126 {\rm GeV}$ are depicted. \label{fig:FT-tanb10}}
\end{figure}

In figure~\ref{fig:FT-tanb10}, we summarize the same fine-tuning measures for the $\tan\beta=10$ case. The other input parameters are same as in figure~\ref{fig:FT-tanb3}. Here, we impose $124\, {\rm GeV} < m_{h_2} < 126\, {\rm GeV}$. There is no parameter region satisfying $m_{h_1}\approx 125$ GeV in this case.
The fine-tuning measures for $\lambda$, $\kappa$ and $A_\kappa$ are below ${\cal O}(1)$. 
The worst tuning ($\Delta$)  is $\Delta^{m_Z^2}_{m_{H_u}^2} \approx 5$ for $m_{h_1}\gtrsim 50$ GeV.  While $\Delta^{m_Z^2}_{m_S^2}$, $\Delta^\mu_{m_{H_u}^2}$ and $\Delta^\mu_{m_S^2}$ are well below the unity with $m_{h_1}\sim 100$ GeV, they increase up to $10-20$ once $m_{h_1} $ decreases down to $40$ GeV.  
Here we do not impose any experimental constraints, however, it is interesting that in fact only the points $m_{h_1} \gtrsim 90$ GeV can survive the LEP Higgs bound with the adopted choice of parameters.  This is because the mixing with the doublet of order 10 \% is necessary for $m_{h_2}$ to reach $125$ GeV. This region exactly matches with the area where the tuning is minimized.

\begin{figure}[tbp]%[h]
\centering
%\begin{center}
\begin{tabular}{l @{\hspace{10mm}} r}
\includegraphics[height=65mm]{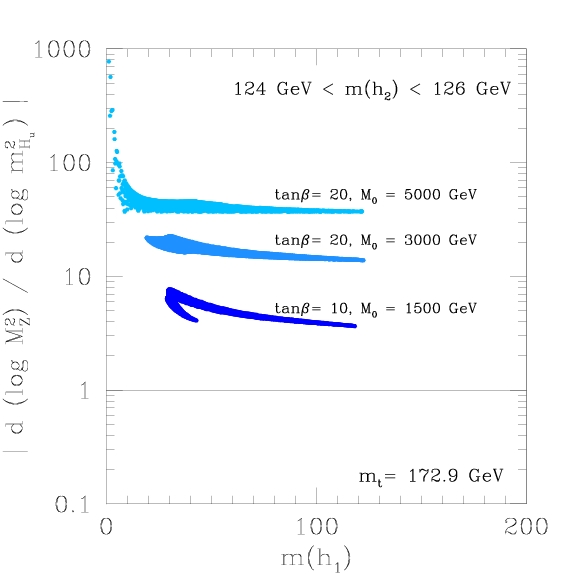} &
\includegraphics[height=65mm]{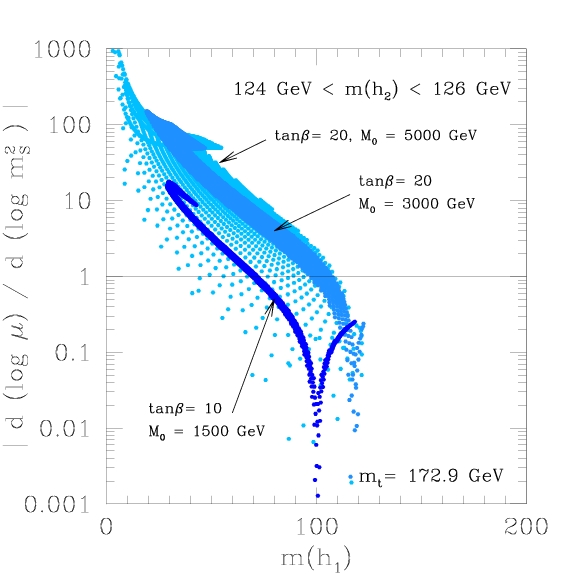} 
\end{tabular}
%\end{center}
\caption{The fine-tuning measures $\Delta^{m_Z^2}_{m_{H_u}^2}$ (left) and $\Delta^\mu_{m_S^2}$ (right) for $M_0=3$ TeV and $M_0=5$ TeV ($\tan\beta=20$) compared with $M_0=1.5$ TeV ($\tan\beta=10$) case. $A_\kappa$ is fixed at $A_\kappa=-100$ GeV. \label{fig:heavy-case}}
\end{figure}

For completeness, in figure~\ref{fig:heavy-case} we plot $\Delta_{m_{H_u}^2}^{m_Z^2}$ and $\Delta_{m_S^2}^\mu$ for $M_0=3$ TeV and $M_0=5$ TeV in addition to the $M_0=1.5$ TeV case. 
The other parameters are fixed at $\tan\beta=20$ and $A_\kappa = -100$ GeV. 
We see that $\Delta_{m_{H_u}^2}^{m_Z^2}$ is $10-20$ for $M_0 = 3$ TeV and $\sim 40$ for $M_0 = 5$ TeV. 
Also we see fine-tuning measure $\Delta_{m_S^2}^\mu\lesssim 10$ if $m_{h_1}\gtrsim 60$ GeV.
It is remarkable that even the $5$ TeV case is acceptable in the standard of the conventional models build around $1$ TeV.
Again the LEP bound for $h_1$ is weak where the fine-tuning measures are minimized ($m_{h_1}\sim 100$ GeV).

\subsection{Higgs couplings and collider signature}

In this section, we discuss the couplings of the Higgs bosons to the SM particles.
They play crucial roles in production and decay of the Higgs bosons at the LHC and future colliders like ILC, FCC-ee(TLEP)/CEPC.
The precision measurements of the signal strength for various production and decay channels will constrain the couplings of the SM-like Higgs boson at {\it O}(1)\% level or below \cite{Dawson:2013bba, Asner:2013psa, Peskin:2013xra, Bechtle:2014ewa,Fujii:2015jha,Gomez-Ceballos:2013zzn}. 
The pattern of deviations from the SM predictions will bring hints for underlying models beyond the SM. 
Also there are chances to detect extra light Higgs bosons if their couplings to the SM particles are strong enough.

The interaction of the CP-even Higgs bosons $h_i\,(i=1-3)$ with the SM fields can be described by the following effective Lagrangian \cite{Carmi:2012in}:
\begin{eqnarray}
{\cal L} &=& 
\sum_{i=1}^3
\left[
C_{V}^i \frac{\sqrt{2} m_W^2}{v} h_i W_\mu^+ W^{-\mu} + C_{V}^i \frac{m_Z^2}{\sqrt{2} v} h_i Z_\mu Z^\mu 
-
\sum_f
C_{f}^i \frac{m_f}{\sqrt{2} v} h_i \overline{f} f 
\right. \nonumber\\
&& 
\left.
\phantom{\sum_{i=1}^3}
+ C_{g}^i \frac{\alpha_s}{12\sqrt{2}\pi v} h_i G^a_{\mu\nu} G^{a\,\mu\nu}+ C_{\gamma}^i \frac{\alpha}{\sqrt{2}\pi v} h_i A_{\mu\nu} A^{\mu\nu} \right],
\end{eqnarray} 
where we assume the custodial symmetry relating the $W$ and $Z$ couplings. The sum of $f$ runs all the SM fermions.
$G^a_{\mu\nu}$ and $A_{\mu\nu}$ are the field strengths of the gluon and photon, respectively.
The coupling constants in the first line
 are normalized so that $C_V^{SM}= C_f^{SM}= 1$  for the SM Higgs boson $h_{\rm SM}$ at the tree level.
In the NMSSM, they are given by
%\begin{eqnarray}
\begin{subequations}
\begin{align}
C_V^i & = {\cal S}_{2i}^\dag \sin\beta + {\cal S}_{1i}^\dag \cos\beta, %\nonumber
\\
C_f^i & = 
\left\{
\begin{array}{cc}
\displaystyle
{\cal S}_{1i}^\dag\, \frac{1}{\cos\beta} & (f = e,\mu,\tau, d,s,b) \\
\displaystyle
{\cal S}_{2i}^\dag\, \frac{1}{\sin\beta} & (f = u,c,t )
\end{array}
\right. ,
\end{align} 
\end{subequations}
%\end{eqnarray}
where ${\cal S}_{ij}$ is the diagonalization matrix for the mass matrix of the CP-even Higgs bosons.
$C_V^i$ is often quoted as $\xi_i$ in the literature and satisfies 
$\sum_{i=1}^3 \xi_i^2 = 1$.
We have already depicted $\xi_1^2$ in the lower right panels in figure~\ref{fig:tanb3} and \ref{fig:tanb10}.
\begin{table}[tbp]%[t]
\centering
%\begin{center}
\begin{tabular}{|c|ccc|}
%\hline
\hline
      & $h$ & $H$ & $h_S$ \\
\hline
$m_{h_i}^2$ & $m_Z^2\cos^2 2\beta + \lambda^2 v^2 \sin^2 2\beta$ & $A_\lambda^2+(m_Z^2-\lambda^2 v^2)\sin^2 2\beta$ & $\lambda^2 v^2 \sin^2 2 \beta$ \\
$C_V^i$ & 1 & 0 & 
$\frac{\lambda v}{A_\lambda}\sin 2\beta$ \\
$C_{d,e}^i$ & 1   & $-\tan\beta$ & 
$\frac{\lambda v}{A_\lambda}$ $\tan\beta$ \\ 
$C_u^i$ & 1   & $\cot \beta$ & 
$\frac{\lambda v}{A_\lambda}$ $\cot\beta$\\
\hline
%\hline
\end{tabular}

%
%\end{center}
\caption{Coupling constants of the CP-even Higgs bosons in the $\kappa = m_S^2 = 0$ limit. \label{tab:s-coupling}}
\end{table}
These couplings are summarized in Table~\ref{tab:s-coupling} in the $\kappa= m_S^2=0$ limit. The singlet-like Higgs $h_S$ seems to be the lightest. However, once we include corrections from non-vanishing $\kappa$ and $m_S^2$, it could be heavier than $h$ for small $\tan\beta$. The light doublet $h$ behaves like the SM Higgs boson within the approximation. Note that $C_V^h$ and $C_u^h$ are stable against the small mixing between $h$ and $H$ induced by the finite $\kappa$ and $m_S^2$, however, $C_{d,e}^h$ could be affected considerably by the $\tan\beta$ enhanced $C_{d,e}^H$ \cite{Choi:2012he, King:2012tr, Badziak:2013bda}.

The coupling constants $C_g^i$ and $C_\gamma^i$ are loop induced.
Generally, if we have a scalar $S$, a fermion $f$ and a (color-less) vector boson $\rho_\mu$ fields with the following couplings with the Higgs bosons \cite{Carmi:2012in},
\begin{equation}
{\cal L} = \sum_{i=1}^3
\left[-C_s^i \frac{\sqrt{2} m_s^2}{v} h_i S^\dag S -C_f^i \frac{m_f}{\sqrt{s} v} h_i \overline{f} f + C_\rho^i \frac{\sqrt{2} m_\rho^2}{v} h_i \rho^\dag_\mu \rho^\mu\right],
\label{eq:loop-induced-coupling}
\end{equation}
their contributions to ${C_g}^i$ and ${C_\gamma}^i$ are given by 
%\begin{eqnarray}
\begin{subequations}
\begin{align}
\delta C_g^i & = \frac{T(r_s)}{2} C_s^i A_s(\tau^i_s) + 2 T(r_f) C_f^i A_f(\tau^i_f),%\nonumber
\\
\delta C_\gamma^i & = \frac{N(r_s) Q_s^2}{24} C_s^i A_s(\tau^i_s) + \frac{N(r_f) Q_f^2}{6} C_f^i A_f(\tau^i_f) -\frac{7 Q_\rho^2}{8} C_\rho^i A_v(\tau^i_\rho),
%\nonumber
\end{align}
\end{subequations}
%\end{eqnarray}
where $T(r_s)$ ($T(r_f)$) denotes the Dynkin index of the scalar (fermion) field in the representation $r_s$ ($r_f$) as defined by $tr(T^a(r) T^b(r)) = T(r) \delta^{ab}$ with the $SU(3)$ generator $T^a(r)$. $N(r_s)$ ($N(r_f)$) is the number of freedom in the representation $r_s$ ($r_f$). $Q_{s,f,\rho}$ are charges of the scalar, fermion and vector fields, respectively.
The mass functions $A_{s,f,v}(\tau)$ are defined as 
%\begin{eqnarray}
\begin{subequations}
\begin{align}
A_s(\tau) & \equiv \frac{3}{\tau^2}[f(\tau)-\tau], %\nonumber
\\
A_f(\tau) & \equiv \frac{3}{2\tau^2}[(\tau-1)f(\tau)+\tau], %\nonumber
\\
A_v(\tau) & \equiv \frac{1}{7\tau^2}[3(2\tau-1)f(\tau)+3\tau+2 \tau^2], %\nonumber
\\
f(\tau) & \equiv 
\left\{
\begin{array}{cc}
\arcsin^2 \sqrt{\tau} & \tau \leq 1 \\
-\frac{1}{4}\left[
\log
\frac{1+\sqrt{1-\tau^{-1}}}{1-\sqrt{1-\tau^{-1}}}
-i \pi
\right]^2 & \tau > 1
\end{array}
\right. ,
\end{align}
\end{subequations}
%\end{eqnarray}
 with $\tau^i_j = \frac{m_{h_i}}{4 m_j^2}$.
Since $A_{s,f,v}(\tau)$ decouple quickly for $\tau>>1$, the standard model contribution is dominated by those of the top quark and the $W$ boson and given by \cite{Carmi:2012in},  
%\begin{eqnarray}
\begin{subequations}
\begin{align}
C_g^{SM} & = \frac{8}{3} A_f(\tau^{SM}_t) \approx 1.03, %\nonumber
\\
C_\gamma^{SM} & = \frac{2}{9} A_f(\tau^{SM}_t) -\frac{7}{8} A_v(\tau^{SM}_W)
 \approx -0.81.
\end{align}
\end{subequations}
%\end{eqnarray}
%
For $\tau<<1$, we obtain $A_{s,f,v}\simeq 1$. Then, we have,
%\begin{eqnarray}
\begin{subequations}
\begin{align}
\delta C_g^i & \simeq \frac{T(r_s)}{2} C_s^i + 2 T(r_f) C_f^i, %\nonumber
\\
\delta C_\gamma^i & \simeq \frac{N(r_s) Q_s^2}{24} C_s^i+ \frac{N(r_f) Q_f^2}{6} C_f^i -\frac{7 Q_\rho^2}{8} C_\rho^i. %\nonumber\\
\end{align}
\end{subequations}
%\end{eqnarray}

Together with the relation,
\begin{eqnarray}
&&C_s^i \frac{\sqrt{2}m_s^2}{v} = \frac{\partial m_s^2}{\partial h_i},\qquad
C_f^i \frac{m_f}{\sqrt{2}v} = \frac{\partial m_f}{\partial h_i},\qquad 
C_v^i \frac{\sqrt{2}m_f^2}{v} = \frac{\partial m_v^2}{\partial h_i}, \nonumber\\
&\to& C_j^i = \frac{v}{\sqrt{2}} \frac{\partial \log m_j^2}{\partial h_i},
\end{eqnarray}
we arrive at 
%\begin{eqnarray}
\begin{subequations}
\begin{align}
\delta C_g^i & = \frac{3 v}{2\sqrt{2}}
\left[
\frac{T(r_s)}{3}\frac{\partial \log\det({\cal M}^2_s)}{\partial h_i} 
+\frac{4T(r_f)}{3}\frac{\partial \log\det({\cal M}^2_f)}{\partial h_i} 
\right], %\nonumber
\\
\delta C_\gamma^i & = \frac{ v}{8\sqrt{2}}
\left[
\frac{N(r_s)Q_s^2}{3}\frac{\partial \log\det({\cal M}^2_s)}{\partial h_i} 
+\frac{4 N(r_f) Q_f^2}{3}\frac{\partial \log\det({\cal M}^2_f)}{\partial h_i} 
\right. \nonumber
\\
& 
\phantom{\frac{v}{8\sqrt{2}}}
\left.
-7 Q_v^2 \frac{\partial \log\det({\cal M}^2_v)}{\partial h_i} 
\right],
\end{align}
\end{subequations}
%\end{eqnarray}
where ${\cal M}^2_{s,f,v}$ is the mass matrix for the corresponding field.
Noticing that the coefficients of the derivatives are given by the steps of the beta function coefficients after integrating out the corresponding fields,
this can be cast into the well known low energy theorem \cite{Ellis:1975ap, Shifman:1979eb, Shifman:1978zn}:
\begin{equation}
C_g^i \frac{\alpha_s}{12\sqrt{2}\pi v} = \frac{1}{4}\frac{\partial \log \alpha_s(\mu)}{\partial h_i},\qquad
C_\gamma^i \frac{\alpha}{\sqrt{2}\pi v} = \frac{1}{4}\frac{\partial \log \alpha(\mu)}{\partial h_i},
\end{equation}
if all the integrated fields are much heavier than the renormalization scale, $\mu$. Here, $h_i$ dependence of $\alpha_s$ and $\alpha$ enters through the mass thresholds. The contribution from the light fields should be included separately.
 
In SUSY models, additional contribution comes from the charged Higgs boson, sfermion and chargino loops.
The charged Higgs contribution to $\delta C_\gamma^{h,S}$ is calculated as 
\begin{equation}
\delta C_\gamma^i \approx
\left\{
\begin{array}{cc}
\displaystyle
\frac{1}{48}\{(-\lambda^2+g_2^2)\sin 2\beta +\frac{1}{2}(g_1^2+g_2^2)\cos^2 2\beta\}\frac{v^2}{m_{h^-}^2} & 
\displaystyle
(i=h)\\
\displaystyle
\frac{1}{48}\{(-\lambda^2+g_2^2)\cos 2\beta +\frac{1}{2}(g_1^2+g_2^2)\sin^2 2\beta\}\frac{v^2}{m_{h^-}^2} & 
\displaystyle
(i=H)\\
\displaystyle
\frac{1}{24}(2\lambda+\kappa \sin 2\beta)\frac{\mu v}{m_{h^-}^2} & 
\displaystyle
(i=S).
\end{array}
\right.
\end{equation}

They are suppressed by $v^2/m_{h^-}^2$, $\mu v/m_{h^-}^2$ and negligible in our model if $m_{h^-}(\approx M_0)$ is above the TeV scale. 

The sfermion and chargino mass matrices are given by 
%\begin{eqnarray}
\begin{subequations}
\begin{align}
{\cal M}^2_{\tilde{u}} & =
\left(
\begin{array}{cc}
m_{Q_i}^2 + y_{u_i}^2 v_u^2
+(v_u^2-v_d^2)\left(\frac{g_1^2}{12}+\frac{g_2^2}{4}\right) 
&
y_{u_i}(A_{u_i} v_u - \mu v_d)\\
y_{u_i}(A_{u_i} v_u - \mu v_d)
&
m_{\overline{U}_i}^2 + y_{u_i}^2 v_u^2
-(v_u^2-v_d^2)\frac{g_1^2}{3} 
\end{array}
       \right), %\nonumber
\\
{\cal M}^2_{\tilde{d}} & =
\left(
\begin{array}{cc}
m_{Q_i}^2 + y_{d_i}^2 v_d^2
+(v_u^2-v_d^2)\left(\frac{g_1^2}{12}+\frac{g_2^2}{4}\right) 
&
y_{d_i}(A_{d_i} v_d - \mu v_u)\\
y_{d_i}(A_{d_i} v_d - \mu v_u)
&
m_{\overline{D}_i}^2 + y_{d_i}^2 v_d^2
+(v_u^2-v_d^2)\frac{g_1^2}{6} 
\end{array}
       \right), %\nonumber
\\
{\cal M}^2_{\tilde{\ell}} & =
\left(
\begin{array}{cc}
m_{L_i}^2 + y_{\ell_i}^2 v_d^2
+(v_u^2-v_d^2)\left(\frac{g_1^2+g_2^2}{4}\right) 
&
y_{\ell_i}(A_{\ell_i} v_d - \mu v_u)\\
y_{\ell_i}(A_{\ell_i} v_d - \mu v_u)
&
m_{\overline{E}_i}^2 + y_{\ell_i}^2 v_d^2
+(v_u^2-v_d^2)\frac{g_1^2}{2} 
\end{array}
       \right),
\end{align}
\end{subequations}
%\end{eqnarray}
\begin{equation}
M_{\chi^-} = 
\left(
\begin{array}{cc}
M_2 & g_2 v_u \\
g_2 v_d & \mu
\end{array}
     \right),
\end{equation}
where $v_{u,d} = \langle H_{u,d} \rangle$.
The determinants of these mass matrices are calculated as 
%\begin{eqnarray}
\begin{subequations}
\begin{align}
\det \left({\cal M}^2_{\tilde{u}} \right) & = 
\prod_i
\left[
m_{Q_i}^2 m_{\overline{U}_i}^2 
+y_{u_i}^2 v_u^2
\left\{m_{Q_i}^2+m_{\overline{U}_i}^2
- \left(A_{u_i}-\mu\cot\beta\right)^2\right\}
\phantom{\frac{g_1^2}{12}} \right. \nonumber
\\
& \left.\phantom{\prod_i}+(v_u^2-v_d^2)\left\{\left(\frac{g_1^2}{12}+\frac{g_2^2}{4}\right)m_{\overline{U}_i^2}-\frac{g_1^2}{3} m_{Q_i}^2\right\}
+{\cal O}(v^4) \right], %\nonumber
\\
\det \left({\cal M}^2_{\tilde{d}} \right) & = 
\prod_i
\left[
m_{Q_i}^2 m_{\overline{D}_i}^2 
+y_{d_i}^2 v_d^2
\left\{m_{Q_i}^2+m_{\overline{D}_i}^2
- \left(A_{d_i}-\mu\tan\beta\right)^2\right\}
\phantom{\frac{g_1^2}{12}} \right.\nonumber\\
& \left.\phantom{\prod_i}+(v_u^2-v_d^2)\left\{\left(\frac{g_1^2}{12}+\frac{g_2^2}{4}\right)m_{\overline{D}_i^2}+\frac{g_1^2}{6} m_{Q_i}^2\right\}
+{\cal O}(v^4) \right], %\nonumber
\\
\det \left({\cal M}^2_{\tilde{\ell}} \right) & = 
\prod_i
\left[
m_{L_i}^2 m_{\overline{E}_i}^2 
+y_{\ell_i}^2 v_d^2
\left\{m_{L_i}^2+m_{\overline{D}_i}^2
- \left(A_{\ell_i}-\mu\tan\beta\right)^2\right\}
\phantom{\frac{g_1^2}{12}} \right.\nonumber\\
&\left.\phantom{\prod_i}+(v_u^2-v_d^2)\left\{\left(\frac{g_1^2+g_2^2}{4}\right)m_{\overline{E}_i^2}+\frac{g_1^2}{2} m_{L_i}^2\right\}
+{\cal O}(v^4) \right], %\nonumber
\\
\det\left({\cal M}_{\chi^-}\right) & = M_2\mu - g_2^2 v_u v_d.
\end{align}
\end{subequations}
%\end{eqnarray}
Thus the squark-slepton contributions to $C_g^i$ and $C_\gamma^i$ are suppressed by $v^2/m_{\tilde{q}, \tilde{\ell}}^2$ at leading order and almost negligible if the squarks and sleptons are above the TeV scale. 
In addition, they are further suppressed if the following conditions are satisfied by the fields coupling via non-vanishing Yukawa interactions,
%\begin{eqnarray}
\begin{subequations}
\begin{align}
m_{Q_i}^2 + m_{\overline{U}_i}^2-(A_{u_i}-\mu \cot\beta)^2 & \approx 0, %\nonumber
\\
m_{Q_i}^2 + m_{\overline{D}_i}^2-(A_{d_i}-\mu \tan\beta)^2 &\approx 0, %\nonumber
\\
m_{L_i}^2 + m_{\overline{E}_i}^2-(A_{\ell_i}-\mu \tan\beta)^2 &\approx 0. 
\end{align}
\end{subequations}
%\end{eqnarray}
It is interesting to observe that this is exactly the prediction of the TeV scale mirage mediation,
\begin{eqnarray}
c_{H_u} = 0, && \qquad c_{\tilde{t}_L}+c_{\tilde{t}_R} = 1, \nonumber\\ 
c_{H_d} = 1, && \qquad c_{\tilde{b}_L}+c_{\tilde{b}_R} = 1, \qquad c_{\tilde{\tau}_L}+c_{\tilde{\tau}_R} = 1, 
\end{eqnarray}
with $\mu \approx M_0/\tan\beta$ \footnote{This suppression also occurs in the type I model of \cite{Choi:2005hd}.}. 
If any of the squarks or sleptons are light (e.g. $c_{\tilde{t}_R}=0$), this provides an exception of the anti-correlation between the fine-tuning of the EW symmetry breaking and the new physics contribution to the Higgs couplings \cite{Farina:2013ssa}.

The chargino contribution can be comparable to the SM contribution if the chargino mass is light enough \cite{Ellwanger:2011aa,Benbrik:2012rm,Carmi:2012in,Choi:2012he}. In the $\kappa=m_S^2=0$ limit, it is calculated as
\begin{equation}
\delta C_\gamma^i = \frac{v}{3\sqrt{2}}\frac{({\cal S}^\dag)_{3i} M_2\lambda}{M_2\mu -g_2^2 v^2 \cos\beta\sin\beta}
\approx \frac{\lambda}{3\sqrt{2}}\frac{v}{\mu}
\left\{
\begin{array}{cc}
(-\lambda) \frac{v}{A_\lambda}\sin 2\beta  & (i=h)\\
(-\lambda) \frac{v}{A_\lambda}\cos 2\beta \left(-\frac{12}{\tan^2\beta} \right) & (i=H)\\
1 & (i=S)
\end{array}
\right.
.
\end{equation}
For $i=H$ we multiplied a suppression factor, $A_s(\tau)\simeq -12/\tan^2\beta$. 
If $\mu\sim v$, $\lambda\sim 1$ and once the finite $\kappa$ and $m_S^2$ induce a mixing of ${\cal O}(10)\%$ between the singlet and the heavy doublet, then ${\cal O}(10)\%$ deviation is expected in $C_\gamma^h$.

The effective Lagrangian describing the CP-odd Higgs bosons is given by 
\begin{eqnarray}
{\cal L} &=& 
\sum_{i=1}^2
\left[
i \sum_f
{C_{f}^\prime}^i \frac{m_f}{\sqrt{2} v} a_i \overline{f} \gamma_5 f \right.\nonumber\\
&&
\left.
\phantom{+\sum_{i=1}^2}
+ {C_{g}^\prime}^i \frac{\alpha_s}{8\sqrt{2}\pi v} a_i \epsilon^{\mu\nu\rho\sigma} G^a_{\mu\nu} G^a_{\rho\sigma}+ {C_{\gamma}^\prime}^i \frac{\alpha}{8\sqrt{2}\pi v} a_i \epsilon^{\mu \nu \rho\sigma} A_{\mu\nu} A_{\rho\sigma}
\right].
\end{eqnarray} 
The renormalizable couplings between the CP-odd Higgs boson and the vector bosons are forbidden by the CP symmetry. 
We omit the loop induced couplings with $W$/$Z$ bosons because they do not give sizable effects in the production and decay of the Higgs bosons.
The coupling constants with the fermions are given by 
\begin{eqnarray}
{C_f^\prime}^i & = 
\left\{
\begin{array}{cc}
\displaystyle
({\cal P}^\dag)_{1i}\, \frac{1}{\cos\beta} & (f = e,\mu,\tau, d,s,b) \\
\displaystyle
({\cal P}^\dag)_{2i}\, \frac{1}{\sin\beta} & (f = u,c,t )
\end{array}
\right. ,
\end{eqnarray}
where ${\cal P}$ is the diagonalization matrix for the mass matrix of the CP-odd Higgs bosons. 
These coupling constants in the $\kappa=0$ limit are summarized in the table~\ref{tab:cpodd-couplings}.
Note that those of $a_1$ are highly suppressed by $\sin\gamma\simeq \lambda v/A_\lambda$ because they require the mixing with the heavy doublet. 
\begin{table}[tbp]%[t]
\centering
%\begin{center}
\begin{tabular}{|c|cc|}
%\hline
\hline
      & $a_1$ & $a_2$ \\
\hline
$m_{h_i}^2$ & 0 & $A_\lambda^2+\lambda^2 v^2$ \\
$C_{d,e}^i$ & $\tan\beta \sin\gamma$ & $\tan\beta \cos\gamma$ \\ 
$C_u^i$ & $\cot \beta \sin\gamma$ & $\cot\beta \cos\gamma$\\
\hline
%\hline
\end{tabular}
\caption{\label{tab:cpodd-couplings}Coupling constants of the CP-odd Higgs bosons in the limit $\kappa=0$, where $\tan\gamma= \lambda v/A_\lambda$.}
%\end{center}
\end{table}
The coupling constants with gluon and photon are calculated as \cite{Djouadi:2005gj, Kalyniak:1985ct},  
%\begin{eqnarray}
\begin{subequations}
\begin{align}
\delta {C_g^{\prime}}^i & = \frac{T(r_f)}{2} {C_f^{\prime}}^i A_f^\prime(\tau^i_f),%\nonumber
\\
\delta {C_\gamma^{\prime}}^i & = \frac{N(r_f) Q_f^2}{2} {C_f^\prime}^i A_f^\prime(\tau^i_f),%\nonumber\\
\end{align}
\end{subequations}
%\end{eqnarray}
where the mass function is given by
\begin{equation}
A^\prime_f(\tau) \equiv 2 \tau^{-1} f(\tau). 
\end{equation}
In the limit $\tau^i_f << 1$, they are approximated as 
%\begin{eqnarray}
\begin{subequations}
\begin{align}
\delta C_g^i & \simeq T(r_f) {C_f^\prime}^i, %\nonumber
\\
\delta C_\gamma^i & \simeq N(r_f) Q_f^2 {C_f^\prime}^i.
\end{align}
\end{subequations}
%\end{eqnarray}
The charged Higgs contribution is forbidden by the CP symmetry and the sfermion contribution is highly suppressed as in the CP-even case.
The chargino contribution to $C_\gamma$ is approximated as 
\begin{eqnarray}
\delta C_\gamma^i &\simeq& \sum_{a=1}^2 {C_{\chi^-_a}^\prime}^i = i\sqrt{2} v \frac{\partial \ln {\rm det}\left({\cal M}_{\chi^-} \right)}{\partial a_i} \simeq -\frac{v}{\mu} \lambda ({\cal P}^\dag)_{4-i\, 3} \nonumber \\
&\simeq&
\left\{
\begin{array}{cc}
-\frac{\lambda v}{\mu} & (i=1) \\
\frac{\lambda^2 v^2}{\mu^2} \sin\beta\cos^2\beta 
\left[-\frac{1}{\tan^2\beta}(\ln\tan^2\beta-i\pi)^2\right]
& (i=2) 
\end{array}
\right.
. 
\end{eqnarray}
For $i=2$, we multiplied a suppression factor $A_s^\prime(\tau)/2$ inside the bracket.
If $\mu\sim v$, $\lambda \sim 1$, ${C_\gamma^\prime}^{1}$ could be comparable with $C_\gamma^{SM}$.

Next we present the results of our numerical calculation on the coupling constants including the effect of finite $\kappa$ and $m_S^2$, the radiative corrections and various phenomenological constraints.
Here we use the scale factors of the Higgs couplings relative to the SM instead of the coupling constants themselves \cite{LHCHiggsCrossSectionWorkingGroup:2012nn,Heinemeyer:2013tqa},
\begin{equation}
\kappa_X^i = \frac{C_X^i}{C_X^{SM}}.
\end{equation}

\begin{figure}[tbp]%[htb]
\centering
%\begin{center}
\begin{tabular}{l @{\hspace{10mm}} r}
\includegraphics[height=65mm]{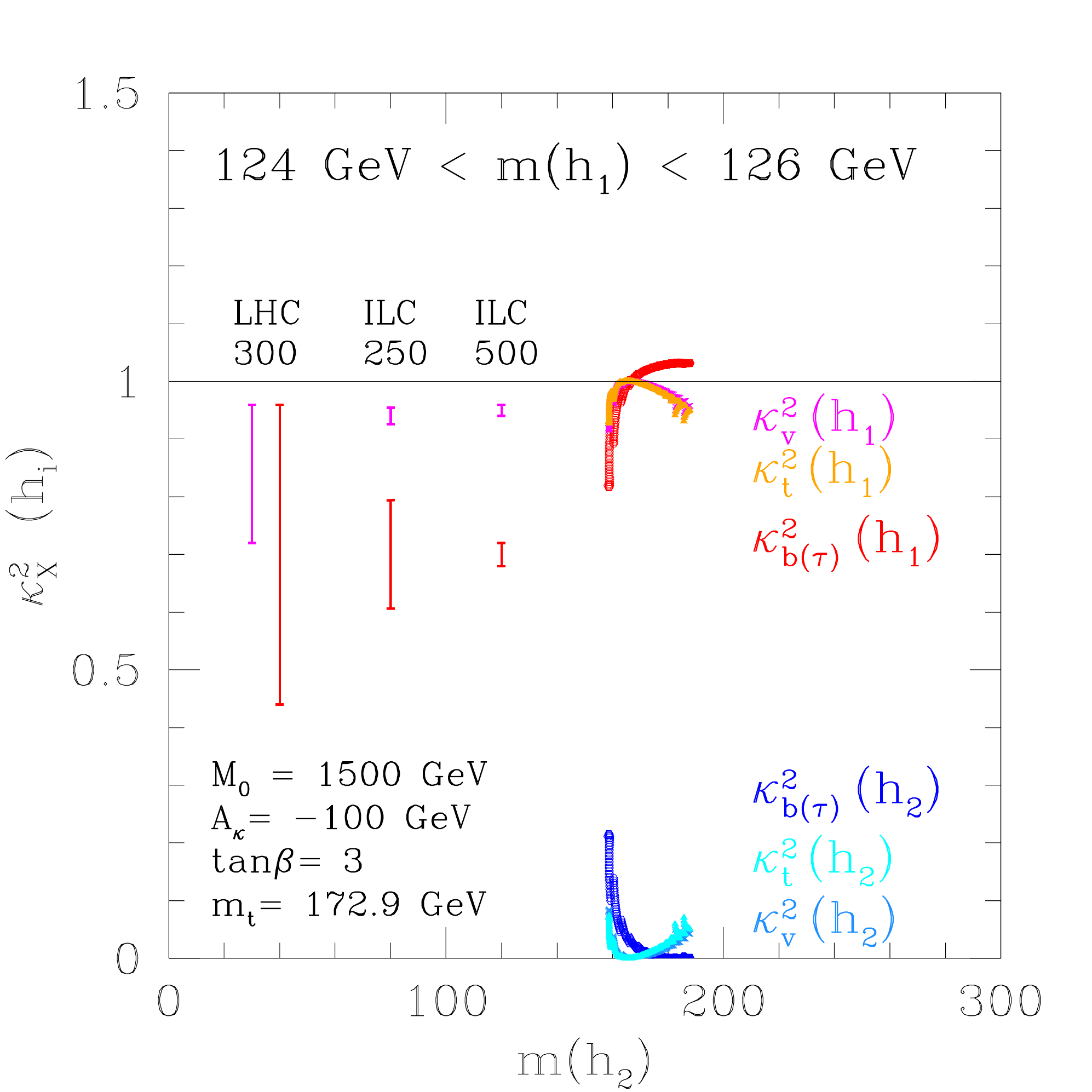} &
\includegraphics[height=65mm]{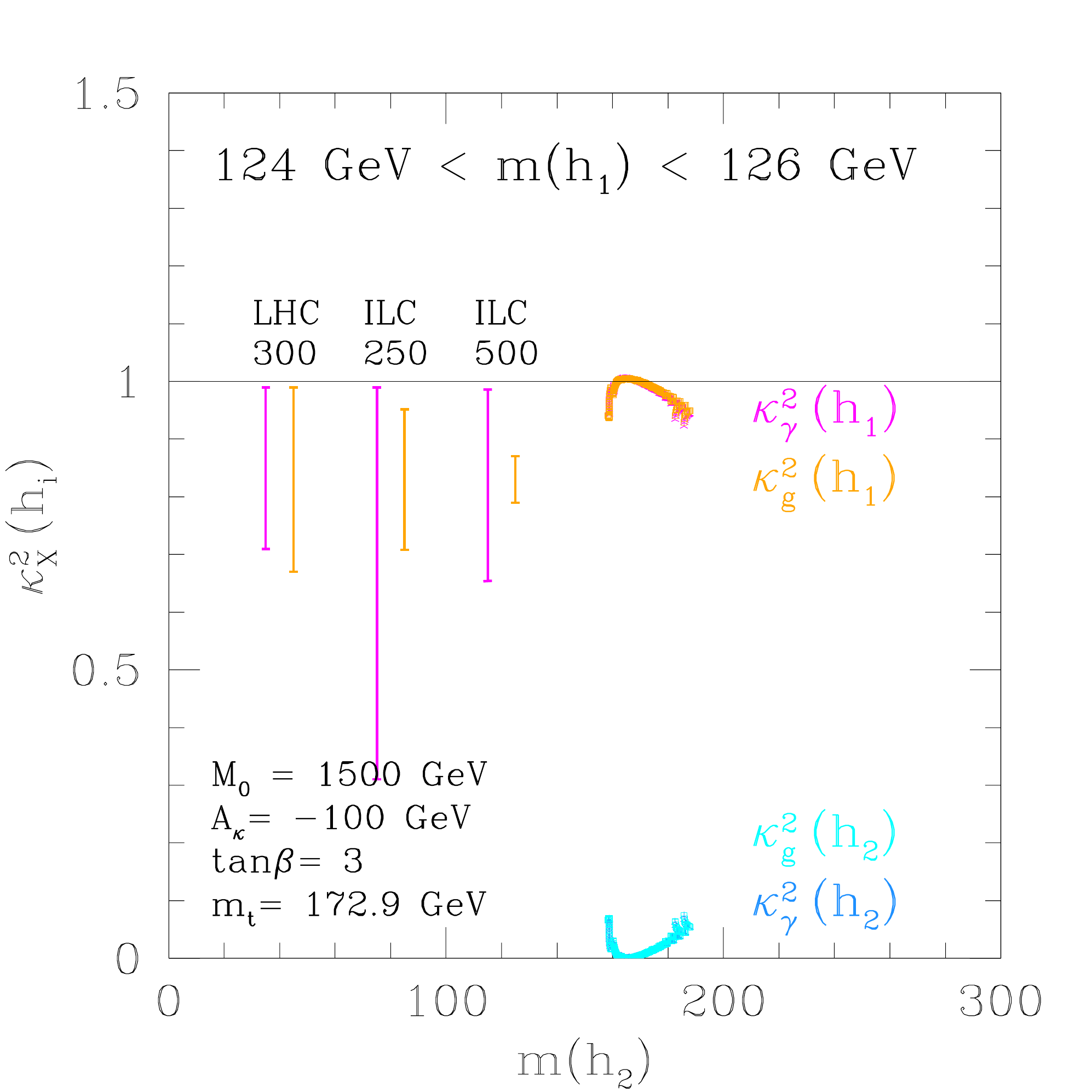} \\ 
\end{tabular}
\caption{\label{fig:cpevencouplings-tanb3} The scale factors of the CP-even Higgs coupling constants for $\tan\beta=3$, $M_0=1500$ GeV and $A_\kappa=-100$ GeV. The vertical lines show the expected precisions of the SM-like Higgs coupling constants for LHC and ILC, $\Delta (\kappa_X)^2 \approx 2 \Delta \kappa_X$.}
%\end{center}
\end{figure}

In figure~\ref{fig:cpevencouplings-tanb3}, we plot the $\kappa$'s for $\tan\beta=3$, $M_0=1500$ GeV and $A_\kappa=-100$ GeV. The horizontal axis is chosen as the mass of the singlet-like Higgs boson which is heavier than the SM-like Higgs boson for small $\tan\beta$. We impose the condition $124\, {\rm GeV} < m_{h_1} < 126\, {\rm GeV}$ and various phenomenological constraints built in the NMSSM Tools package.  
The parameter region with $124\, {\rm GeV} < m_{h_2} < 126\, {\rm GeV}$ is excluded by the LEP Higgs boson search \cite{Schael:2006cr}.
The vertical lines show the expected precision of the coupling constants for LHC ($300\,fb^{-1}$) \cite{Dawson:2013bba}, ILC ($250$ GeV) and ILC ($250$ GeV+$500$ GeV) \cite{Asner:2013psa} 
(The position of the center is arbitrary).
We also quote the numbers in table~\ref{tab:projection}. 
We plot the precision of $\kappa_Z^2$ for $\kappa_V^2$ and take conservative values for the LHC expectation.

\begin{table}[tbp]%[htb]
\centering
%\begin{center}
\scalebox{0.8}{
\begin{tabular}{|c|cccccc|}
%\hline
\hline
Facility & LHC & HL-LHC & ILC(250) & ILC(500) & ILC(1000) & ILC(LumUP)\\
$\sqrt{s}$ (GeV) & 1,400 & 1,400 & 250 & 250+500 & 250+500+1000 &250+500+1000\\
$\int {\cal L} dt$ ($fb^{-1}$)& 300/expt & 3000/expt & 250 & 250+500 & 250+500+1000 & 1150+1600+2500\\
\hline 
$\kappa_\gamma$     & 5-7\%   & 2-5\% & 17\%  & 8.3\% & 3.8\% & 2.3\% \\
$\kappa_g$         & 6-8\%   & 3-5\% & 6.1\% & 2.0\% & 1.1\% & 0.7\% \\
$\kappa_W$         & 4-6\%   & 2-5\% & 4.7\% & 0.4\% & 0.3\% & 0.2\% \\
$\kappa_Z$         & 4-6\%   & 2-4\% & 0.7\% & 0.5\% & 0.5\% & 0.3\% \\
$\kappa_\ell$       & 6-8\%   & 2-5\% & 5.2\% & 1.9\% & 1.3\% & 0.7\% \\
$\kappa_b$ & 10-13\% & 4-7\% & 4.7\% & 1.0\% & 0.6\% & 0.4\% \\
$\kappa_t$ & 14-15\% & 7-10\%& 6.4\% & 2.5\% & 1.3\% &  0.9\% \\
\hline
%\hline
\end{tabular}
}
\caption{Expected precisions on the Higgs coupling for LHC \cite{Dawson:2013bba} and ILC \cite{Asner:2013psa}. Two numbers for LHC represent different assumptions on the systematic error.
The seven parameter HXSWG benchmark parametrization \cite{LHCHiggsCrossSectionWorkingGroup:2012nn,Heinemeyer:2013tqa} is assumed.\label{tab:projection}}
%\end{center}
\end{table}

\begin{figure}[tbp]%[htbp]
\centering
%\begin{center}
\begin{tabular}{l @{\hspace{10mm}} r}
\includegraphics[height=65mm]{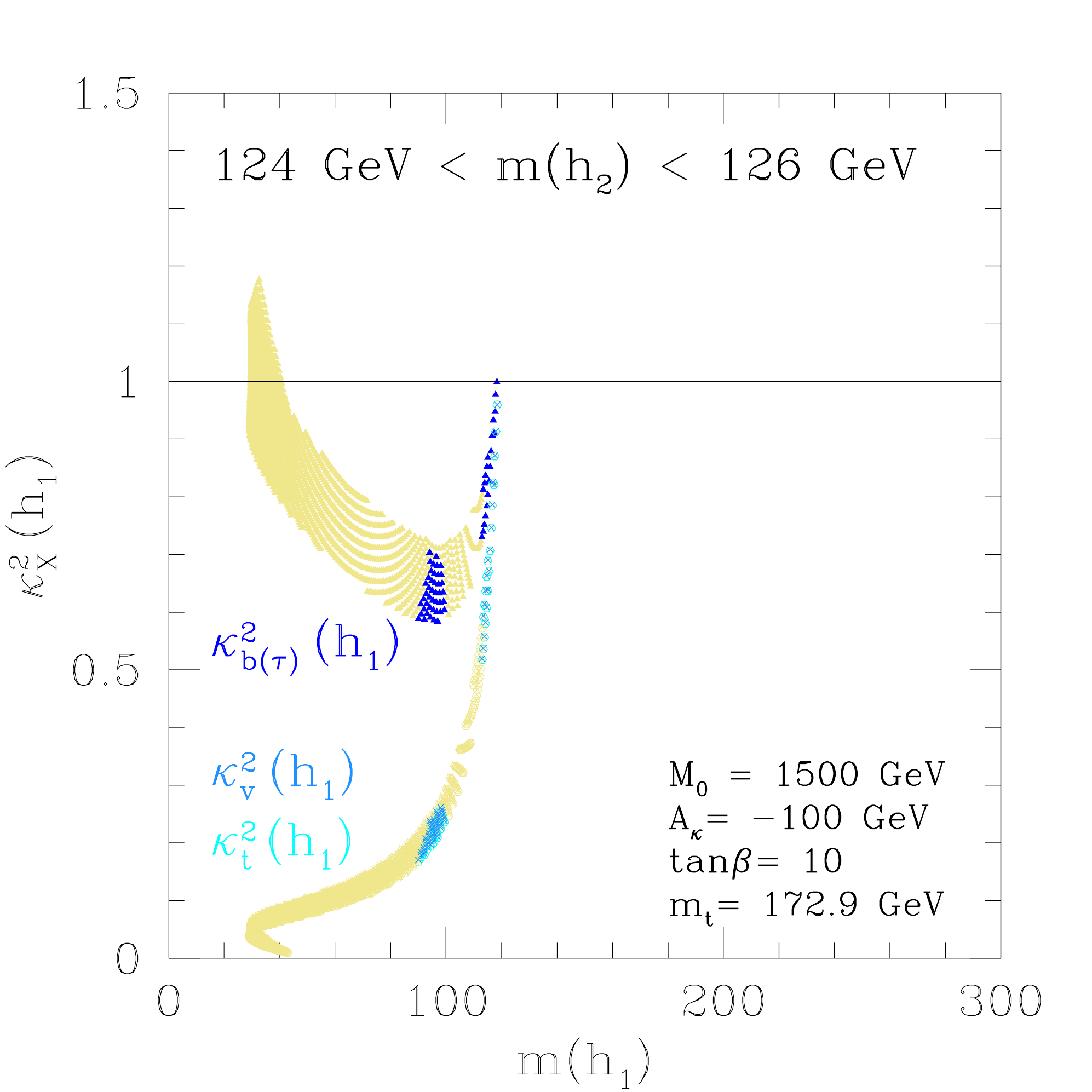} &
\includegraphics[height=65mm]{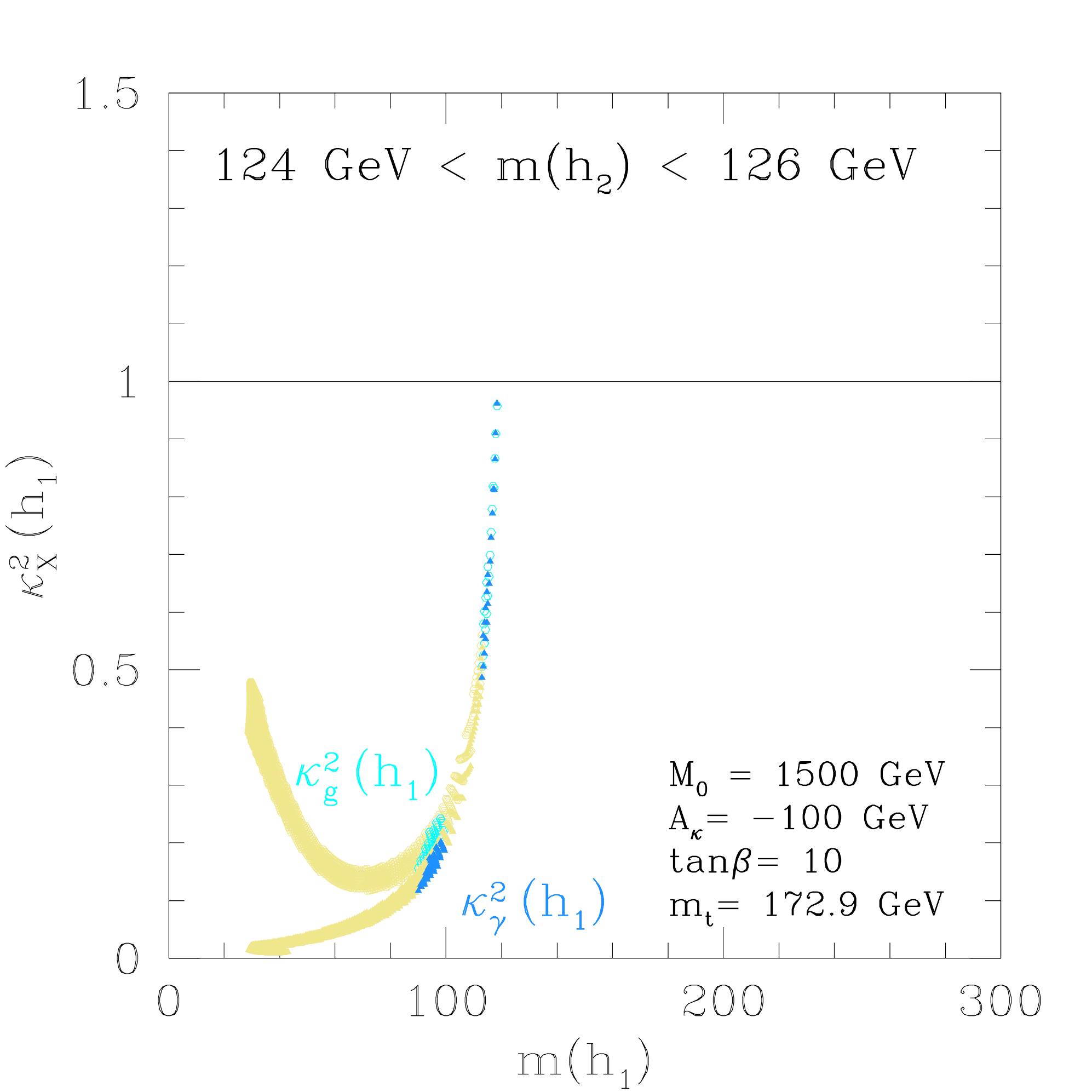} \\
\rule{0cm}{10mm} & \\
\includegraphics[height=65mm]{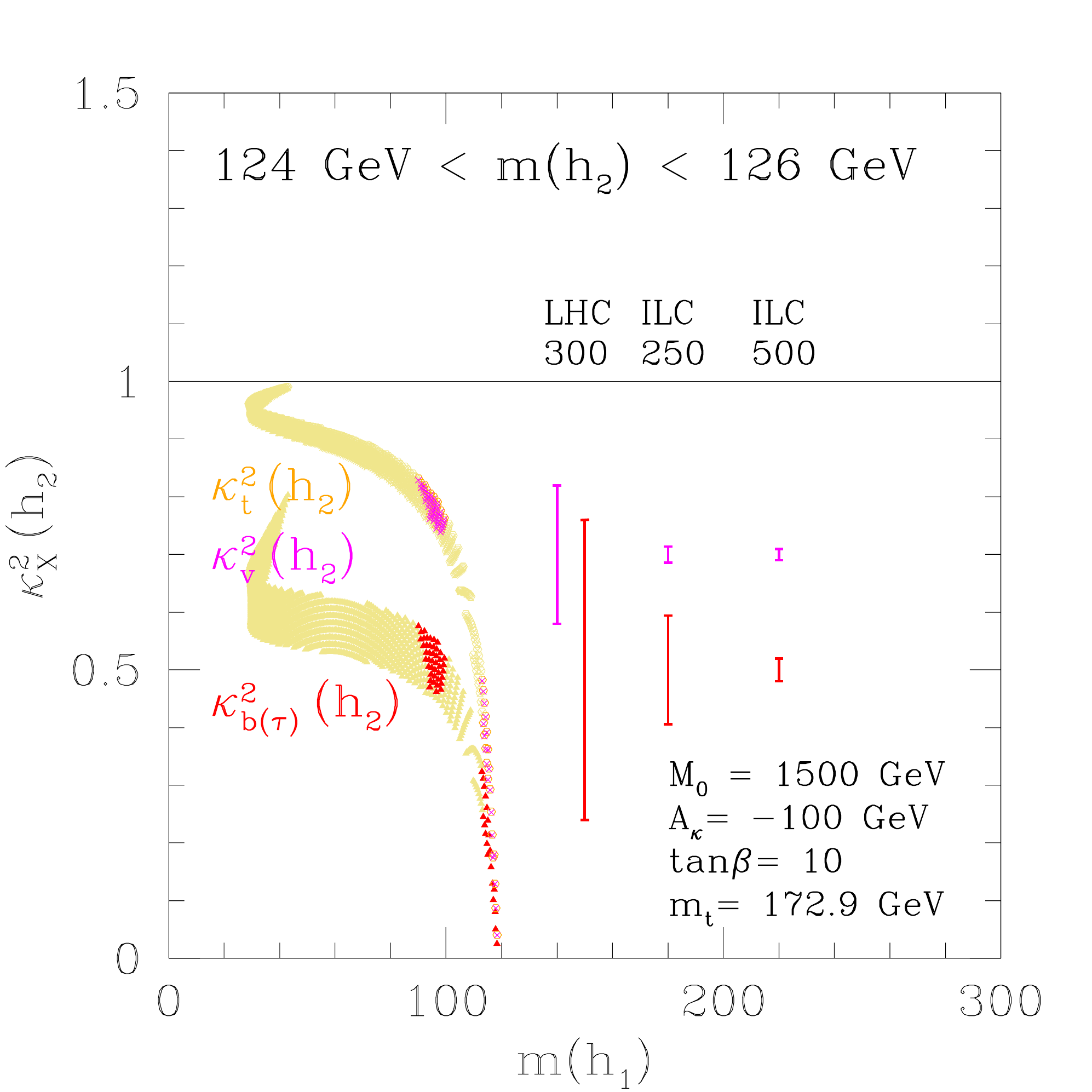} &
\includegraphics[height=65mm]{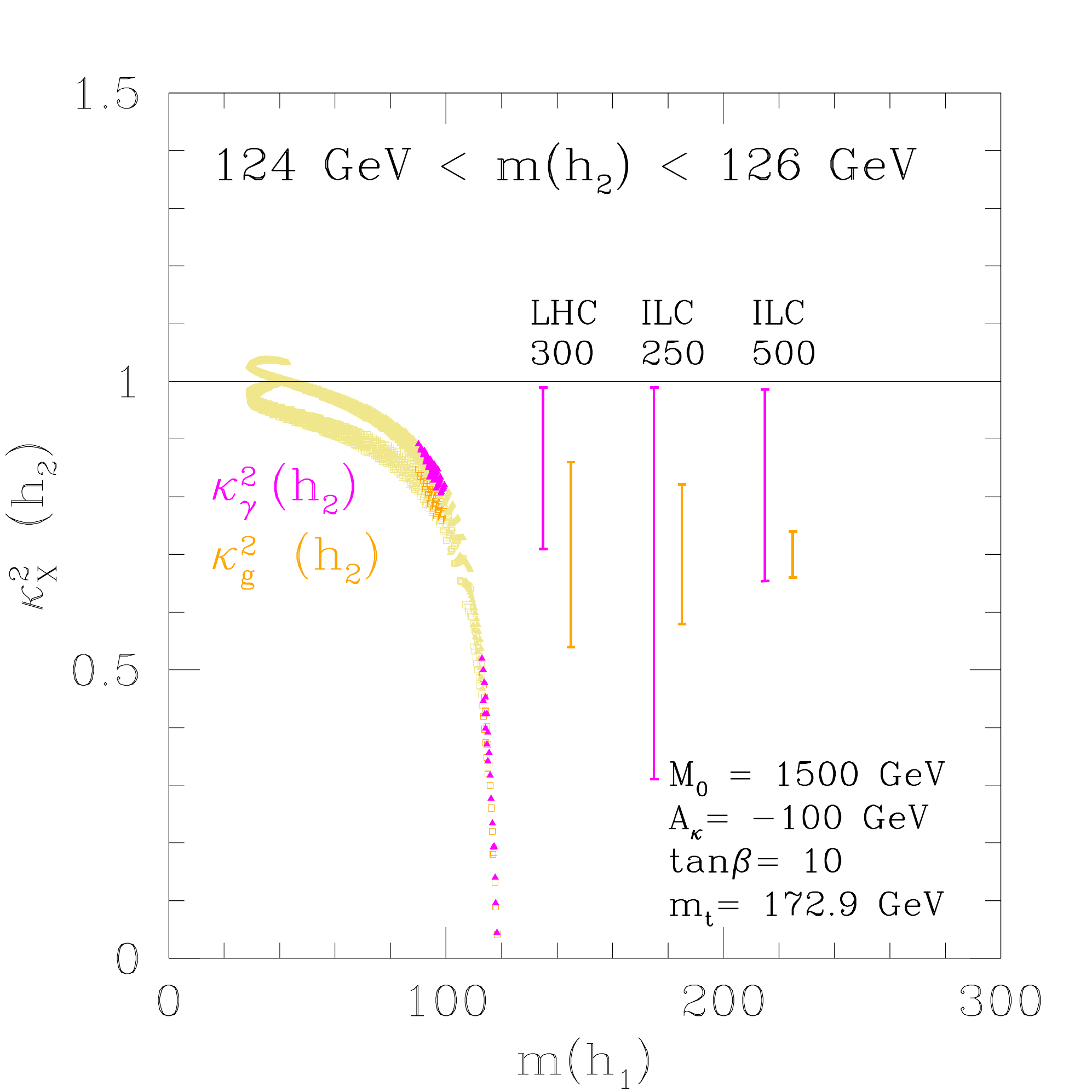} \\ 
\end{tabular}
\caption{\label{fig:cpevencouplings-tanb10}The scale factors of the CP-even Higgs coupling constants for $\tan\beta=10$, $M_0=1500$ GeV and $A_\kappa=-100$ GeV.
The vertical lines show the expected precisions of the SM-like Higgs coupling constants for LHC and ILC, $\Delta (\kappa_X)^2 \approx 2 \Delta \kappa_X$.
$(\kappa_V^1)^2$ and $(\kappa_t^1)^2$ ($(\kappa_V^2)^2$ and $(\kappa_t^2)^2$) in the upper (lower) left panel almost overlap each other.
}
%\end{center}
\end{figure}

The pseudo Nambu-Goldstone nature of the SM-like Higgs boson well suppresses the mixing with the singlet and the deviation from the SM is at most $20$ \%.
This suppression is also important for not reducing the lighter SM-like Higgs boson mass due to the mixing.
The approximate sum rule $(\kappa_X^1)^2+(\kappa_X^2)^2=1$ means that the deviation mainly comes from the mixing with the singlet and the SUSY contribution is negligible except for $b\,(\tau)$ for which the contamination of the heavy doublet has $\tan\beta$ enhanced coupling.
The singlet-like Higgs boson is hidden from the collider search and an interesting target for the precision measurement at the LHC and ILC.
In this parameter region, the prediction does not change significantly even if we adopt a conservative bound, $120\, {\rm GeV} < m_{h_1} < 130\, {\rm GeV}$, taking into account the theoretical error \cite{Staub:2015aea}.

In figure~\ref{fig:cpevencouplings-tanb10}, we plot the same scale factors (normalized by the SM Higgs couplings) for $\tan\beta=10$. 
Increasing $\tan\beta$, the singlet-like Higgs boson is now lighter than the SM-like Higgs boson and we impose $124\, {\rm GeV} < m_{h_2} < 126\, {\rm GeV}$.
We again choose the horizontal axis as the singlet-like Higgs mass.
The tree-level contribution to the SM-like Higgs mass is now not effective with larger $\tan\beta$ and small mixing with the singlet is favorable to achieve $125$ GeV. 
However, a strong constraint exists for the light singlet from the LEP Higgs boson search if it couples with the SM particles through the mixing with the SM-like Higgs boson \cite{Schael:2006cr}.
The khaki shaded region is mostly excluded by this constraint.
The region $90\,{\rm GeV} \lesssim m_{h_1} \lesssim 100\,{\rm GeV}$ survives due to a small excess in the LEP measurement \cite{Schael:2006cr,Belanger:2012tt,Kobayashi:2012ee,Drees:2005jg,Dermisek:2005gg}.
Around $m_{h_1} \simeq 120$ GeV, the two bosons almost degenerate and the mixing is maximized, but the LEP bound is weak in this region.
For large $\lambda$, the new radiative corrections to the SM-like Higgs mass, eq.\eqref{eq:nmssm-higgs-radiative-correction} could relax the mixing with the singlet (around $m_{h_1}\simeq 40$ GeV). 
In this parameter region, the lightest neutralino becomes lighter than
 the half of the $Z$ boson mass and the decay channel, $Z \to 2 \chi^0_1$ opens.
Then the region is excluded by the measurement of the $Z$ boson width \cite{ALEPH:2005ab}.
The mixing with the singlet leads to large deviations in the coupling constants of the SM-like Higgs boson observable at the LHC. 
On the other hand, searching the light singlet boson missed by LEP is important physics at ILC.  The sum rule holds again except for $b(\tau)$ and $g$, for which the mixing with the heavy doublet is not negligible due to the $\tan\beta$ enhancement. The small excess in $(\kappa_\gamma^2)^2$ around $m_{h_1}\simeq 30$ GeV is contribution from the chargino interfering with that of the SM.
The surviving region expands once we raise the SUSY breaking scale $M_0$ and increase the top radiative correction to the SM-like Higgs mass.
Also note that the prediction is sensitive to the theoretical error in the Higgs mass calculation. If we take a conservative bound, e.g. $120\, {\rm GeV} < m_{h_2} < 130\, {\rm GeV}$, much lighter $h_1$ ($m_{h_1} \gtrsim 20$ GeV) is allowed because the singlet-doublet mixing is suppressed. For such $m_{h_1}$, the $h_1$ couplings to the SM particles and the deviations of $h_2$ couplings from those of the SM Higgs boson are highly suppressed except for $b$ and $\tau$.

\begin{figure}[tbp]%[htbp]
\centering
%\begin{center}
\begin{tabular}{l @{\hspace{10mm}} r}
\includegraphics[height=65mm]{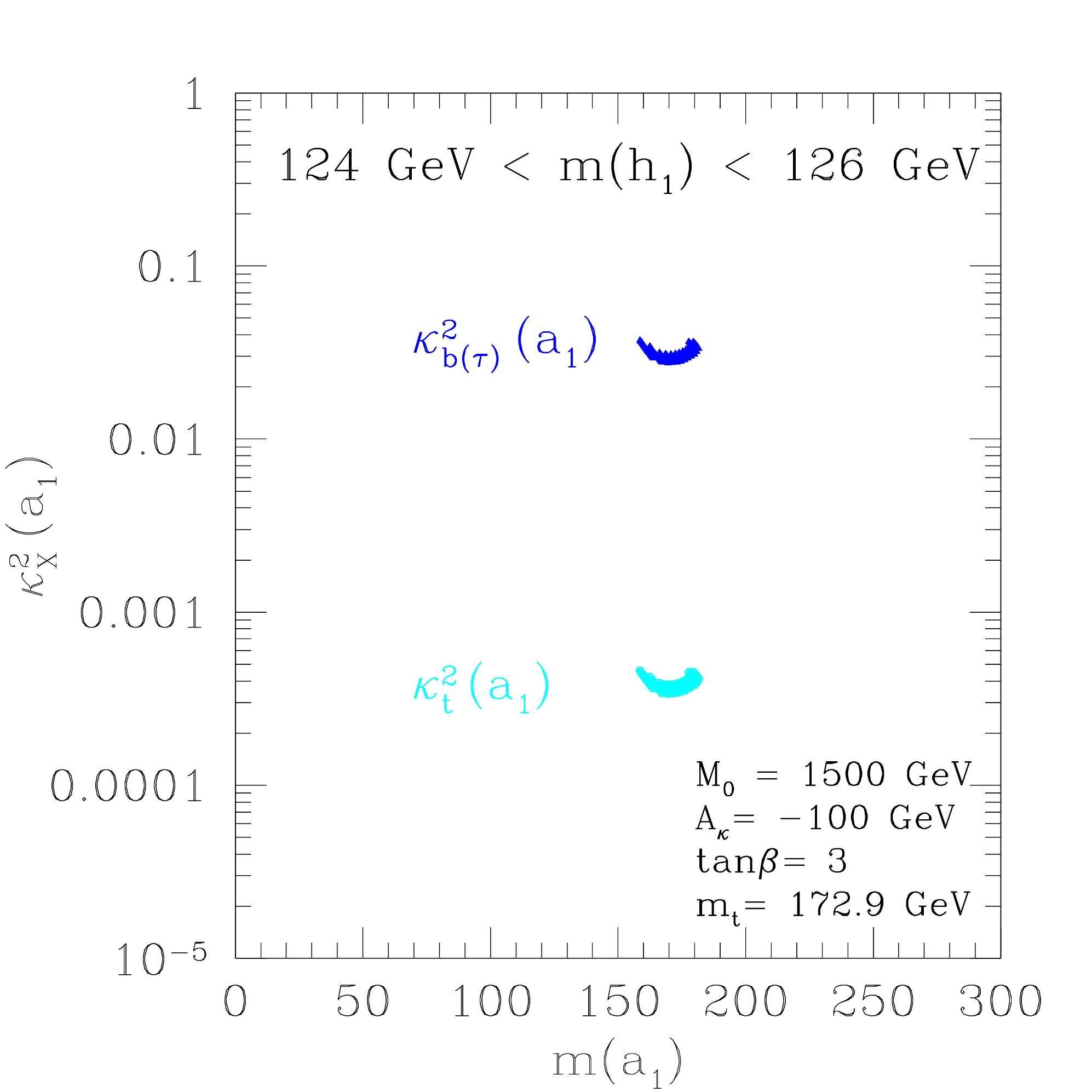} &
\includegraphics[height=65mm]{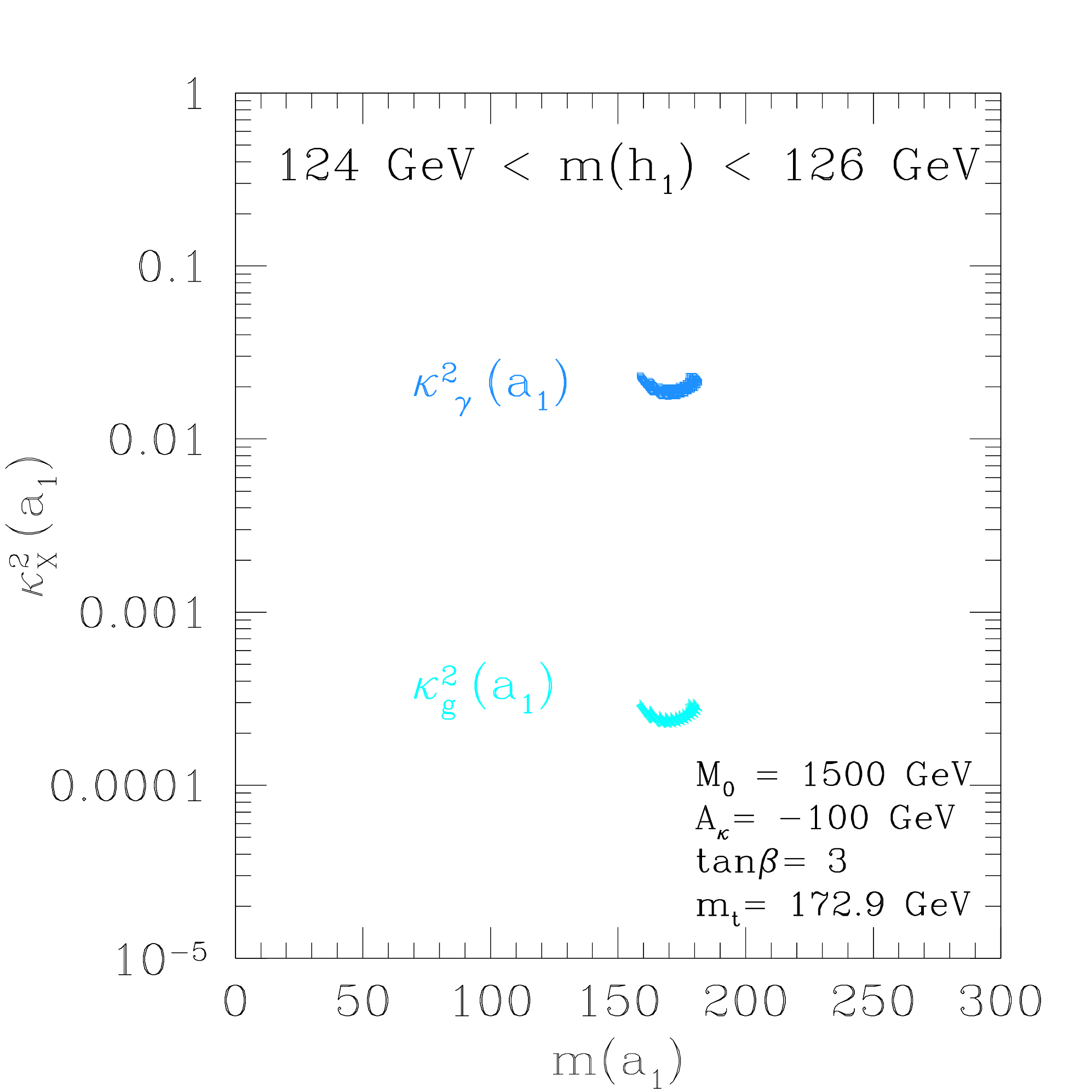} \\
\rule{0cm}{10mm} & \\
\includegraphics[height=65mm]{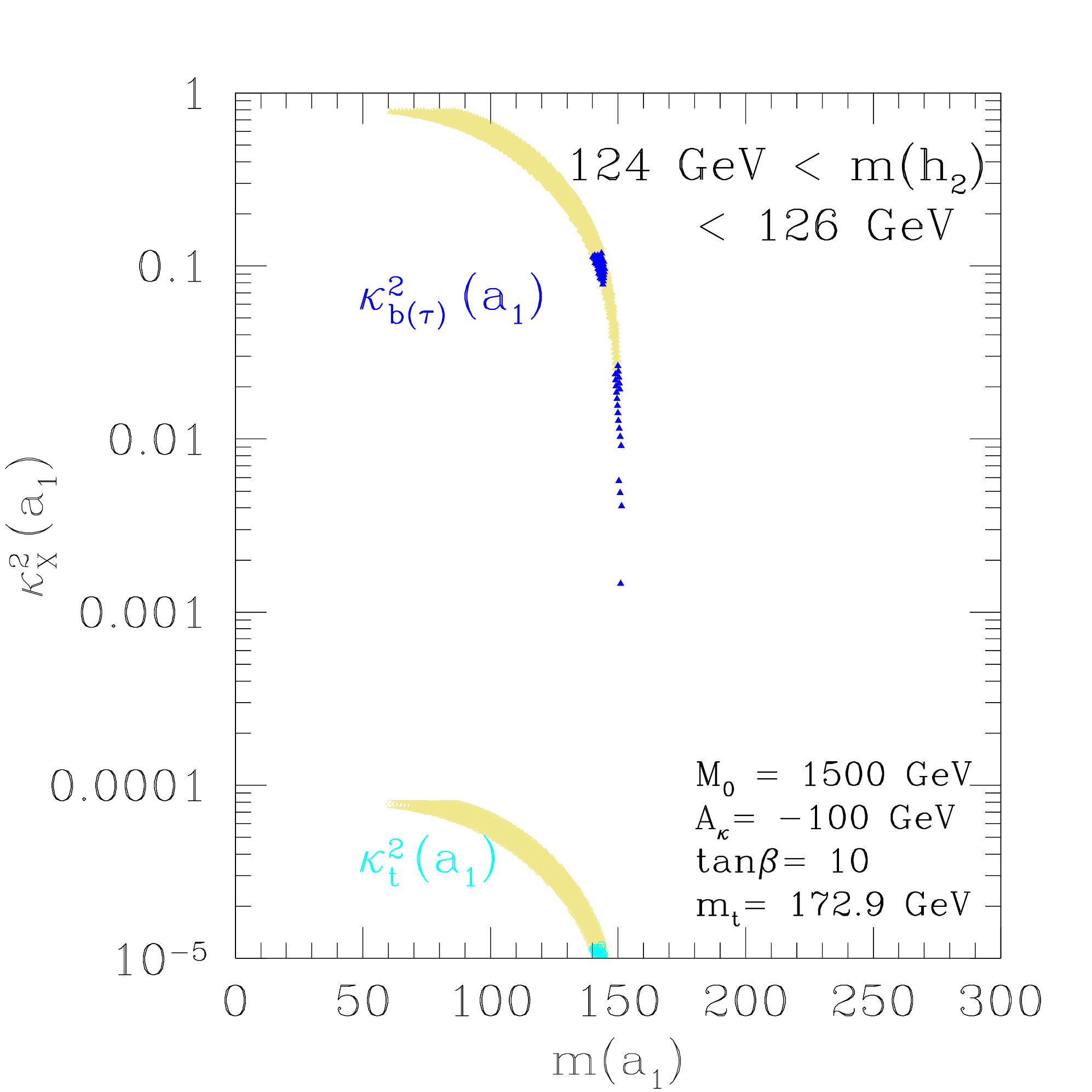} &
\includegraphics[height=65mm]{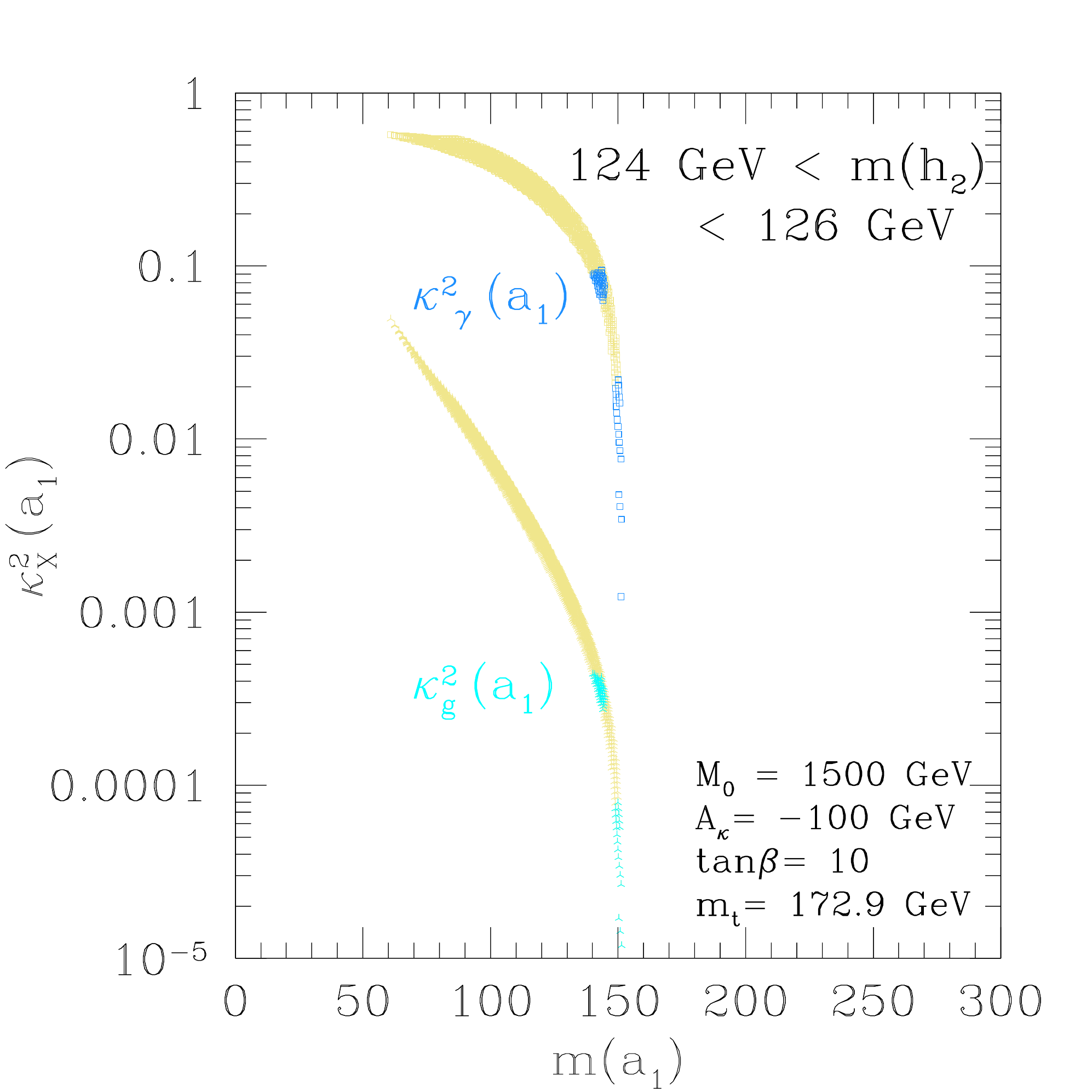} \\ 
\end{tabular}
\caption{\label{fig:cpoddcouplings}The scale factors for the CP-odd Higgs boson coupling constants for $\tan\beta=3$ (upper panel), $\tan\beta=10$ (lower panel), $M_0=1500$ GeV and $A_\kappa=-100$ GeV.}
%\end{center}
\end{figure}

In figure~\ref{fig:cpoddcouplings}, we show the scale factors of the lightest CP-odd Higgs boson for the same parameter set with the CP-even bosons. The horizontal axis is chosen as the mass of $a_1$, which is rather sensitive to the choice of the PQ symmetry breaking parameter, $A_\kappa$. 
$({\kappa_t^\prime}^1)^2$ and $({\kappa_g^\prime}^1)^2$ are doubly suppressed by $\tan^{-2}\beta$ and $\sin^2\gamma$ and negligible.
The direct production of $a_1$ at the LHC is almost hopeless and the gamma-gamma option of ILC may help if $({\kappa_\gamma^\prime}^1)^2$ is large enough.

\subsection{Dark Matter}

The existence of the dark matter (DM) is one of the few compelling evidences of physics beyond the SM.
In SUSY, the Lightest Supersymmetric Particle (LSP) is stable and good candidate of the dark matter, once we assume the R-parity to forbid the dimension 4 baryon/lepton number violating operators to prevent rapid nucleon decay. 
If LSP is weakly interacting massive particle (WIMP), its annihilation cross section predicts the thermal relic abundance in the ballpark of the observed DM density. This ''WIMP miracle'' scenario provides a valuable constraint on the SUSY models in light of the recent progress of precision cosmology (for more details, see e.g. \cite{Jungman:1995df}).
However, in the mirage mediation, the oscillation of the modulus field after the inflation once dominates the energy density of the universe.
The reheating temperature of the modulus field is ${\it O}(100)$ MeV.
It is much larger than the temperature at which the Big Bang Nucleosynthesis (BBN) takes place, however, lower than the decoupling temperature of LSP.
Huge entropy produced by the modulus decay dilutes the preexistent 
thermal LSP abundance to negligible level.
The modulus decay produces non-thermal LSP.
In particular, daughter gravitino later decays at the decay temperature ${\it O}(10)$ MeV and overproduces LSP (moduli-induced gravitino problem \cite{Nakamura:2006uc,Endo:2006zj}).
This non-thermal overproduction of DM does not mean the model is excluded, if we consider a specific non-minimal cosmological scenario or modify the DM sector of the model. For example,
\begin{enumerate}
\item Suppressing the modulus oscillation or reducing the gravitino branching ratio
(e.g. enhanced symmetry \cite{Dine:1995kz}, adiabatic oscillation \cite{Linde:1996cx, Nakayama:2011wqa}, alignment \cite{Dine:2006ii}).
\item Violating R-parity and introducing another source of DM (e.g. axion \cite{Bose:2013fqa})  or introducing thermally decoupled light LSP (e.g. axino \cite{Nakamura:2008ey}).
\item Diluting the LSP after the gravitino decay (e.g. by thermal inflation \cite{Lyth:1995hj, Lyth:1995ka, Choi:2012ye}, Q-ball \cite{Kawasaki:2007yy} or unstable domain wall \cite{Hattori:2015xla}).  
\end{enumerate} 
Explicit model building is beyond the scope of this paper. In this section, we   just present the prediction of our model with the minimal DM sector assuming
 the first or the third scenario. 
The calculation is performed based on \texttt{micrOMEGAs} 
 built in \texttt{NMSSMTools} \cite{Ellwanger:2005dv}.

In our model, LSP is given by mixture of the higgsino and singlino.
Generally, the thermal relic of the pure higgsino of {\it O}(100) GeV
 cannot saturate the observed DM density. However, mixing with the 
singlino reduces the annihilation cross section and enables to explain
 the observed amount of DM (Well-tempered neutralino)  in the first scenario
as in the case of bino and higgsino in MSSM \cite{Kobayashi:2012ee,
 ArkaniHamed:2006mb}. (There is no constraint in the third scenario.)

The searches of DM are classified into the direct and indirect detections.
The former detects the scattering of target nucleus by DM
 at underground laboratories.
The latter observes the annihilation/decay products of DM in celestial/astrophysical circumstances such as the center of the earth/sun or galaxies.

\begin{figure}[tbp]%[htbp]
\centering
%\begin{center}
\begin{tabular}{l @{\hspace{10mm}} r}
\includegraphics[height=65mm]{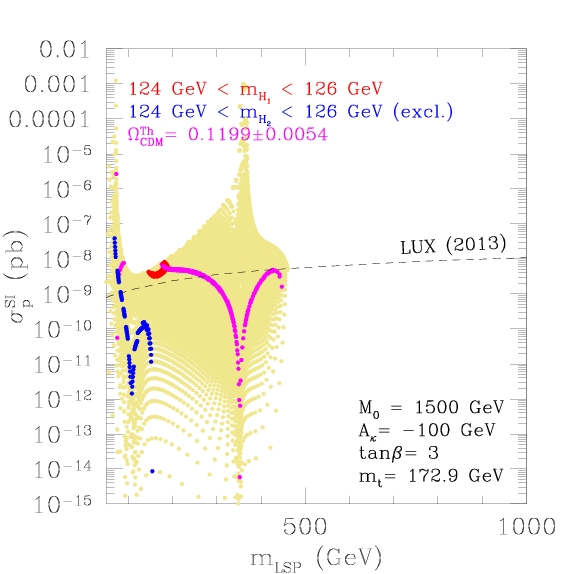} &
\includegraphics[height=65mm]{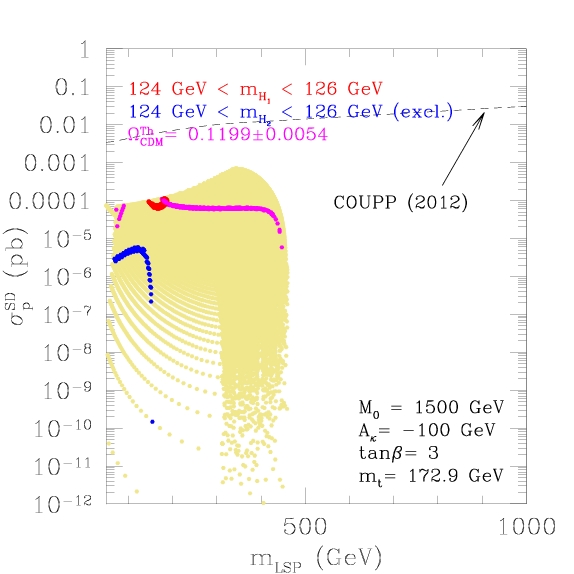}
\\
\rule{0cm}{10mm} & \\
\includegraphics[height=65mm]{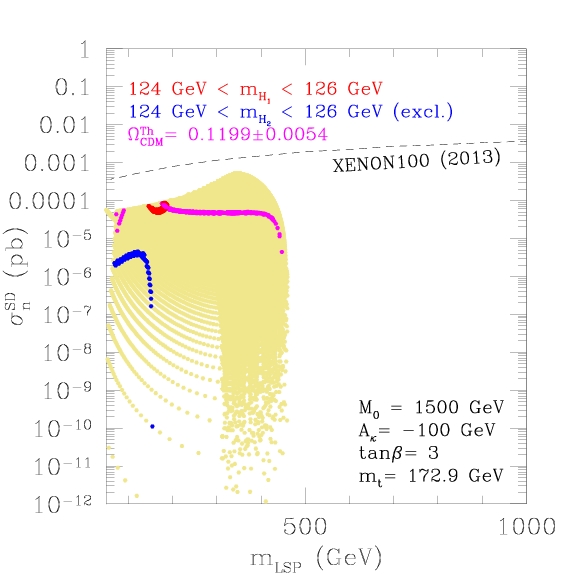}
 & 
\includegraphics[height=65mm]{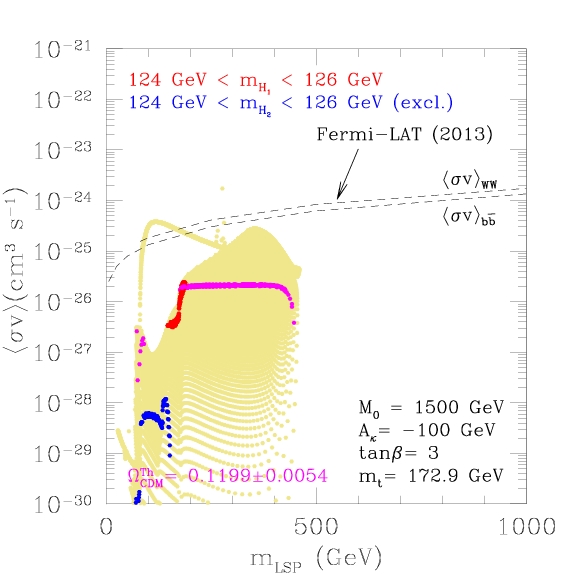} \\ 
\end{tabular}
%\end{center}
\caption{
The predictions for the direct and indirect searches
 of the dark matter for $\tan\beta=3$, $M_0=1500$ GeV and $A_\kappa=-100$ GeV.  \label{fig:DMtanb3}}
\end{figure}

We first discuss the prediction for the direct detection. 
In the direct detection experiments,
 the scattering between (slow) neutralino and target nucleus proceeds through
 the scalar or pseudo-vector interactions due to its Majorana nature.
The former gives the spin-independent cross section
 and the latter gives the spin-dependent cross section \cite{Jungman:1995df}.
In our model, the scalar interaction mainly comes from the exchange of the CP-even Higgs bosons and the pseudo-vector interaction mainly comes from the exchange of the $Z$ boson.
The spin-independent nuclear cross section is enhanced by the square of the mass number because the process is coherent, while the spin-dependent cross section does not have such a factor. Thus the experiments are much sensitive to the spin-independent cross section than the spin-dependent ones at nucleon level.
On the other hand, the $Z$ boson interacts with nucleon much stronger than the Higgs bosons. Therefore both cross sections could give meaningful constraints on our model, depending on the amount of higgsino component in LSP.

In the upper left panel of figure~\ref{fig:DMtanb3}, we show the spin-independent cross section of LSP for $\tan\beta=3$ case. The red region indicates the region where the lightest CP-even boson corresponds to the SM Higgs boson and the blue region indicates the region where the second lightest CP-even boson is the SM-like Higgs boson (although excluded by the other phenomenological constraints). In the pink region, the thermal abundance saturates the observed value of DM density. 
Around $m_{LSP} \approx 100, 350$ GeV, the scattering amplitudes mediated by the lightest and second lightest CP-even bosons cancel each other and the cross section vanishes \cite{Cerdeno:2004xw, Kang:2012sy}. 
The neutralino-Higgs boson coupling involves $\lambda$, therefore the red region is on the upper edge of the distribution because almost maximum $\lambda$ is required to raise the Higgs mass to $125$ GeV. 
With the particular model parameters chosen in the figure, the red region overlaps with the pink region where the thermal abundance saturates the DM, however, is almost excluded by the LUX experiment \cite{Akerib:2013tjd}. 
If we increases $M_0$, the cross section decreases while the pink region remains almost same.
In the upper right and lower left panels, we present the spin-dependent cross sections for proton and neutron, respectively. 
The red region is again on the edge of the distribution because the singlino-higgsino mixing terms in the mass matrix are proportional to $\lambda$.
The current experimental bounds are 1-2 order above the model prediction \cite{COUPP, Aprile:2013doa}.

Next we discuss the indirect detection. We consider the constraint by the gamma ray from the dwarf spheroidal galaxies observed by Fermi satellite \cite{Ackermann:2013yva}.
The annihilation of the neutralinos mainly produces $W^+ W^-$, $Z^0 Z^0$ pairs, $t \bar{t}$, $b \bar{b}$ pairs, $h_i a_j$ and $h_i Z^0_j$ pairs if they are kinematically available. Other channels are strongly suppressed due to the Majorana nature of the neutralino and CP conservation \cite{Jungman:1995df}. 
In the lower right panel of figure~\ref{fig:DMtanb3}, we show the total annihilation cross section of the neutralino and the experimental bound assuming the decay proceeds through $W^+W^-$ and $b \bar{b}$ pairs.
In our model, $t\bar{t}$ decay dominates the process above the threshold.
The bound is one order larger than the prediction assuming that the gamma ray spectrum from $t\bar{t}$ is not much different from those of $WW$, $b\bar{b}$ pairs.

Here we impose $124\, {\rm GeV} < m_{h_1} < 126\, {\rm GeV}$,  however, the conservative Higgs mass bound, $120\, {\rm GeV} < m_{h_1} < 130\, {\rm GeV}$ does not change the results for $\tan\beta=3$ significantly.

\begin{figure}[tbp]%[htbp]
\centering
%\begin{center}
\begin{tabular}{l @{\hspace{10mm}} r}
\includegraphics[height=65mm]{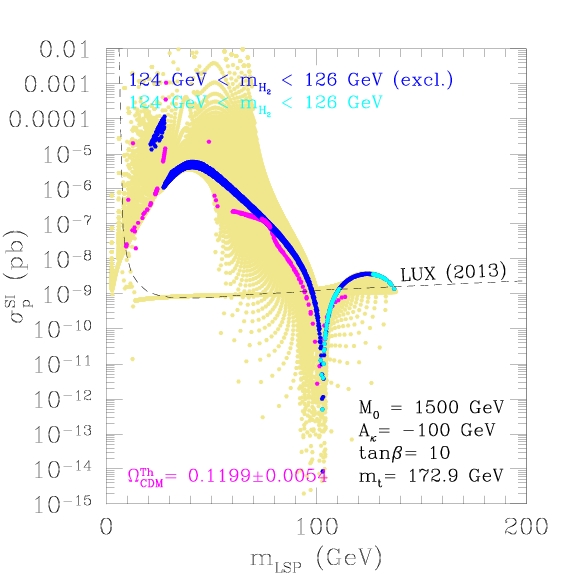} &
\includegraphics[height=65mm]{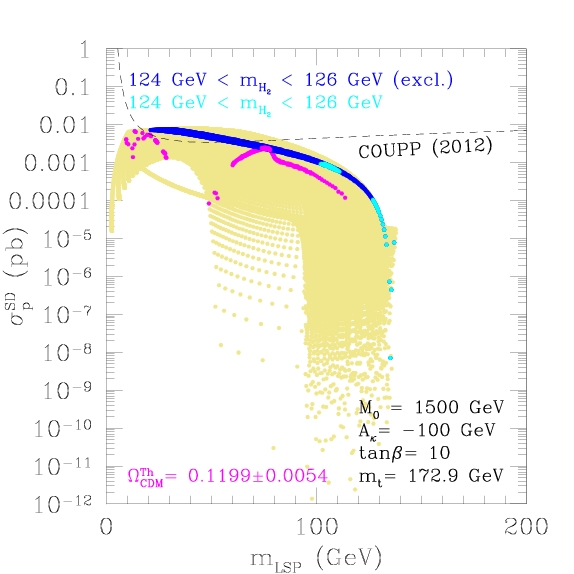}\\
\rule{0cm}{10mm} & \\
\includegraphics[height=65mm]{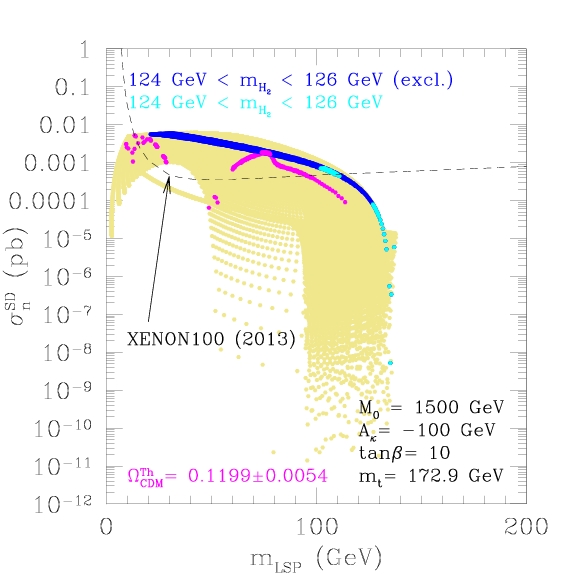} & 
\includegraphics[height=65mm]{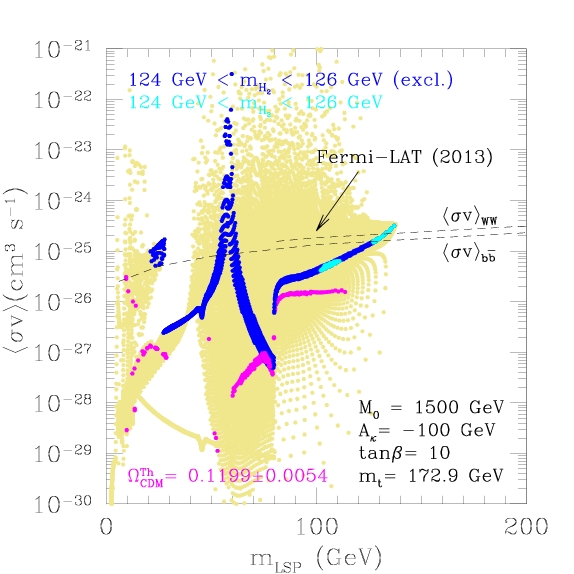} \\ 
\end{tabular}
%\end{center}
\caption{The predictions for the direct and indirect searches
 of the dark matter for $\tan\beta=10$,　$M_0=1500$ GeV and $A_\kappa=-100$ GeV.\label{fig:DMtanb10}}
\end{figure}

In figure~\ref{fig:DMtanb10}, we present the same set of the predictions for $\tan\beta=10$ case.
The blue and cyan regions indicate the region where the second lightest CP-even boson corresponds to the SM Higgs boson. The blue region is already excluded by the phenomenological constraints explained in the previous section.
In the pink region, the thermal LSP abundance saturates the observed DM density. In the surviving cyan region, the annihilation cross section is larger than the value saturating the observed DM density.
The spin-independent cross section (upper left panel) shows a dip around $m_{LSP}\approx 100$ GeV due to the cancellation between the amplitudes mediated by the two CP even bosons.
While the spin-dependent cross sections (upper right and lower left panels) do not show such a behavior because only one diagram dominates the process.
They are enhanced with $\tan\beta$ because $\mu \approx M_0/\tan\beta$ and the higgsino component in LSP increases with $\tan\beta$.
The cyan region is marginally excluded by the spin-independent cross section and the spin-dependent cross section for neutron. Discounting the constraints considering the astrophysical, hadronic and nuclear physics ambiguities in the calculation (see e.g. recent analysis in \cite{Marcos:2015dza}), it is interesting that the surviving region is exactly located around the dip structure
 where the fine-tuning of the EW symmetry breaking is also minimized. 
Once observed, this feature can be checked by the experiment using the target nucleus insensitive to the spin-dependent cross section (enriched even-even nucleus) or combining the results of different target nuclei.
In the lower right panel, the resonance structure in the annihilation cross section appears due to the pseudo-scalar mediated $b \bar{b}$ decay, however, this region is excluded by the phenomenological constraints. Above the $W^+W^-$, $Z^0Z^0$ threshold, the process is dominated by $W^+W^-$ and $Z^0Z^0$ decays. The surviving region is about a factor $2$ below the current bound.
Once we raise the SUSY scale, $M_0$, the constraint from the LEP Higgs boson search becomes weaker until the radiative correction overshoots 125 GeV and the cyan regions in the figures expand, while the EW fine-tuning deteriorates as shown in figure~\ref{fig:heavy-case}.  
The conservative Higgs mass bound, $120\, {\rm GeV} < m_{h_1} < 130\, {\rm GeV}$ opens a lighter LSP region ($m_{LSP} \simeq 50$ GeV). However, this region is excluded by the upper bound of the spin-independent cross section, if the dominant component of the DM consists of the lightest neutralino. 

In this section, we discussed the DM prediction of the model assuming the LSP saturates the observed DM abundance. We stress that the novel structure of the EW symmetry breaking in our model is not directly connected to the DM physics and the model will not be excluded even if the prediction of the minimal model is not confirmed by the experiments.

\section{Conclusion}

We have studied the NMSSM with the TeV scale mirage mediation.
Our choice of the modular weights \eqref{eq:ci} realizes a little hierarchy
 $m_{H_u}^2\simeq M_0^2/8\pi^2<< M_0^2$ at the EW scale 
and automatic cancellation between 
$\mu^2$ and $m_{H_d}^2/\tan^2 \beta$ in \eqref{eq:mZ}.
This leads to a natural EW symmetry breaking with
 ${\it O}(10)$\% tuning for the TeV scale stop mass.
The $\mu$ term can be as large as $\sqrt{m_Z M_0}$ without introducing a fine-tuning despite conventional wisdom tells that it lies around the EW scale.
The Higgs boson mass is lifted to $125$ GeV by the new quartic coupling (small $\tan\beta)$ or small mixing with the singlet (moderate $\tan\beta$)
 in addition to the radiative corrections (from the top and singlino sectors).
The suppression of the singlet-doublet mixing is important for raising the Higgs boson mass in the first case and hiding the singlet from the LEP Higgs boson search in the second case.
In the NMSSM, if $\kappa=0$ and $m_S^2=0$, the doublet Higgs VEVs spontaneously break an approximate scale symmetry \eqref{eq:scale}.
The resultant pseudo Nambu-Goldstone boson (mass eigenstate) almost consists
of the doublets and corresponds to the SM-like Higgs
with a suppressed singlet component.
There is another scale symmetry if $\kappa=0$ and $m_{H_u}^2=0$ in which $S$ and $H_u$ are interchanged. The corresponding Nambu-Goldstone boson is the singlet-like Higgs boson with suppressed $H_u$ component.
The mixing with the suppressed components must break the two symmetries and disappear if $H_d$ decouples. Thus it is highly suppressed although the symmetries are broken at the EW scale.
%Our model inherits this feature even for $m_S^2 \simeq m_Z^2$ 
%thanks to the relation $\mu \approx A_\lambda/\tan\beta$ with 
%$m_{H_d}^2 = A_\lambda^2$.
To evaluate the fine-tuning, we introduced a simple formula for the fine-tuning measure having a clear physical meaning which calls for careful treatment of the radiative corrections. 
Using it, we performed a comprehensive analysis of the fine-tuning in the EW symmetry breaking which shows the model is tuned at most ${\it O}(10)\%$ for $M_0=1.5$ TeV. 
For moderate $\tan\beta$, the least tuned region is realized with the singlet Higgs boson mass around $100$ GeV which overlaps with the anomaly found in the LEP Higgs boson search.  

We further investigated the coupling constants of the Higgs bosons in our model. 
The radiative corrections from the SUSY particles and heavy Higgs bosons 
are highly suppressed due to the little hierarchy. 
Thus the deviation from the SM occurs through the suppressed mixing between the SM-like and singlet Higgs bosons. 
The numerical calculation showed that $20$ \% deviations are possible
 in the squares of the scale factors, $(\kappa_X^i)^2$
 and the sum rule $(\kappa^1_X)^2+(\kappa^2_X)^2 =1$
 holds well except for those of $b$ and $\tau$. 
The mixing with the heavy doublet is also important for $b$ and $\tau$
due to $\tan\beta$ enhancement and an ${\it O}(1)$ deviation is possible 
for their couplings to the SM-like Higgs boson.
Future precision measurements at the LHC and the ILC will provide
 good opportunities to explore these deviations and
 the hidden singlet-like scalar.

We also studied prospects for the direct/indirect dark matter searches
 assuming the minimal dark matter sector in our model.
The spin-independent neutralino-nucleus scattering is mainly mediated by
 the SM-like and singlet-like Higgs bosons. 
The two amplitudes interfere destructively in some parameter regions.
This suppresses the cross section below the current experimental bound.
While the spin-dependent scattering is dominated by the $Z$ boson exchange
and does not show such a feature. The surviving regions
 from various phenomenological constraints are almost excluded
 for $\tan\beta=3$ and on the verge of exclusion for $\tan\beta=10$.
The suppression of the spin-independent cross section in the surviving 
region is characteristic signature to identify the model.    
The bound on the neutralino annihilation cross section from
 the gamma ray emission by the dwarf spheroidal galaxy is 
 an order above the prediction and will soon start to explore our model.

\appendix

\section{Soft SUSY breaking terms}
\label{app:soft}

Here we give explicitly soft SUSY breaking terms induced 
by the mirage mediation mechanism in the NMSSM.

In the mirage mediation, 
the soft parameters at the scale  just  below  $M_{GUT}$ are given by
\begin{eqnarray}
M_a(M_{GUT}) &=& M_0 + \frac{m_{3/2}}{8\pi^2} b_a g_a^2, \nonumber \\
A_{ijk}(M_{GUT}) &=& (c_i + c_j + c_k)M_0 - (\gamma_i + \gamma_j + \gamma_k)\frac{m_{3/2}}{8\pi^2}, \nonumber \\
m_i^2(M_{GUT}) &=& c_i M_0^2 - \dot{\gamma_i}(\frac{m_{3/2}}{8\pi^2})^2
				- \frac{m_{3/2}}{8\pi^2} M_0 \theta_i,
\end{eqnarray}
where
\begin{eqnarray}
b_a &=& -3\mathrm{tr}(T_a^2(\mathrm{Adj})) + \sum_i \mathrm{tr}(T_a^2(\phi^i)), \nonumber \\
\gamma_i &=& 2\sum_a g_a^2 C_2^a(\phi^i) - \frac{1}{2} \sum_{jk} |y_{ijk}|^2, \nonumber \\
\theta_i &=& 4\sum_a g_a^2 C_2^a(\phi^i) - \sum_{jk} a_{ijk} |y_{ijk}|^2, \nonumber \\
\dot{\gamma_i} &=& 8\pi^2 \frac{d\gamma_i}{d \ln \mu_R}.
\end{eqnarray}
Here, $T_a^2(\mathrm{Adj})$ and $T_a^2(\phi^i)$  denote 
Dynkin indices of the adjoint representation and the representation 
of matter fields $\phi^i$.
We have assumed
$\omega_{ij}=\Sigma_{kl}y_{ijk} y_{jkl}^*$ to be diagonal.

Within the framework of the NMSSM, the $\beta$-function coefficients, anomalous 
dimensions and other coefficients in the above equations are obtained as 
\begin{eqnarray}
b_3 &=& -3,\  b_2 = 1,\  b_1 = 11, \nonumber \\
\gamma_{H_u} &=& \frac{3}{2}g_2^2 + \frac{1}{2}g_1^2 - 3y_t^2 - \lambda^2,  \nonumber \\
\gamma_{H_d} &=& \frac{3}{2}g_2^2 + \frac{1}{2}g_1^2  - \lambda^2,  \nonumber \\
\gamma_{S} &=& -2\kappa^2 - 2\lambda^2,  \nonumber \\
\gamma_{Q_a} &=& \frac{8}{3}g_3^2 + \frac{3}{2}g_2^2 + \frac{1}{18}g_1^2 - (y_t^2 + y_b^2) \delta_{3a},  \nonumber \\
\gamma_{U_a} &=& \frac{8}{3}g_3^2 + \frac{8}{9}g_1^2 - 2y_t^2 \delta_{3a},  \nonumber \\
\gamma_{D_a} &=& \frac{8}{3}g_3^2 + \frac{2}{9}g_1^2 - 2y_b^2 \delta_{3a},  \nonumber \\
\gamma_{L_a} &=& \frac{3}{2}g_2^2 + \frac{1}{2}g_1^2 - y_{\tau}^2 \delta_{3a}, \nonumber \\
\gamma_{E_a} &=& 2g_1^2 - 2y_{\tau}^2 \delta_{3a},
\end{eqnarray}

\begin{eqnarray}
\theta_{H_u} &=& 3g_2^2 + g_1^2 - 6y_t^2 a_{H_uQ_3U_3^c} -2\lambda^2a_{H_uH_dS},  \nonumber \\
\theta_{H_d} &=& 3g_2^2 + g_1^2 - 6y_b^2 a_{H_dQ_3D_3^c} - 2y_{\tau}^2 a_{H_dL_3E_3^c} -2\lambda^2a_{H_uH_dS},  \nonumber \\
\theta_{S} &=& -2\lambda^2a_{H_uH_dS} -\kappa^2 a_{SSS},  \nonumber \\
\theta_{Q_a} &=& \frac{16}{3}g_3^2 + 3g_2^2 + \frac{1}{9}g_1^2 - 2(y_t^2 a_{H_uQ_3U_3^c} + y_b^2 a_{H_dQ_3D_3^c}) \delta_{3a}, \nonumber \\
\theta_{U_a} &=& \frac{16}{3}g_3^2 + \frac{16}{9}g_1^2 - 4y_t^2 a_{H_uQ_3U_3^c}\delta_{3a},  \nonumber \\
\theta_{D_a} &=& \frac{16}{3}g_3^2 + \frac{4}{9}g_1^2 - 4y_b^2 a_{H_dQ_3D_3^c}\delta_{3a}, \nonumber \\
\theta_{L_a} &=& 3g_2^2 + g_1^2 - 2y_{\tau}^2 a_{H_dL_3E_3^c}\delta_{3a}, \nonumber \\
\theta_{E_a} &=& 4g_1^2 - 4y_{\tau}^2 a_{H_dL_3E_3^c}\delta_{3a},
\end{eqnarray}

\begin{eqnarray}
\dot{\gamma}_{H_u} &=& \frac{3}{2}g_2^4 + \frac{11}{2}g_1^4 - 3y_t^2 b_{y_t} -\lambda^2 b_\lambda,  \nonumber \\
\dot{\gamma}_{H_d} &=& \frac{3}{2}g_2^4 + \frac{11}{2}g_1^4 - 3y_b^2 b_{y_b} - y_{\tau}^2 b_{y_\tau} -\lambda^2 b_\lambda, \nonumber \\
\dot{\gamma}_{S} &=& -2\kappa^2 b_\kappa - 2\lambda^2 b_\lambda,  \nonumber \\
\dot{\gamma}_{Q_a} &=& -8g_3^4 + \frac{3}{2}g_2^4 + \frac{11}{18}g_1^4 - (y_t^2 b_{y_t} + y_b^2 b_{y_b}) \delta_{3a},  \nonumber \\
\dot{\gamma}_{U_a} &=& -8g_3^4 + \frac{88}{9}g_1^4 - 2y_t^2 b_{y_t} \delta_{3a},  \nonumber \\
\dot{\gamma}_{D_a} &=& -8g_3^4 + \frac{22}{9}g_1^4 - 2y_b^2 b_{y_b} \delta_{3a},  \nonumber \\
\dot{\gamma}_{L_a} &=& \frac{3}{2}g_2^4 + \frac{11}{2}g_1^4 - y_{\tau}^2 b_{y_\tau} \delta_{3a}, \nonumber \\
\dot{\gamma}_{E_a} &=& 22g_1^4 - 2y_{\tau}^2 b_{y_\tau} \delta_{3a},
\end{eqnarray}
where
\begin{eqnarray}
b_{y_t} &=& -\frac{16}{3}g_3^2 -3g_2^2 - \frac{13}{9}g_1^2 + 6y_t^2 + y_b^2 + \lambda^2,  \nonumber \\
b_{y_b} &=& -\frac{16}{3}g_3^2 -3g_2^2 - \frac{7}{9}g_1^2 + y_t^2 + 6y_b^2 + y_\tau^2 + \lambda^2, \nonumber \\
b_{y_\tau} &=& -3g_2^2 - 3g_1^2 + 3y_b^2 + 4y_\tau^2 + \lambda^2,  \nonumber \\
b_{\kappa} &=&  6\lambda^2 + 6\kappa^2,  \nonumber \\
b_{\lambda} &=& -3g_2^2 - g_1^2 + 3y_t^2 + 3y_b^2 + y_\tau^2 + 4\lambda^2 + 2\kappa^2.
\end{eqnarray}
Here, $Q_a, U_a, D_a, L_a,$  and  $E_a$ denote left-handed quark, right-handed up-sector quark, 
right-handed down-sector quark, left-handed lepton, and right-handed lepton fields, respectively, 
and the index $a$ denotes the generation index.
We have included effects due to Yukawa couplings, 
$y_t, y_b,$ and $y_\tau$, only for the third generations.

\acknowledgments
%\section*{Acknowledgments}

We would like to thank Kiwoon Choi very much for helpful comments and discussions.
T.~K. is supported in part by a Grant-in-Aid for Scientific Research No.~25400252  from the Ministry of 
Education, Culture, Sports, Science and Technology of Japan. 
K.~O. is supported in part by a Grant-in-Aid for Scientific Research No.~21740155 and No.~18071001 from the MEXT of Japan. 
T.~S. is partially supported by a Grant-in-Aid for Young Scientists (B) No.~23740190 and a Grant-in-Aid for Scientific Research No.~15K17645 from the MEXT Japan.
Numerical computation in this work was partly carried out at the Yukawa Institute Computer Facility.

\end{document}